\documentclass[12pt]{article}

\usepackage{amsfonts,amssymb,epsfig,amsmath}
\usepackage{caption,subcaption}
\usepackage{color}

\renewcommand{\baselinestretch}{1.2}

\def\det{{\rm det}}

\newcommand{\be}{\begin{eqnarray}}
\newcommand{\ee}{\end{eqnarray}}
\newcommand{\nn}{\nonumber}
\newcommand{\bn}{\begin{enumerate}}
\newcommand{\en}{\end{enumerate}}

\begin{document}

\makeatletter \@addtoreset{equation}{section} \makeatother
\renewcommand{\theequation}{\thesection.\arabic{equation}}
\renewcommand{\thefootnote}{\alph{footnote}}

\begin{titlepage}

\begin{center}
\hfill {\tt KIAS-P18003}\\
\hfill {\tt SNUTP17-004}\\

\vspace{2cm}

{\Large\bf 6d strings and exceptional instantons}

\vspace{2cm}

\renewcommand{\thefootnote}{\alph{footnote}}

{\large Hee-Cheol Kim$^{1,2}$, Joonho Kim$^3$, Seok Kim$^4$,
Ki-Hong Lee$^4$ and Jaemo Park$^2$}

\vspace{0.7cm}

\textit{$^1$Jefferson Physical Laboratory, Harvard University, Cambridge,
MA 02138, USA.}\\

\vspace{0.2cm}

\textit{$^2$Department of Physics, Postech, Pohang 790-784, Korea.}\\

\vspace{0.2cm}

\textit{$^3$School of Physics, Korea Institute for Advanced Study,
Seoul 130-722, Korea.}\\

\vspace{0.2cm}

\textit{$^4$Department of Physics and Astronomy \& Center for
Theoretical Physics,\\
Seoul National University, Seoul 151-747, Korea.}\\

\vspace{0.7cm}

E-mails: {\tt  heecheol1@gmail.com, joonhokim@kias.re.kr, \\
skim@phya.snu.ac.kr, khlee11812@gmail.com, jaemo@postech.ac.kr}

\end{center}

\vspace{1cm}

\begin{abstract}

We propose new ADHM-like methods to compute the Coulomb branch
instanton partition functions of 5d and 6d
supersymmetric gauge theories, with certain
exceptional gauge groups or exceptional matters. We study $G_2$ theories with
$n_{\bf 7}\leq 3$ matters in ${\bf 7}$,
and $SO(7)$ theories with $n_{\bf 8}\leq 4$ matters in the spinor representation
${\bf 8}$. We also study the elliptic genera of self-dual
instanton strings of 6d SCFTs with exceptional gauge groups or matters, including
all non-Higgsable atomic SCFTs with rank $2$ or $3$ tensor branches. Some of them
are tested with topological vertex calculus. We also explore a D-brane-based method
to study instanton particles of 5d $SO(7)$ and $SO(8)$ gauge
theories with matters in spinor representations, which
further tests our ADHM-like proposals.

\end{abstract}

\end{titlepage}

\renewcommand{\thefootnote}{\arabic{footnote}}

\setcounter{footnote}{0}

\renewcommand{\baselinestretch}{1}

\tableofcontents

\renewcommand{\baselinestretch}{1.2}

\section{Introduction}

Instantons are semi-classical representations of nonperturbative
quantum phenomena. In Yang-Mills gauge theories, instantons
given by self-dual gauge fields on $\mathbb{R}^4$ play important roles in
various contexts. For gauge theories in higher dimensions, $d>4$, they
can be solitonic objects rather than vacuum tunneling,
being particles in 5d and strings in 6d.

Self-dual Yang-Mills instanton solutions have continuous parameters,
which form a moduli space. Understanding the moduli space
dynamics is often an important step towards better understanding
the physics of instantons. For gauge theories with classical gauge groups,
the self-dual solutions and the moduli space
dynamics can be described by the ADHM formalism
\cite{Atiyah:1978ri}. The ADHM formalism provides gauge theories
on the worldvolume of instantons (1d for particles, 2d for strings), which at
low energy reduce to the nonlinear sigma models on the instanton moduli space.
In string theory, such gauge theories come from open fundamental strings
between D$p$-D$(p\!+\!4)$ branes.
This is why only classical gauge groups admit such constructions. Including
matters to gauge theories
also affects the moduli space dynamics of instantons. Open strings
can engineer matters in the fundamental or rank $2$ product
representations. This gives the notion of
`classical matters,' whose
inclusion to the instanton moduli space dynamics still admits ADHM description.
But matters in other representations, like higher rank product representations
or spinor representations of $SO(N)$, cannot be engineered using open strings.
We shall call them `exceptional matters.'

Yang-Mills instantons also play crucial roles in
supersymmetric gauge theories. Among others, in 4d $\mathcal{N}=2$
theories, the Seiberg-Witten solution \cite{Seiberg:1994rs}
in the Coulomb branch acquires all order multi-instanton contributions.
It can be microscopically derived
by computing the instanton partition function \cite{Nekrasov:2002qd}
on $\mathbb{R}^4$ with the so-called Omega deformation. The uplifts of this partition
function to 5d $\mathcal{N}=1$ gauge theories on $\mathbb{R}^4\times S^1$,
and to 6d $\mathcal{N}=(1,0)$ gauge theories on $\mathbb{R}^4\times T^2$,
are also important observables of 5d/6d SCFTs. The computation of
this partition function in \cite{Nekrasov:2002qd} relies on the
ADHM method, applicable only to classical gauge theories. However, exceptional
gauge theories are also important in various situations. For instance,
one often uses various $(p,q)$ 7-branes wrapped on 2-cycles to engineer
6d $\mathcal{N}=(1,0)$ SCFTs, which admit exceptional gauge groups and matters
rather generically.

In this paper, we develop ADHM-like formalisms of instantons for a
small class of exceptional gauge groups or matters, which can be used to study
the instanton partition functions in the Coulomb phase. Namely, we provide
1d/2d gauge theories which we suggest to describe
certain aspects of exceptional instanton particles and strings.
We managed to find such formalisms
for $G_2$ theories with matter hypermultiplets in ${\bf 7}$, and for
$SO(7)$ theories with exceptional matters in the spinor representation
${\bf 8}$.
We also expect their 0d reductions to describe exceptional instantons of
4d gauge theories, although we do not study them here. In 5d, we can describe
$G_2$ theories
with $n_{\bf 7}\leq 3$ matters in ${\bf 7}$, and $SO(7)$ instantons with
$n_{\bf 8}\leq 4$ matters in ${\bf 8}$. In 6d, gauge anomaly cancelations
restrict our set-up to $n_{\bf 7}=1$ and $n_{\bf 8}=2$.

Our constructions have the following features.
The correct instanton moduli space has $G_r$ isometry for gauge group
$G_r$ of rank $r$. However, our gauge theories realize the symmetry only in
a subgroup $H_r$ with same rank. We always take $H_r$ to be a classical
group, say $H_r=SU(r+1)$. The moduli space of $G_r$ instantons contains
that of $H_r\subset G_r$ instantons as a subspace.
Our 1d/2d gauge theories only realize the latter correctly.
Away from this subspace, only the dimensions of moduli space agree.
So we \textit{do not expect} that our gauge theories capture
the full moduli space dynamics of the exceptional $G_r$ instantons.
In the Coulomb branch, the instanton size and the $G_r$ gauge orbit parts of
the moduli space are lifted to a set of isolated points (which are non-degenerate
after Omega deformation). We propose that our gauge theories
correctly compute the Coulomb branch observables of $G_r$ instantons. In fact
the SUSY partition functions of our models in the Coulomb branch
exhibit full $G_r$ symmetry. More precisely, since we turn on $r$ Coulomb
VEVs, we find that the Weyl symmetry of $H_r$ enhances to that of $G_r$.
Since the UV gauge theory flows to the non-linear sigma model on the
moduli space, this is a kind of IR symmetry enhancement.

Our ADHM-like descriptions are found based on trials-and-errors. We
start from $H_r$ classical ADHM construction, and add more worldvolume
matters and interactions to suit the physics.
In particular, they are motivated by \cite{Kim:2016foj},
where 2d gauge theories were found
for the instanton strings of non-Higgsable 6d $SU(3)$ gauge theories.
Although $SU(3)$ is a classical group, 6d non-Higgsable $SU(3)$ gauge
theory cannot be engineered by using just D-branes. Rather, it is engineered by
using mutually non-local 7-branes wrapping a 2-cycle, in precisely the
same way as engineering exceptional gauge theories. Incidently, the naive ADHM
description for $SU(3)$ instantons is sick,
by having 2d gauge anomalies. The correct gauge theories for the
$SU(3)$ instanton strings were found in \cite{Kim:2016foj}. Its reduction
to 1d provides a novel alternative ADHM-like description for $SU(3)$ instantons.
The ADHM-like descriptions of this paper extend these results,
related by Higgsings $SO(7)\stackrel{\textrm{VEV of }{\bf 8}}{\longrightarrow}
G_2\stackrel{\textrm{VEV of }{\bf 7}}{\longrightarrow} SU(3)$.

In 6d, $G_2$ and $SO(7)$ theories with matters are somewhat important, as
they appear in the `atomic' constituents of 6d SCFTs \cite{Heckman:2013pva,Heckman:2015bfa}.\footnote{There are closely related 
but slightly different notions in the literature, such as `atomic' SCFTs, 
`minimal' SCFTs, non-Higgsable clusters,
and so on. We are not very careful about the distinctions here.}
Roughly speaking, in the list of atomic 6d SCFTs, there are $9$
of them with rank $1$ tensor branches. They are constructed by putting F-theory
on an elliptic Calabi-Yau 3-fold, whose base is given by the
$O(-n)\rightarrow\mathbb{P}^1$ bundle with $n=1,2,\cdots,8,12$
\cite{Morrison:1996na,Morrison:1996pp,Witten:1996qb}.
Also, there are $3$ more atomic SCFTs with higher rank tensor branches
called `32,' `322' and `232' \cite{Morrison:2012np}.
Constructing more complicated 6d SCFTs
is basically forming `quivers' of these atoms
\cite{Morrison:2012np,Heckman:2015bfa,Heckman:2013pva}.
The last $3$ atomic SCFTs with higher rank tensor branches contain
$G_2$ or $SO(7)$ gauge theories with half-hypermultiplets
in ${\bf 7}$, ${\bf 8}$, respectively. See our section 4 for a summary.
Our descriptions of $G_2$ and $SO(7)$ instantons allow us
to study the instanton strings of such SCFTs.
Collecting recent studies and new findings of this paper,
one now has 2d gauge theory methods to study the strings of the following
6d atomic SCFTs: $O(-1)$ theory \cite{Kim:2014dza,Kim:2015fxa}, $O(-2)$ theory
\cite{Haghighat:2013gba}, $O(-3)$ theory \cite{Kim:2016foj}, $O(-4)$ theory
\cite{Haghighat:2014vxa}, and all three higher rank theories (this paper).
Quivers of these 2d theories are also
explored in \cite{Haghighat:2013gba,Gadde:2015tra,Kim:2015fxa,Kim:2016foj}.
Among others, these gauge theories can be used to study the elliptic genera
of the strings. These elliptic genera were also studied
using other approaches, including the topological vertex methods
\cite{Hayashi:2017jze} and the modular bootstrap-like
approach \cite{DelZotto:2016pvm,DelZotto:2017mee,klp}.

We test the BPS spectra of our ADHM-like models using various
alternative approaches. Among others, in section 3, we develop a
D-brane-based approach to study exceptional instanton particles in 5d
gauge theories. This approach is applicable to
$G_2$ instantons, and $SO(7)$ or $SO(8)$ instantons with matters
in spinor representations. For $G_2$ and $SO(7)$ cases, the Witten indices
computed from this approach test our ADHM-like
proposals. Also, a 5d description for the circle compactified
`$232$' SCFT \cite{Morrison:2012np} has been found in \cite{Hayashi:2017jze},
using 5-brane webs. This allows
us to compute the BPS spectrum of its strings using topological vertices.
We do this calculus in section 4 and find agreement
with our new gauge theories.

We leave a small technical remark on $SO(7)$ instantons.
The canonical ADHM description for $k$ $SO(7)$ instantons (without matters in
${\bf 8}$) is given by an $Sp(k)$ gauge theory. Its partition function is
given by a contour integral \cite{Nekrasov:2004vw,Hwang:2014uwa}, yielding a
complicated residue sum. Unlike $SU(N)$ instanton partition functions,
in which case closed form expressions for the residue sums are
known \cite{Nekrasov:2002qd}, such expressions have been unknown for $SO(N)$.
In this paper, using our alternative ADHM-like description,
we find a closed form residue sum expression for $SO(7)$.

The rest of this paper is organized as follows. In section 2, we sketch the
basic ideas. We then present the ADHM-like descriptions of instantons for $SO(7)$
theories with matters in ${\bf 8}$, and $G_2$ theories with matters in ${\bf 7}$.
In section 3, we explore alternative D-brane descriptions to study certain
exceptional instanton particles, and use them to test our proposals.
In section 4, we construct the 2d gauge theories for the strings of 6d atomic
SCFTs with rank $2$ or $3$ tensor branches, with $G_2\times SU(2)$,
$G_2\times SU(2)\times\{\}$, $SU(2)\times SO(7)\times SU(2)$ gauge groups.
We test the last one with topological vertices. Section 5 concludes with
various remarks.

We note that a paper \cite{hkly} closely related to some part of our work
will appear.

\section{Exceptional instanton partition functions}

Our proposal is based on the following ideas:
(1) We are interested in the Coulomb phase partition functions of
exceptional instantons, not in the symmetric phase.
(2) In the Coulomb phase, the instanton moduli space is lifted by massive
parameters, to saddle points lying within the moduli space of
instantons with classical subgroups.
(3) Thus we only seek for a formalism to study the massive fluctuations around
the last saddle points, accomplished by extending ADHM formalisms
for classical instantons. We elaborate on these ideas in some detail.

\hspace*{-0.65cm}{\bf \underline{Coulomb phase}:}
We are interested in the gauge theory
in the Coulomb branch. Suppose that the
gauge group $G_r$ has rank $r$. We turn on
nonzero VEV $v$ of the scalar in the vector multiplet, which
breaks $G_r$ to $U(1)^r$. In 6d, vector multiplet does not contain
scalars. In this case, we consider the theory compactified on circle, with nonzero
holonomy playing the role of Coulomb VEV. In the symmetric phase, instantons
develop a moduli space, part of which being gauge orientations and instanton
sizes. In the Coulomb
phase, there appears nonzero potential on the instanton moduli space, proportional to
$v^2$. This potential lifts the size and orientation 0-modes.
There are extra $4k$ position moduli of $k$ instantons on $\mathbb{R}^4$,
which will also be lifted in the Omega background.
The moduli space is then completely lifted
to points. So we expect that it suffices to understand the quantum dynamics of
instantons near these points.

\begin{table}[t!]
\begin{center}
$$
\begin{array}{c|c|c}
	\hline
    G_r&H_r&\textrm{branching rules}\\
    \hline
    G_2&SU(3)&{\bf 14}\rightarrow{\bf 8}\oplus{\bf 3}\oplus\overline{\bf 3}\ ,\
    {\bf 7}\rightarrow{\bf 3}\oplus\overline{\bf 3}+{\bf 1}\\
    \hline F_4&SO(9)&{\bf 52}\rightarrow{\bf 36}\oplus{\bf 16}\ ,\
    {\bf 26}\rightarrow{\bf 1}\oplus{\bf 9}\oplus{\bf 16}\\
    \hline E_7&SU(8)&{\bf 133}\rightarrow{\bf 63}\oplus{\bf 70}\ ,\
    {\bf 56}\rightarrow{\bf 28}\oplus\overline{\bf 28}\\
    \hline E_8&SU(9)&{\bf 248}\rightarrow{\bf 80}\oplus{\bf 84}\oplus\overline{\bf 84}\\
    \hline E_8&SO(16)&{\bf 248}\rightarrow{\bf 120}\oplus{\bf 128}\\
    \hline\hline
    SO(7)&SU(4)&{\bf 21}\rightarrow{\bf 15}\oplus{\bf 6}\ ,\
    {\bf 8}\rightarrow{\bf 4}\oplus\overline{\bf 4}\\
    \hline
\end{array}
$$
\caption{Possible choices of $H_r$ for various $G_r$, when $H_r$
is a simple group}\label{subgroup}
\end{center}
\end{table}
\hspace*{-0.65cm}{\bf \underline{ADHM on a subspace}:}
The second idea is that one can use the ADHM formalism of instantons
when $G_r$ is a classical group. In $d$ dimensional gauge theory, the ADHM
formalism can be understood as a $d-4$ dimensional gauge theory living on the
instanton solitons. For classical $G_r$, the low energy moduli space of $d-4$
dimensional gauge theory is the
instanton moduli space, so one expects in IR to get non-linear sigma
models on the instanton moduli space. When $G_r$ is exceptional, no such
formalisms are known. However, it is often possible to find a classical
subgroup $H_r\subset G_r$ of the given exceptional group $G_r$ with same rank.
Then, we try to describe
the (massive) quantum fluctuations around the saddle points by expanding
the $H_r$ ADHM formalism, adding more $d-4$ dimensional fields.
This is where we need educated guesses, in the spirit of model buildings.
We want a subgroup $H_r$ with same rank as $G_r$, partly because
we wish our formalism to see all $U(1)^r$ in the Coulomb
phase. Possible $G_r$ and $H_r$ are given
in Table \ref{subgroup}, when $H_r$ is a simple group.
To study `exceptional matters' of
$SO(7)$, we shall also consider $H=SU(4)$ for $G=SO(7)$.

For example, consider the case with $H_{r=N-1}=SU(N)$. The $SU(N)$
ADHM description of $k$ instantons has $U(k)$ gauge symmetry,
and the following fields,
\begin{eqnarray}
  {\rm chiral}&:&(q,\psi)\in({\bf k},\overline{\bf N})\ ,\ \ (\tilde{q},\tilde\psi)
  \in(\bar{\bf k},{\bf N})\ ,\ \ (a,\Psi),\ (\tilde{a},\tilde\Psi)\in({\bf adj},{\bf 1})
  \nonumber\\
  {\rm vector}\sim{\rm Fermi}&:&(A_\mu,\lambda_0)\in({\bf adj},{\bf 1})\ ,\ \
  (\lambda)\in({\bf adj},{\bf 1})\ .
\end{eqnarray}
The fields are organized into 2d $\mathcal{N}=(0,2)$
supermultiplets, and we have shown the representations in $U(k)\times SU(N)$.
Fields in a parenthesis denote bosonic/fermionic ones in a
multiplet, while $(\lambda)$ denotes a Fermi multiplet.
These fields combine to $\mathcal{N}=(0,4)$ vector multiplet
and hypermultiplets. The instanton moduli space is obtained from the scalar fields,
subject to the complex ADHM constraint and the D-term constraint
(real ADHM constraint)
\begin{equation}\label{SUN-ADHM}
  q\tilde{q}+[a,\tilde{a}]=0\ ,\ \
  qq^\dag-\tilde{q}^\dag\tilde{q}+[a,a^\dag]+[\tilde{a},\tilde{a}^\dag]=0\ ,
\end{equation}
and after modding out by the $U(k)$ gauge orbit. More precisely, the non-linear
sigma model on the instanton moduli space is obtained from the gauged linear
sigma model at low energy.
This part is the standard ADHM construction of $SU(N)$ instantons.
Now we should add extra light fields, including
more scalars to describe $G_r$ instantons' extra moduli.
$d$ dimensional vector multiplet in $G_r$ decomposes in $H_r$ as
\begin{equation}
  {\bf adj}(G)\rightarrow{\bf adj}(H)\bigoplus_i{\bf R}_i(H)\ ,
\end{equation}
where ${\bf R}_i(H)$ are suitable representations of $H_r$
in Table \ref{subgroup}. Vector multiplet in
${\bf adj}(H)$ induces the standard instanton moduli, described in
UV by the above ADHM description. Vector multiplets in ${\bf R}_i$
introduce further moduli, whose real dimension is $4kT({\bf R}_i)$.
$T({\bf R})$ is the Dynkin index of ${\bf R}$.
When ${\bf R}_i$ is a fundamental representation or rank $2$
product representations, we managed to find the extra
fields. We are technically motivated by the mathematical
constructions of \cite{Shadchin:2005mx}, but will simply present them
as our `ansatz' for the UV uplift of these zero modes.
From Table \ref{subgroup}, one finds that ${\bf R}_i$'s are product
representations with ranks less than or equal to $2$ only for
$G_2\supset SU(3)$ and $SO(7)\supset SU(4)$.
For these, the adjoint representations of $G_r$ decompose as
\begin{eqnarray}\label{rank-2}
  SU(3)\subset G_2&:&{\bf 14}\rightarrow{\bf 8}\oplus{\bf 3}\oplus\bar{\bf 3}
  ={\bf 8}\oplus{\bf 3}\oplus{\bf anti}({\bf 3}\otimes{\bf 3})\nonumber\\
  SU(4)\subset SO(7)&:&{\bf 21}\rightarrow{\bf 15}\oplus{\bf 6}\ .
\end{eqnarray}
We shall present extra chiral and Fermi multiplets with suitable interactions
in the next subsections, which extends the moduli space in
$\sum_i 4kT({\bf R}_i)$ new directions.

\hspace*{-0.65cm}{\bf \underline{When it fails}:} We made similar trials with
other exceptional gauge groups and matters, which failed.
It may be worthwhile to briefly report the reasons of failure.
A typical reason is that the UV theory has extra branch of
moduli space that does not belong to our instanton moduli space at low energy.
Namely, apart from the exceptional instanton's moduli space,
one sometimes has extra branch which cannot be lifted by supersymmetric
potentials.

For instance, we tried to extend the ADHM construction of $H=SU(8),SU(9)$ to
get those for $G=E_7,E_8$.
In these cases, ${\bf R}_i$'s are product representations
of rank $4$ and $3$, respectively. We made several trials to realize the
extra moduli with right dimensions, especially without unwanted extra
branches of moduli space which may spoil the instanton calculus.
We however failed to get the precise descriptions, despite finding models
which partly exhibit the right physics of instantons and instanton strings.
See section 5 for more discussions.

We also suspect that some choice of $H_r\subset G_r$ may miss certain
small instanton saddle points. To clearly understand this issue, we should
find more examples than we have now.

There are simpler examples in which our new formalism fails.
For instance, we consider our alternative
$SO(7)$ ADHM (section 2.1) and try to add zero modes
from matters in ${\bf 7}$ (vector representation).
Zero modes of ${\bf 7}$ are well known in the standard
$SO(N)$ ADHM, which form an $Sp(k)$ fundamental Fermi multiplet.
In our $SU(4)\times U(k)$ formalism, ${\bf 7}$ is regarded as
${\bf 7}\rightarrow{\bf 6}+{\bf 1}$. We can ignore the singlet if
the gauge orientation of instantons is along $SU(4)$.
${\bf 6}$ is the rank $2$ anti-symmetric representation.
According to \cite{Shadchin:2005mx}, and in D-brane engineerings, the ADHM
fields induced by matters in bulk anti-symmetric representation include
scalars in rank $2$ symmetric representation of $U(k)$. This creates an
extra branch of moduli space which is unphysical in the instanton calculus,
but is present only in the UV uplifts. Even in ADHM models
engineered by string theory, there are often such extra branches.
In \cite{Hwang:2014uwa}, the contributions from these branches
are factored out, mainly guided by string theory. However, including
this extra branch in our $SO(7)$ ADHM-like model, we find it difficult to properly
identify and separate the extra contributions. Similarly, we cannot do an ADHM-like
calculus for the 5d $G_2$ $\mathcal{N}=1^\ast$ theory.

Now we explain examples that turn out to work.

\subsection{$SO(7)$ instantons and matters in ${\bf 8}$}

The adjoint representation of $SO(7)$ decomposes in $SU(4)$
as ${\bf 21}\rightarrow{\bf 15}+{\bf 6}$. We first
seek for an alternative ADHM-like formalism of pure $SO(7)$ instantons,
extending $SU(4)$ ADHM.
We explain it as the quantum mechanics of instanton particles
in 5d $\mathcal{N}=1$ Yang-Mills theory.

The quantum mechanics for $k$ $SU(4)$ instantons has $U(k)$ gauge symmetry.
It has following fields: $\mathcal{N}=(0,4)$ $U(k)$ vector multiplet, consisting
of 1d reduction of 2d gauge fields $A_\mu=(A_0,\varphi=A_1)$, and fermions $\lambda_0,\lambda$; hypermultiplets with bosonic fields
$q_i,\tilde{q}^i$ in $({\bf k},\bar{\bf 4})+(\bar{\bf k},{\bf 4})$,
where $i=1,2,3,4$; hypermultiplets with bosonic fields
$a,\tilde{a}$ in $({\bf adj},{\bf 1})$. In IR, one imposes
\begin{equation}
  D\sim qq^\dag-\tilde{q}^\dag\tilde{q}+[a,a^\dag]+[\tilde{a},\tilde{a}^\dag]=0
  \ ,\ \ J_\lambda\sim q\tilde{q}+[a,\tilde{a}]=0
\end{equation}
by D-term or $J$-term potentials in the $\mathcal{N}=(0,2)$
language. See, \cite{Witten:1993yc,Tong:2014yna,Kim:2016foj} for the notations and
reviews. These constraints and modding out by $U(k)$ gauge orbit eliminate
$2k^2$ complex variables from $2k^2+8k$ components of
$q,\tilde{q},a,\tilde{a}$. So one finds
$8k$ complex moduli.

With extra vector multiplet fields in ${\bf 6}$,
there are extra bosonic zero modes.
A vector multiplet in rank $2$ antisymmetric representation of $SU(N)$ induces
$2kT({\bf anti}_2)=k(N\!-\!2)$ complex bosonic zero modes in $k$ instanton background.
So we should add extra fields in UV and modify interactions, to
get extra $2k$ complex bosonic modes at $N=4$.
We find that the following extra fields, taking
the forms of $\mathcal{N}=(0,2)$ chiral or Fermi multiplets, yield
the right physics\footnote{Technically, we took the equivariant index
of the so-called universal bundle \cite{Shadchin:2005mx},
and made an antisymmetrized product of two of them (for ${\bf 6}$):
this is the character for the fields we list in (\ref{SO7-extra}). Since
we lack physical explanations of this procedure, we just present them as
our ansatz for the UV theory.} (only bosonic fields shown for the chiral
multiplets):
\begin{eqnarray}\label{SO7-extra}
  \textrm{chiral }\phi_i&:&(\bar{\bf k},\bar{\bf 4})_{J=\frac{1}{2}}
  \nonumber\\
  \textrm{chiral }b,\tilde{b}&:&(\overline{\bf anti}_2,{\bf 1})_{J=\frac{1}{2}}\nonumber\\
  \textrm{Fermi }\hat\lambda&:&({\bf sym}_2,{\bf 1})_{J=0}\nonumber\\
  \textrm{Fermi }\check{\lambda}&:&({\bf sym}_2,{\bf 1})_{J=-1}\ .
\end{eqnarray}
${\bf sym}_2$, ${\bf anti}_2$ denote rank $2$ (anti-)symmetric representations
of ${\bf k}$, and the charge $J$
in the subscript will be explained shortly.
We introduced extra $4k+2\cdot\frac{k^2-k}{2}$ complex bosonic fields.
Using the extra Fermi fields $\hat\lambda$, $\check\lambda$, we introduce the
following interactions. As noted in \cite{Kim:2016foj}, the desired interactions
should be non-holomorphic in the chiral multiplet fields, which is possible
only with $\mathcal{N}=(0,1)$ SUSY. Therefore, we regard all these fields
as $(0,1)$ superfields, as explained in \cite{Kim:2016foj}, and turn on
the following $\mathcal{N}=(0,1)$ superpotential,
\begin{equation}\label{SO7-(0,1)}
  J^{(0,1)}_{\hat\lambda}\sim 
  (\phi_i q^{i\dag})_S+(ba^\dag+\tilde{b}\tilde{a}^\dag)_S
  \ ,\ \ J^{(0,1)}_{\check\lambda}\sim (\phi_i\tilde{q}^i)_S+
  (\tilde{b}a-b\tilde{a})_S\ .
\end{equation}
The subscripts $S$ denote symmetrization of the $k\times k$ matrices.
We want (\ref{SO7-(0,1)}) to be the only source of breaking $(0,2)$ SUSY to
$(0,1)$ in the classical action. The D-term is given by
\begin{equation}
  D\sim qq^\dag-\tilde{q}^\dag\tilde{q}-\phi^\dag\phi+[a,a^\dag]
  +[\tilde{a},\tilde{a}^\dag]-2b^\dag b-2\tilde{b}^\dag\tilde{b}\ .
\end{equation}
Then, since $|J|^2$ appear in the bosonic potential for each $J$,
one imposes at low energy extra $k^2+k$ complex constraints from the new
superpotentials. Collecting all, one finds
\begin{equation}
  3\cdot 4k+2\cdot k^2+2\cdot\frac{k^2-k}{2}-2k^2-2\cdot\frac{k^2+k}{2}
  =10k
\end{equation}
complex bosonic zero modes. This agrees with the dimension $2kc_2$ of
$SO(7)$ instanton moduli space, where $c_2(SO(2N\!+\!1))=2N\!-\!1$.
The $SU(4)$ ADHM quantum mechanics with extra charged fields given by
(\ref{SO7-extra}) is our proposed ADHM-like formalism for $SO(7)$ instantons.

\begin{table}[t!]
\centering\vspace{0.3cm}
\begin{tabular}{c|cccc}
\hline fields  & $U(k)$ & $SU(4)$ & $U(1)_J$ & $SU(2)_l$ \\ \hline
$(q_i,\tilde{q}^i)$  & $({\bf k},\bar{\bf k})$ & $(\bar{\bf 4},{\bf 4})$
& $\frac{1}{2}$ & ${\bf 1}$ \\
$(a,\tilde{a})$ & ${\bf adj}$ & ${\bf 1}$ & $\frac{1}{2}$ & ${\bf 2}$ \\
$(\lambda_0,\lambda)$ & {\bf adj} & ${\bf 1}$ & $(0,-1)$ & ${\bf 1}$ \\
\hline
$\phi_i$  & $\bar{\bf k}$ & $\bar{\bf 4}$ & $\frac{1}{2}$ & ${\bf 1}$ \\
$(b,\tilde{b})$ & $\overline{\bf anti}_2$ & ${\bf 1}$ & $\frac{1}{2}$ & ${\bf 2}$ \\
$(\hat\lambda,\check\lambda)$ & ${\bf sym}_2$ & ${\bf 1}$ & $(0,-1)$ & ${\bf 1}$ \\
\hline
\end{tabular}
\caption{Charges/representations of fields in our $SO(7)$ ADHM-like model}
\label{SO7-symmetry}
\end{table}
We explain the symmetries of this model. All symmetries we explain below
are compatible with the superpotentials. It first has $SU(4)$
symmetry. There is also a $U(1)_J$ symmetry, whose charges $J$ we already
listed above when we introduced fields. There is also $SU(2)_l$,
which rotates $a,\tilde{a}$ and also $b,\tilde{b}$ as doublets.
The charges and representations are summarized in Table \ref{SO7-symmetry}.
This system has only $\mathcal{N}=(0,1)$ supersymmetry and $SU(4)$ global
symmetry in UV. We assert that they enhance to $(0,4)$ SUSY and $SO(7)$
in IR, when we compute the Coulomb branch partition
functions. $SO(7)$ enhancement will be visible as $SO(7)$ character
expansions of the partition functions. In the context of SUSY
enhancement, we claim that $U(1)_J$ enhances to
$SU(2)_r\times SU(2)_R$, where $SO(4)=SU(2)_r\times SU(2)_l$ rotates
the spatial $\mathbb{R}^4$ on which particles can move, and $SU(2)_R$ is the
5d R-symmetry. $J$ is identified as $J=\frac{J_r+J_R}{2}$, where $J_r,J_R$ are
the Cartans of $SU(2)_r$, $SU(2)_R$. One Cartan is not visible in UV.
The index of our models will agree with different
computations whose settings manifestly preserve $(0,4)$ SUSY.

We study the moduli space, with and without 5d Coulomb VEV.
For technical reasons, let us just consider the case with $k=1$.
At $k=1$, the fields $b,\tilde{b}$ are absent, and
$a,\tilde{a}$ are free fields for the center-of-mass motion.
First consider the symmetric phase at $v=0$.
One should solve the following equations:
\begin{equation}
  |q_i|^2-|\tilde{q}^i|^2-|\phi_i|^2=0
  \ ,\ \ q_i\tilde{q}^i=0\ ,\ \ \phi^{\dag i} q_i=0, \ \
  \tilde{q}^i\phi_i=0\ .
\end{equation}
At $\phi_i\!=\!0$, this is the equation for the $SU(4)$ instantons.
This subspace is the cone over $SU(4)/U(2)$. Away from $\phi_i\!=\!0$,
although the dimension of the moduli space is same as the relative moduli
space of an $SO(7)$ instanton, the two moduli spaces are different.
The proper $SO(7)$ instanton moduli space is
the cone over the $SO(7)/(SU(2)\times SO(3))$ coset,
whose metric is given by the homogeneous metric. However, we find
no $SO(7)$ isometry on our moduli space.

In the Coulomb branch and with the Omega deformation,
the moduli space lifts to isolated points, on the
$SU(4)\subset SO(7)$ instanton moduli space. To see this,
we expand the studies of \cite{Kim:2011mv}.
The Coulomb VEV $v_i$ ($i=1,\cdots,4$)
satisfying $\sum_i v_i=0$ couples to the 1d fields as follows.
Let us denote by $\varphi\equiv A_1$ the scalar in
the 1d vector multiplet. $v$ is a traceless diagonal matrix, with
eigenvalues $v_i$. Nonzero $v$ changes the coupling to $\varphi$
as follows,
\begin{equation}
  |\varphi q|^2+|\tilde{q}\varphi|^2+|\phi\varphi|^2\rightarrow
  |\varphi q-qv|^2+|v\tilde{q}-\tilde{q}\varphi|^2+|v\phi+\phi\varphi|^2\ .
\end{equation}
This is because $\varphi$, $v$ are scalars in the
1d vector multiplet of $U(k)\times SU(4)$, where $v$ is a background field,
and fields couple to them according to
their representations in $U(k)\times SU(4)$. Note that the relative $-$
signs for $q,\tilde{q}$ appear because they are in the bifundamental representations
$({\bf k},\bar{\bf 4})$ or its conjugate, while relative $+$ sign for $\phi$ is
because it is in $(\bar{\bf k},\bar{\bf 4})$. We set
all complete square terms to zero at energies lower than 1d gauge coupling.
One should also minimize the following D-term potential at $k=1$,
\begin{equation}
  V\leftarrow \left(|q|^2-|\tilde{q}|^2-|\phi|^2-\xi\right)^2\ .
\end{equation}
Here, since we have a $U(k)$ gauge theory, we have turned on a Fayet-Iliopoulos (FI) parameter, $\xi$, which we take to be positive $\xi>0$ for
convenience. $\xi$ can also be taken to be negative, without changing the
Coulomb phase partition function, as we shall see below. However, physics is
easier to interpret with $\xi>0$. So we set (at $k=1$)
\begin{equation}\label{eigensystem}
  v_iq_i=\varphi q_i\ ,\ \
  v_i\tilde{q}^i=\varphi \tilde{q}^i\ ,\ \
  v_i\phi_i=-\varphi\phi_i
\end{equation}
where $i=1,2,3,4$ indices are not summed over, and
\begin{equation}\label{SO7-FI}
  |q_i|^2-|\tilde{q}^i|^2-|\phi_i|^2=\xi>0\ .
\end{equation}
The equations (\ref{eigensystem}) are eigenvector equations for the matrix
$v$, whose eigenvalues are $\varphi$ for $q,\tilde{q}$ to be nonzero,
and $-\varphi$ for $\phi$ to be nonzero. From (\ref{SO7-FI}), one should
have $q_i\neq 0$, which means that $\varphi$ is set equal to one of the
$v_i$'s. Then one can have nonzero $q_i$ at the saddle point, whose value is
tuned to meet (\ref{SO7-FI}). At generic values of $v_i$'s, one should set
$\phi_i=0$, meaning that we are forced to stay in the $SU(4)$ instanton
moduli space.\footnote{At this stage, $\tilde{q}^i$ can also
be nonzero by solving the same eigenvector equation as $q_i$. However, as shown in
the appendix of \cite{Kim:2011mv}, the eigenvector equations for
$q_i$ and $\tilde{q}^i$ become different with nonzero Omega background
parameter. Therefore, in the fully Omega-deformed
background, only $q_i$ is nonzero.} So in the Coulomb branch calculus,
$\phi$ provides massive degrees of freedom living on the $SU(4)$
instanton moduli space.


The Witten index of the quantum mechanics preserving $(0,4)$ SUSY is defined by
\begin{equation}\label{index-definition}
  Z_k(\epsilon_{1,2},v_i)={\rm Tr}_k\left[(-1)^F
  e^{-\epsilon_1(J_1+J_R)}e^{-\epsilon_2(J_2+J_R)}
  e^{-v_iq_i}e^{-m_aF_a}\right]\ ,
\end{equation}
where trace is over states in the $k$ instanton sector. $J_1,J_2$ are the two
Cartans of $SO(4)$ which rotate the spatial $\mathbb{R}^4$, where they rotate
mutually orthogonal $\mathbb{R}^2$ factors. They are related to
$J_{l,r}$ by $J_r=\frac{J_1+J_2}{2}$, $J_l=\frac{J_1-J_2}{2}$. $J_R$ is the Cartan of
$SU(2)_R$ coming from the 5d R-symmetry.
Note that only the combination $J_r+J_R=2J$ appears, so our UV model can fully
detect them.
$q_i$ are the $r$ electric charges in $U(1)^r\subset G_r$, which is $SO(7)$ here.
$F_a$ denote other flavor symmetries, which is absent now but introduced for
later purpose. The measures are
chosen to commute with two Hermitian supercharges
$Q^{A\dot\alpha}=Q^{+\dot{-}},Q^{-\dot{+}}$.
See, e.g. \cite{Hwang:2014uwa} for the notations.
These two supercharges are mutually Hermitian conjugate, which we write as
$Q,Q^\dag$. They form a pair of fermionic oscillators, pairing a set of bosonic
and fermionic states. Such a pair of states is not counted in the index, as
their contributions cancel due to the factor $(-1)^F$. Such a Hilbert
space interpretation will hold with as little as $(0,2)$ SUSY. In our UV
$(0,1)$ system, we abstractly interpret the partition function as a
SUSY path integral of the Euclidean QFT on $T^2$. $1$ Hermitian SUSY in UV
is enough to derive the formula for $Z_k$ available in the literatures.
With IR SUSY enhancement, $Z_k$ acquires the interpretation of an index.

For gauge theories, this index can be evaluated by a residue sum
\cite{Hwang:2014uwa,Hori:2014tda,Cordova:2014oxa} (see also
\cite{Benini:2013nda,Benini:2013xpa}). The formula was discussed in the
context of $(0,2)$ theories, but it applies with 1 Hermitian supercharge
as well \cite{Kim:2016foj}. In our model, the contour integral
takes the following form\footnote{The overall signs of $Z_k$ are
fixed by requiring agreement with the index for the $Sp(k)$ ADHM theory \cite{Nekrasov:2004vw}.}:
\begin{eqnarray}\label{SO7-integral}
  Z_{k}&=&\frac{1}{k!}\oint\prod_{I=1}^k\frac{d\phi_I}{2\pi i}\cdot
  \frac{\prod_{I\neq J}2\sinh\frac{\phi_{IJ}}{2}\cdot
  \prod_{I,J}2\sinh\frac{2\epsilon_+-\phi_{IJ}}{2}}
  {\prod_{I=1}^k\prod_{i=1}^4 2\sinh\frac{\epsilon_+\pm(\phi_I-v_i)}{2}
  \cdot\prod_{I,J}2\sinh\frac{\epsilon_{1,2}+\phi_{IJ}}{2}}
  \nonumber\\
  &&\times\frac{\prod_{I\leq J}\left(2\sinh\frac{\phi_I+\phi_J}{2}
  \cdot 2\sinh\frac{\phi_I+\phi_J-2\epsilon_+}{2}\right)}
  {\prod_I\prod_i2\sinh\frac{\epsilon_+-\phi_I-v_i}{2}
  \cdot\prod_{I<J}2\sinh\frac{\epsilon_{1,2}-\phi_I-\phi_J}{2}}\ .
\end{eqnarray}
Here, $\phi_{IJ}\equiv\phi_I-\phi_J$, and $2\sinh$ factors with repeated
signs or subscripts (like $\pm$ or $\epsilon_{1,2}$) are all multiplied.
The $SU(4)$ chemical potentials satisfy $\sum_{i=1}^4v_i=0$. We also used
$\epsilon_\pm\equiv\frac{\epsilon_1\pm\epsilon_2}{2}$.
The integrand on the first line comes from the
$SU(4)$ ADHM fields $q,\tilde{q},a,\tilde{a}$ and $U(k)$ vector multiplet fermions.
The second line comes from the extra fields.

The integral can be performed as follows.
The nonzero residue contributing to $Z_k$ is called
the JK residue. To define this, one first picks up
an auxiliary vector $\eta$ in the $k$ dimensional charge space (`conjugate' to
the integral variables $\phi_I$). Possible poles in the integrand are given by
hyperplanes of the form $\rho_\alpha\cdot\phi+\cdots =0$, where the expression on the
left hand side comes from the argument of the sinh factors
$2\sinh\frac{\rho_\alpha\cdot\phi+\cdots}{2}$ in the denominator of
(\ref{SO7-integral}).
One can in general pick $d(\geq k)$ charge vectors
$\rho_\alpha$, $\alpha=1,\cdots,d$ and hyperplanes to specify a pole.
In our systems, all relevant poles satisfy $d=k$.
With chosen $\eta$, JK-Res may be nonzero only if $\eta$ is
spanned by the $k$ charge vectors $\rho_{1},\rho_{2},\cdots,\rho_{k}$
with positive coefficients. Here, the choice
$\eta=(1,\cdots,1)$ simplifies the evaluation \cite{Hwang:2014uwa}.
Since the charges appearing in the denominator of the second line
are all negative in (\ref{SO7-integral}), one can show (combined with
the fact that charges on the first line take the form of $e_I$ or
$e_I-e_J$) that JK-Res should always be zero by definition if one of the charges
from the second line are chosen in $\rho_\alpha$. This implies that the poles
with nonzero residues are always chosen from the first line only, which are
already classified in \cite{Nekrasov:2002qd,Flume:2002az,Bruzzo:2002xf,Hwang:2014uwa}.
The pole locations for $\phi_I$ are classified by
the colored Young diagrams with $k$ boxes, meaning a collection of $4$
Young diagrams $Y=(Y_1,\cdots,Y_4)$ whose box numbers sum to $k$.
Let us denote by $s=(m,n)$ the box of a Young diagram $Y_i$, which is the
box on the $m$'th row and $n$'th column of $Y_i$. $s$ running over possible
$k$ boxes replaces $I=1,\cdots,k$ index of $\phi_I$.
We specify the pole location associated with $Y$ as $\phi(s)$. The result is
\cite{Nekrasov:2002qd,Flume:2002az,Bruzzo:2002xf,Hwang:2014uwa}
\begin{equation}\label{phi(s)}
  \phi(s)=v_i-\epsilon_+-(n-1)\epsilon_1-(m-1)\epsilon_2\ \ ,\ \ \
  s=(m,n)\in Y_i\ \ \ (i=1,\cdots,4)\ .
\end{equation}
(This corrects a typo in \cite{Hwang:2014uwa}, exchanging $m\leftrightarrow n$.)
Had there been only the first line in (\ref{SO7-integral}), the residues
were computed in \cite{Flume:2002az,Bruzzo:2002xf,Hwang:2014uwa}.
Plugging in $\phi(s)$ into the second line of (\ref{SO7-integral}),
one obtains an extra factor for each residue.
The residue sum is given by
\begin{eqnarray}\label{SO7-residue}
  Z_k&=&\sum_{\vec{Y};|\vec{Y}|=k}\prod_{i=1}^4\prod_{s\in Y_i}
  \frac{2\sinh(\phi(s))\cdot 2\sinh(\phi(s)-\epsilon_+)}
  {\prod_{j=1}^4 2\sinh\frac{E_{ij}(s)}{2}\cdot
  2\sinh\frac{E_{ij}(s)-2\epsilon_+}{2}\cdot 2\sinh\frac{\epsilon_+-\phi(s)-v_j}{2}}\nonumber\\
  &&\times \prod_{i\leq j}^4\prod_{s_{i,j}\in Y_{i,j};s_i<s_j}
  \frac{2\sinh\frac{\phi(s_i)+\phi(s_j)}{2}\cdot
  2\sinh\frac{\phi(s_i)+\phi(s_j)-2\epsilon_+}{2}}
  {2\sinh\frac{\epsilon_{1,2}-\phi(s_i)-\phi(s_j)}{2}}
\end{eqnarray}
where
\begin{equation}\label{Eij}
  E_{ij}(s)=v_i-v_j-\epsilon_1h_i(s)+\epsilon_2(v_j(s)+1)\ .
\end{equation}
Here and below, $s_i\!<\!s_j$ means ($i<j$) or ($i\!=\!j$ and $m_i\!<\!m_j$) or
($i\!=\!j$ and $m_i\!=\!m_j$ and $n_i\!<\!n_j$).
$h_i(s)$ denotes the distance from $s$ to
the right end of the diagram $Y_i$ by moving right. $v_j(s)$ denotes the
distance from $s$ to the bottom of the diagram $Y_j$ by moving down.
See, e.g. \cite{Kim:2011mv}. (\ref{SO7-residue}) is our proposal for the
partition function of $k$ $SO(7)$ instantons. This is quite novel for
the following reason. $SO(7)$ instantons have standard ADHM formulation,
using $Sp(k)$ gauge theories for $k$ instantons.
The pole classification is unknown for the $Sp(k)$ index.
On the other hand, (\ref{SO7-residue}) is an explicit formula.

Before adding matters in ${\bf 8}$,
we first check that (\ref{SO7-residue}) is indeed the
correct $SO(7)$ instanton partition function.
We checked the equivalence of (\ref{SO7-integral}), or (\ref{SO7-residue}),
and the index of $Sp(k)$ ADHM gauge theory
\cite{Nekrasov:2004vw,Hwang:2014uwa}, up to $k\leq 3$ (turning off all
chemical potentials except $\epsilon_+$ at $k=3$).
Here we explain the case with $k=1$ in detail, which
is already nontrivial. For the purpose of illustration, we
directly start from the contour integral. At $k=1$, one finds
\begin{eqnarray}
  \left(2\sinh\frac{\epsilon_{1,2}}{2}\right)Z_1
  =\oint d\phi\frac{2\sinh\epsilon_+}{\prod_{i=1}^4 2\sinh\frac{\epsilon_+\pm(\phi-v_i)}{2}}
  \cdot\frac{2\sinh\phi\cdot 2\sinh(\phi-\epsilon_+)}
  {\prod_{i=1}^42\sinh\frac{\epsilon_+-\phi-v_i}{2}}
\end{eqnarray}
from our model. Taking the residues at $\phi=v_i-\epsilon$, for $\eta>0$,
one finds
\begin{equation}
  \left(2\sinh\frac{\epsilon_{1,2}}{2}\right)Z_1=\sum_{i=1}^4\frac{1}
  {\prod_{j(\neq i)}2\sinh\frac{v_{ij}}{2}\cdot 2\sinh\frac{2\epsilon_+-v_{ij}}{2}}
  \cdot\frac{2\sinh(2\epsilon_+-v_i)}
  {\prod_{j(\neq i)}2\sinh\frac{2\epsilon_+-v_i-v_j}{2}}\ .
\end{equation}
This is a special case of (\ref{SO7-residue}).
To check this result is correct, we study the $SO(7)$ single instanton
partition function obtained from the standard $Sp(1)$ ADHM formalism
\cite{Nekrasov:2004vw,Hwang:2014uwa}
\begin{equation}
  \left(2\sinh\frac{\epsilon_{1,2}}{2}\right)Z_1^{\rm standard}=
  \frac{1}{2}\oint d\phi\frac{2\sinh\epsilon_+\cdot 2\sinh(\epsilon_+\pm \phi)
  \cdot 2\sinh(\pm \phi)}
  {\prod_{a=1}^3 2\sinh\frac{\epsilon_+\pm \phi\pm u_a}{2}\cdot
  2\sinh\frac{\epsilon_+\pm \phi}{2}}\ .
\end{equation}
Residues are taken at $\phi=\pm u_a-\epsilon_+$ and $-\epsilon_+$ for $\eta>0$,
but the last residue is $0$. $u_a$ and $v_i$ are related by
\begin{equation}\label{SO7SU4}
  v_1=\frac{u_1+u_2+u_3}{2}\ ,\ \ v_2=\frac{u_1-u_2-u_3}{2}\ ,\ \
  v_3=\frac{-u_1+u_2-u_3}{2}\ ,\ \ v_4=\frac{-u_1-u_2+u_3}{2}\ .
\end{equation}
The residue sum is given by
\begin{equation}
  \left(2\sinh\frac{\epsilon_{1,2}}{2}\right)
  Z_1^{\rm standard}=\frac{1}{2}\sum_{a=1}^3\sum_{s=\pm}\frac{2\cosh\frac{su_a}{2}
  \cdot 2\cosh\frac{2\epsilon_+-su_a}{2}\cdot 2\sinh(\pm(su_a-\epsilon_+))}
  {2\sinh(su_a)\cdot 2\sinh(\epsilon_+\!-\!su_a)
  \prod_{b(\neq a)}2\sinh\frac{su_a\pm u_b}{2}\cdot
  2\sinh\frac{2\epsilon_+-su_a\pm u_b}{2}}.
\end{equation}
Despite very different looks, one can show (say, by using computer) that
\begin{equation}
  Z_1(v_i)=Z_1^{\rm standard}(u_a)
\end{equation}
after the identification (\ref{SO7SU4}). This identity and similar ones at higher
$k$'s imply that $Z_k$ exhibits
$SU(4)\rightarrow SO(7)$ enhancement,
since $Z_k^{\rm standard}$ has manifest $SO(7)$
Weyl symmetry.

Now we discuss the inclusion of ADHM fields coming from the hypermultiplet
matters in ${\bf 8}$. We continue to study the instanton particles of 5d SYM.
${\bf 8}$ decomposes in $SU(4)$ as
\begin{eqnarray}
  {\bf 8}\rightarrow{\bf 4}+\bar{\bf 4}\ .
\end{eqnarray}
In the original ADHM formalism of $SO(7)$ instantons, it is unclear how to
UV-uplift the fermion zero modes caused by these hypermultiplets in the
instanton background. One may even feel it impossible, since the
standard $SO(7)$ ADHM cannot see $4\pi$ rotations in $Spin(7)$.
However, viewing it as $SU(4)$ instantons with certain
extensions, each
hypermultiplet in ${\bf 4}$ (or $\bar{\bf 4}$) induces a Fermi multiplet
which is fundamental (or anti-fundamental) in $U(k)$. So in our new
description, we naturally guess that the effect of $n_{\bf 8}$ hypermultiplets
is adding $n_{\bf 8}$ pairs of Fermi multiplets of
the following form:
\begin{equation}
  \Psi_a,\tilde\Psi_a\ :\ ({\bf k},{\bf 1})+
  (\bar{\bf k},{\bf 1})\ \ \ \ (a=1,\cdots, n_{\bf 8})\ .
\end{equation}
It has been known \cite{Intriligator:1997pq} that the 5d $SO(7)$ SYM has a UV
completion to a 5d SCFT for $n_{\bf 8}\leq 4$.
Recently, it was discussed that 5d SCFTs can exist till $n_{\bf 8}\leq 6$
\cite{Zafrir:2015uaa}. See also \cite{Jefferson:2017ahm}. Our construction
provides good descriptions of instantons for $n_{\bf 8}\leq 4$.
It will be easiest to explain this point after we discuss the index below.
The flavor symmetry for $\Psi_a,\tilde{\Psi}_a$
may naively appear to be $U(2n_{\bf 8})$. This is because
we do not have any superpotential for these Fermi fields. They interact
with other fields through gauge coupling only, so that one can rotate
$\Psi_a,\tilde\Psi_a^\dag$ with $U(2n_{\bf 8})$. However, these fermions can
couple to 5d background bulk fields, including the
hypermultiplet fields in ${\bf 8}$. (Even in ADHM models based on D-brane
engineerings, it sometimes happens that the soliton quantum mechanics is
ignorant on the bulk symmetry, in a similar manner.)
These couplings will only preserve
$U(n_{\bf 8})\subset Sp(n_{\bf 8})$.
See the beginning of the next subsection for this coupling
to the bulk fields.

Adding these fermions, our ADHM-like description
can be easily generalized. Namely, the extra Fermi fields are given
standard kinetic term, whose derivatives are
covariantized with 1d $U(k)$ vector multiplet fields.
Its Witten index $Z_k^{n_{\bf 8}}(\epsilon_{1,2},v_i,m_a)$ with $a=1,\cdots,n_{\bf 8}$ is defined with extra factors $e^{-m_aF_a}$ inserted in its definition, where
$F_a$ are the Cartans of $Sp(n_{\bf 8})$.
The contour integral expression for the Witten index takes
the form of (\ref{SO7-integral}), with the following extra integrand multiplied
for the new Fermi fields:
\begin{equation}\label{SO7-spinor-integrand}
  \prod_{a=1}^{n_{\bf 8}}\prod_{I=1}^k 2\sinh\frac{m_a+\phi_I}{2}
  \cdot 2\sinh\frac{m_a-\phi_I}{2}\ .
\end{equation}
The extra factor (\ref{SO7-spinor-integrand})
does not create new poles at finite $\phi$, but may create new
poles at infinity $\phi_I\rightarrow\pm\infty$.
We first discuss the last possibility.

Here, first note that $\phi_I$ originate
from the eigenvalues of the 1d $U(k)$ vector multiplet fields,
$\phi\equiv\varphi+iA_\tau$, where $A_\tau$ is the vector potential
on the Euclidean time.
The contour integrand $Z_{\textrm{1-loop}}$ comes from 1-loop
path integral of 1d fields in the background of constant $\phi_I$. So
$V(\phi)\sim -\log Z_{\textrm{1-loop}}$ is the 1-loop potential
energy for $\phi_I$. Before multiplying (\ref{SO7-spinor-integrand}),
the integrand of (\ref{SO7-integral}) converges
to zero at $|\phi_I|\rightarrow\infty$ for any $I$, since there are
more bosonic fields than fermionic fields. More concretely, consider
the case with $k\!=\!1$. One obtains
$Z_{\textrm{1-loop}}^{n_{\bf 8}=0}\sim e^{-4|\phi|}$, implying that
the linear potential $V(\phi)=4|\phi|$ confines the eigenvalues
to $\phi=0$. In other words, although $\phi$ classically develops
a continuum to $\phi\rightarrow\pm\infty$,
1-loop effect lifts this continuum by an attractive force.
In ADHM models with brane engineering, this can be visualized as the instantons
being attracted to the locations of 5d SCFTs \cite{Hwang:2014uwa}.
The $U(k)$ vector multiplet fields are clearly extra degrees of freedom
that enter while making a UV completion of the nonlinear sigma model.
If there is a continuum created by $\phi$,
this represents states that do not belong to 5d QFT.
The confinement from $V(\phi)=4|\phi|$ signals that such obvious extra
states may not be present in the quantum system.

Now we extend these studies to $n_{\bf 8}>0$. At $k=1$, one obtains
$V(\phi)=(4-n_{\bf 8})|\phi|$. So at $n_{\bf 8}\leq 3$, the quantum potential
still confines the instanton. At $n_{\bf 8}=4$, $\phi$ generates a flat direction.
This branch has extra states which is an artifact of UV completion, not
belonging to the 5d QFT Hilbert space. So strictly speaking, $n_{\bf 8}\leq 3$ is
the bound in which our ADHM-like model is reliable. Fortunately, there are
well developed empirical ways of factoring out such extra states' contribution
to the index. So we believe that our approach will be useful
till $n_{\bf 8}=4$. At $n_{\bf 8}\geq 5$,
the quantum potential is repulsive, and it is not clear whether one can use
this theory to study 5d QFT at all. (However, see \cite{Yun:2016yzw}
for some progress.) In the contour integral like (\ref{SO7-integral}) or
its extension with (\ref{SO7-spinor-integrand}), the absence of continuum means
the absence of poles at infinity. This implies that
the choice of $\eta$ in the JK-residue evaluation does not change
the final result \cite{Hwang:2014uwa,Hori:2014tda}. This is the case
for $n_{\bf 8}\leq 3$.

For $n_{\bf 8}\leq 3$, the pole classification that
we explained earlier for pure $SO(7)$ instantons still holds, labeled by
$SU(4)$ colored young diagrams. We only need to multiply the value of (\ref{SO7-spinor-integrand})
at the pole to the residue. The resulting index is given by
\begin{eqnarray}\label{SO7-spinor-residue}
  Z_k&=&\sum_{\vec{Y};|\vec{Y}|=k}\prod_{i=1}^4\prod_{s\in Y_i}
  \frac{2\sinh(\phi(s))\cdot 2\sinh(\phi(s)-\epsilon_+)}
  {\prod_{j=1}^4 2\sinh\frac{E_{ij}(s)}{2}\cdot
  2\sinh\frac{E_{ij}(s)-2\epsilon_+}{2}\cdot 2\sinh\frac{\epsilon_+-\phi(s)-v_j}{2}}\\
  &&\times \prod_{i\leq j}^4\prod_{s_{i,j}\in Y_{i,j};s_i<s_j}
  \frac{2\sinh\frac{\phi(s_i)+\phi(s_j)}{2}\cdot 2\sinh\frac{2\epsilon_+-\phi(s_i)-\phi(s_j)}{2}}
  {2\sinh\frac{\epsilon_{1,2}-\phi(s_i)-\phi(s_j)}{2}}
  \cdot\prod_{i=1}^4\prod_{s\in Y_i}\prod_{a=1}^{n_{\bf 8}}
  2\sinh\frac{m_a\pm\phi(s)}{2}\ .\nonumber
\end{eqnarray}
The partition functions (\ref{SO7-spinor-residue}) will be tested in sections
3 and 4 at $n_{\bf 8}=1,2$ using alternative descriptions, which include no guess
works but are more elaborate in calculations.
For instance, the indices at $k=1$ divided by the center-of-mass factor
$\hat{Z}_1\equiv \left(2\sinh\frac{\epsilon_{1,2}}{2}\right)Z_1$ are given by
\begin{eqnarray}
  \hat{Z}_1^{n_{\bf 8}=1}&=&\prod_{i<j} \frac{t^2}{(1-t^2 b_i^\pm b_j^\pm)}
  \left[\chi_{\bf 2}^{Sp(1)}f_0(v)+f_1(v)\right]\label{SO7-Nf=1}\\
  \hat{Z}_1^{n_{\bf 8}=2}&=&\prod_{i<j} \frac{t^2}{(1-t^2 b_i^\pm b_j^\pm)}
  \left[\chi_{\bf 5}^{Sp(1)}f_0(v)+\chi_{\bf 4}^{Sp(2)}f_1(v)+f_2(v)\right]
  \label{SO7-Nf=2}\\
  \hat{Z}_1^{n_{\bf 8}=3}&=&\prod_{i<j} \frac{t^2}{(1-t^2 b_i^\pm b_j^\pm)}
  \left[\chi_{\bf 14^\prime}^{Sp(3)}f_0(v)+\chi_{\bf 14}^{Sp(3)}f_1(v)
  +\chi_{\bf 6}^{Sp(3)}f_2(v)+f_3(v)\right]\label{SO7-Nf=3}
\end{eqnarray}
where
\begin{eqnarray}\label{SO-expansion}
  f_0(v)&=&\chi_{\bf 9}^{SU(2)} + \chi_{\bf 7}^{SU(2)} (\chi_{\bf 7}^{SO(7)} + 1) + \chi_{\bf 5}^{SU(2)} (-\chi_{\bf 35}^{SO(7)} +\chi_{\bf 7}^{SO(7)} + 1)\\
  &&+ \chi_{\bf 3}^{SU(2)} (-\chi_{\bf 35}^{SO(7)}+\chi_{\bf 27}^{SO(7)} + 1) + \chi_{\bf 105}^{SO(7)} - \chi_{\bf 21}^{SO(7)} + \chi_{\bf 7}^{SO(7)}\nonumber\\
  f_1(v)&=&-\chi_{\bf 8}^{SU(2)} \chi_{\bf 8}^{SO(7)}
  - \chi_{\bf 6}^{SU(2)} \chi_{\bf 8}^{SO(7)} + \chi_{\bf 4}^{SU(2)} \chi_{\bf 112}^{SO(7)}	 - \chi_{\bf 2}^{SU(2)} \chi_{\bf 168}^{SO(7)}\nonumber\\
  f_2(v)&=&\chi_\mathbf{7}^{SU(2)}\chi_\mathbf{35}^{SO(7)} -\chi_\mathbf{5}^{SU(2)}\big(\chi_\mathbf{7}^{SO(7)}\!-\!\chi_\mathbf{35}^{SO(7)}\!
   +\!\chi_\mathbf{105}^{SO(7)}\big)
   -\chi_\mathbf{3}^{SU(2)}\big(\chi_\mathbf{21}^{SO(7)}+\chi_\mathbf{27}^{SO(7)}-
   \chi_\mathbf{35}^{SO(7)}\nonumber\\
   &&-\chi_\mathbf{77}^{SO(7)}+\chi_\mathbf{168^\prime}^{SO(7)}+1\big)
   -\chi_\mathbf{7}^{SO(7)}+\chi_\mathbf{21}^{SO(7)}
   +\chi_\mathbf{27}^{SO(7)}-\chi
   _\mathbf{105}^{SO(7)}+\chi_\mathbf{189}^{SO(7)}+\chi_\mathbf{330}^{SO(7)}\nonumber\\
   f_3(v)&=& - \chi_{\bf 112^\prime}^{SO(7)}\chi_{\bf 6}^{SU(2)}+ (\chi_{\bf 48}^{SO(7)}-\chi_{\bf 112^\prime}^{SO(7)}+\chi_{\bf 512}^{SO(7)})
   \chi_{\bf 4}^{SU(2)} - (\chi_{\bf 112^\prime}^{SO(7)}+\chi_{\bf 448}^{SO(7)})
   \chi_{\bf 2}^{SU(2)}\ .\nonumber
\end{eqnarray}
Here $t\equiv e^{-\epsilon_+}$. $\chi_{\bf R}^{SU(2)}$ is the character
of ${\rm diag}[SU(2)_R\times SU(2)_r]$ in representation ${\bf R}$,
in the convention $\chi_{\bf 2}^{SU(2)}=t+t^{-1}$.
$b_i\equiv e^{-v_i}$, and $(1-t^2b_{i}^\pm b_j^\pm)$ means that
all $4$ factors with different signs are multiplied. 
The convention on representations (e.g. primes) all follows 
\cite{Feger:2012bs}. The numerators are invariant under
$SO(7)\times Sp(n_{\bf 8})$ Weyl symmetry, being character sums. Since the
denominators are products with all possible $\pm$ signs, they are also invariant
under $SO(7)$ Weyl group which flips $b_i\rightarrow b_i^{-1}$. So
$\hat{Z}_1^{n_{\bf 8}}$ is invariant under the Weyl group
of $SO(7)\times Sp(n_{\bf 8})$.

We expect our quantum mechanics to work also
at $n_{\bf 8}=4$. Here, the 1d Coulomb branch with nonzero $\phi_I$ has
a continuum. There may appear extra contribution from
this continuum to the index \cite{Hwang:2014uwa}, apart from (\ref{SO7-spinor-residue}).
(For conceptual simplicity, we consider the problem at zero FI parameter $\xi=0$.)
The extra contribution from the 1d Coulomb continuum
is neutral in $SO(7)\times Sp(4)$. This is because the extra states in the 1d Coulomb branch come from the region
with large $\phi_I$, where all $U(k)$ charged
fields acquire large masses. The charged fields are those which see
$SO(7)\times Sp(4)$. So the extra continuum does not see these charges.
Here we shall only test the $SO(7)\times Sp(4)$ symmetry enhancements at
$n_{\bf 8}=4$. So we simply ignore the extra contribution, and show that
(\ref{SO7-spinor-residue}) exhibits $SO(7)\times Sp(4)$ Weyl symmetry.
The result at $k=1$,
showing $SO(7)\times Sp(4)$ Weyl symmetry, is given by
\begin{equation}
  \hspace*{-.3cm}
  \hat{Z}_1^{n_{\mathbf{8}}=4}=\prod_{i<j} \frac{t^2}{(1-t^2 b_i^\pm b_j^\pm)}
  \left[\chi_{\bf 42}^{Sp(4)}f_0(v)+\chi_{\bf 48}^{Sp(4)}f_1(v)
  +\chi_{\bf 27}^{Sp(4)}f_2(v)+\chi_{\bf 8}^{Sp(4)}f_3(v)
  +f_4(v)\right]
\end{equation}
with $f_{0,1,2,3}(v)$ given by (\ref{SO-expansion}), and
\begin{eqnarray}
  \hspace*{-1cm}f_4(v)\!&\!=\!&\!-\!\chi_{\bf 13}^{SU(2)}
  \!+\!\chi_{\bf 11}^{SU(2)} (\chi_{\bf 21}^{SO(7)}\!-\!\chi_{\bf 7}^{SO(7)})
  +\chi_{\bf 9}^{SU(2)} (\chi_{\bf 7}^{SO(7)}\!-\!
  \chi_{\bf 21}^{SO(7)}\!-\!\chi_{\bf 27}^{SO(7)}\!+\!\chi_{\bf 35}^{SO(7)}
  \!+\!\chi_{\bf 105}^{SO(7)}\!-\!\chi_{\bf 189}^{SO(7)}\!-\!1)\nonumber \\
  \hspace*{-1.5cm}&&
  -\chi_{\bf 7}^{SU(2)} \big(2 \chi_{\bf 7}^{SO(7)}-2 \chi_{\bf 21}^{SO(7)}-2 \chi_{\bf 27}^{SO(7)}+\chi_{\bf 35}^{SO(7)}+\chi_{\bf 77}^{SO(7)}+2 \chi_{\bf 105}^{SO(7)}-\chi_{\bf 168^\prime}^{SO(7)}-\chi_{\bf 189}^{SO(7)}
  -\chi_{\bf 294}^{SO(7)}\nonumber \\
  \hspace*{-1.2cm}&&
  -\chi_{\bf 330}^{SO(7)}+\chi_{\bf 378}^{SO(7)}-1\big)
  -\chi_{\bf 5}^{SU(2)} \big(-3 \chi_{\bf 7}^{SO(7)}+3 \chi_{\bf 21}^{SO(7)}+3
  \chi_{\bf 27}^{SO(7)}-\chi_{\bf 35}^{SO(7)}
  -2 \chi_{\bf 77}^{SO(7)}-4 \chi_{\bf 105}^{SO(7)}\nonumber \\
  \hspace*{-1.2cm}&&
  +2 \chi_{\bf 168^\prime}^{SO(7)}+\chi_{\bf 182}^{SO(7)}
  +2 \chi_{\bf 189}^{SO(7)}-\chi_{\bf 294}^{SO(7)}+2 \chi_{\bf 330}^{SO(7)}
  -\chi_{\bf 378}^{SO(7)}-\chi_{\bf 616}^{SO(7)}-\chi_{\bf 693}^{SO(7)}
  +\chi_{\bf 1617}^{SO(7)}+1\big)\nonumber\\
  \hspace*{-1.2cm}&&
  -\chi_{\bf 3}^{SU(2)} \big(3 \chi_{\bf 7}^{SO(7)}-4 \chi_{\bf 21}^{SO(7)}-4 \chi_{\bf 27}^{SO(7)}+2 \chi_{\bf 35}^{SO(7)}
  +3 \chi_{\bf 77}^{SO(7)}+6 \chi_{\bf 105}^{SO(7)}
  -3 \chi_{\bf 168^\prime}^{SO(7)}-\chi_{\bf 182}^{SO(7)}\nonumber \\
  \hspace*{-1.2cm}&&
  -2 \chi_{\bf 189}^{SO(7)}-\chi_{\bf 294}^{SO(7)}
  -4 \chi_{\bf 330}^{SO(7)}+\chi_{\bf 378}^{SO(7)}+2 \chi_{\bf 616}^{SO(7)}
  +2 \chi_{\bf 693}^{SO(7)}+\chi_{\bf 819}^{SO(7)}-\chi_{\bf 825}^{SO(7)}-\chi_{\bf 1560}^{SO(7)}-2\big)\nonumber\\
  \hspace*{-1.2cm}&&
  +\big(4 \chi_{\bf 7}^{SO(7)}\!-\!4 \chi_{\bf 21}^{SO(7)}\!-\!5
  \chi_{\bf 27}^{SO(7)}\!+\!2 \chi_{\bf 35}^{SO(7)}\!+\!4 \chi_{\bf 77}^{SO(7)}
  \!+\!7 \chi_{\bf 105}^{SO(7)}\!-\!3 \chi_{\bf 168^\prime}^{SO(7)}\!-\!
  \chi_{\bf 182}^{SO(7)}\!-\!3 \chi_{\bf 189}^{SO(7)}
  \!-\!5 \chi_{\bf 330}^{SO(7)}\nonumber \\
  \hspace*{-1.2cm}&&+2 \chi_{\bf 378}^{SO(7)}
  \!+\!2 \chi_{\bf 616}^{SO(7)}\!+\!3 \chi_{\bf 693}^{SO(7)}\!+\!
  2 \chi_{\bf 819}^{SO(7)}\!-\!\chi_{\bf 1617}^{SO(7)}\!-\!
  \chi_{\bf 1911}^{SO(7)}-1\big)\ .
\end{eqnarray}

Now we consider the instanton strings of 6d super-Yang-Mills theories with
$SO(7)$ gauge group and matters in ${\bf 8}$. The number $n_{\bf 8}$ of
hypermultiplets cannot be arbitrary, due to gauge anomalies
\cite{Bershadsky:1997sb,Heckman:2015bfa}. Without
matters in other representations,
one should have $n_{\bf 8}=2$ \cite{Heckman:2015bfa}.
Incidently, the 6d consistency requirement
$n_{\bf 8}=2$ is also reflected in our ADHM-like construction, uplifted
to 2d for instanton strings. This comes from 2d $U(k)$ gauge anomaly
cancelation. First consider the
$SU(k)$ anomaly, proportional to $D_{\bf R}=\pm 2T({\bf R})$ for
right/left moving fermions. From $D_{\bf k}=1$, $D_{\bf adj}=2k$,
$D_{{\bf sym}_2}=k+2$, $D_{{\bf anti}_2}=k-2$, one obtains
\begin{equation}
  -2\cdot 2k+2\cdot 4\cdot 1+2\cdot 2k+ 4\cdot 1+ 2\cdot(k-2)-2\cdot(k+2)
  -n_{\bf 8}\cdot 2\cdot 1=2(2-n_{\bf 8})\ .
\end{equation}
These terms come from fermions in the multiplets
$(\lambda_0,\lambda)$, $(q,\tilde{q})$, $(a,\tilde{a})$, $\phi$, $(b,\tilde{b})$,
$(\hat\lambda,\check\lambda)$, $(\Psi_a,\tilde\Psi_a)$, respectively.
The $SU(k)$ anomaly cancels only at $n_{\bf 8}=2$.
The overall $U(1)$ anomaly is proportional to the square of charges.
The net anomaly is given by
\begin{equation}
  2\cdot 4k\cdot 1^2 + 4k\cdot 1^2+2\cdot\frac{k^2-k}{2}\cdot 2^2
  -2\cdot\frac{k^2+k}{2}\cdot 2^2-n_{\bf 8}\cdot 2k\cdot 1^2=2k(2-n_{\bf 8})\ .
\end{equation}
This again cancels at $n_{\bf 8}=2$. So our ADHM-like quiver consistently
uplifts to 2d at $n_{\bf 8}=2$.

As a basic test of our 2d gauge theories, we study the 't Hooft anomalies
of global symmetries. The full 2d symmetry is expected to be
$SO(7)\times Sp(2)\times SU(2)_l\times SU(2)_r\times SU(2)_R$. From our
UV description, we can only study $SU(4)\times U(2)\times SU(2)_l\times U(1)_J$.
There is an alternative way of computing the anomalies
on the strings, using anomaly inflow \cite{Kim:2016foj,Shimizu:2016lbw}.
By comparing two calculations, we shall provide a test of our gauge theories.

Using the inflow method, the 2d anomaly can be computed as follows.
We first compute the anomaly polynomial 8-form of the 6d SCFT with a
tensor multiplet, $SO(7)$ vector multiplet, and half-hypermultiplets in
$\frac{1}{2}({\bf 8},{\bf 4})$ of $SO(7)\times Sp(2)$.
The anomaly polynomial in the tensor branch consists of
1-loop contribution $I_{\textrm{1-loop}}$, coming from massless
tensor/vector/hyper-multiplets, and the classical
Green-Schwarz contribution $I_{GS}$ \cite{Green:1984bx,Sagnotti:1992qw}.
The two contributions should partly cancel for the terms containing
$SO(7)$ gauge fields \cite{Intriligator:2014eaa,Ohmori:2014kda}.
$I_{\textrm{1-loop}}$ is given by
\begin{eqnarray}
  I_{\textrm{1-loop}}&=&-\frac{3}{32}\left[{\rm Tr}(F^2_{SO(7)})\right]^2
  +\frac{1}{16}{\rm Tr}(F^2_{SO(7)})
  \left[2{\rm tr}_{\bf 4}(F_{Sp(2)}^2)-20c_2(R)-p_1(T)\right]
  +\cdots\nonumber\\
  &=&-\frac{3}{2}\left[\frac{1}{4}{\rm Tr}(F^2_{SO(7)})+
  \frac{1}{12}\left(20c_2(R)+p_1(T)-2{\rm tr}_{\bf 4}(F_{Sp(2)}^2)
  \right)\right]^2+\cdots\ ,
\end{eqnarray}
where $\cdots$ denote terms independent of the
$SO(7)$ field strength $F_{SO(7)}$.
Following \cite{Ohmori:2014kda}, we use the notation
${\rm Tr}\equiv\frac{1}{h^\vee}{\rm tr}_{\bf adj}$, and
${\rm tr}_{\bf adj}(F^4)=-{\rm tr}_{\rm fund}(F^4)+3({\rm Tr}(F^2))^2$,
${\rm tr}_{\bf 8}(F^4)=-\frac{1}{2}{\rm tr}_{\rm fund}(F^4)+\frac{3}{8}({\rm tr}(F^2))^2$, ${\rm tr}_{\bf 8}(F^2)={\rm Tr}(F^2)$,
${\rm tr}_{\bf adj}(F^2)=5{\rm Tr}(F^2)$ for $SO(7)$.
To cancel the 1-loop $SO(7)$ anomaly, one should have the
following Green-Schwarz 8-form \cite{Ohmori:2014kda}:
\begin{equation}
  I_{GS}=\frac{3}{2}I^2\ ,\ \
  I\equiv \frac{1}{4}{\rm Tr}(F^2_{SO(7)})+
  \frac{1}{12}\left(20c_2(R)+p_1(T)-2{\rm tr}_{\bf 4}(F_{Sp(2)}^2)
  \right)\ .
\end{equation}
This takes the form of $I_{GS}=\frac{1}{2}\Omega^{ij}I_iI_j$ with $i,j$
running over just $1$, so that $I_1=I$ and $\Omega^{11}=3$.
$\Omega^{11}$ may be fixed from the fact that it comes from
$O(-3)\rightarrow\mathbb{P}^1$ geometry in F-theory, with self intersection
number of $\mathbb{P}^1$ being $3$.
Knowing $I_i$ appearing in $I_{GS}=\frac{1}{2}\Omega^{ij}I_iI_j$,
one can determine the 2d anomaly 4-form on the strings, from inflow.
The formula is \cite{Kim:2016foj,Shimizu:2016lbw}
\begin{equation}\label{inflow-anomaly}
  I_4=-\Omega^{ij}k_i\left[I_j+\frac{1}{2}k_j\chi(T_4)\right]\ ,
\end{equation}
where $k_i$ is the string number in the $i$'th gauge group (or $i$'th
tensor multiplet). We decomposed the 6d tangent bundle
$T$ to $T_2\times T_4$, along/normal to the strings.
From this formula, one finds
\begin{equation}\label{2d-anomaly}
  I_4=-\frac{3}{2}k^2\chi(T_4)-3k\left[\frac{1}{4}{\rm Tr}(F^2_{SO(7)})+
  \frac{5}{3}c_2(R)+\frac{p_1(T)}{12}-\frac{1}{6}{\rm tr}_{\bf 4}(F_{Sp(2)}^2)
  \right]
\end{equation}
for the $SO(7)$ instanton strings at $n_{\bf 8}=2$, with topological number $k$.

Now we compute $I_4$ from our gauge theory.
A chiral fermion's anomaly 4-form is given by
\begin{equation}\label{2d-anomaly-multiplet}
  I_4=\pm\left[\frac{1}{2}{\rm tr}(F^2)+\frac{p_1(T_2)}{24}\right]\ ,
\end{equation}
where $\pm$ signs are for left/right-moving fermions, respectively, in our convention.
$F$ collectively denotes all background gauge fields for the global symmetries acting
on the fermion. Here it is for $SU(4)\times U(1)_J\times SU(2)_l\times U(2)_F$.
We can only study the anomalies of the symmetries
surviving in UV, and check the consistency with (\ref{2d-anomaly}).
Fermi and vector multiplets have left-moving fermions,
while chiral multiplets have right-moving fermions. Each multiplet
contributes to terms of the form (\ref{2d-anomaly-multiplet}) with a suitable sign.
Firstly, contributions from fields neutral in $SU(4)\times U(2)$
are already computed in \cite{Kim:2016foj}:
\begin{eqnarray}
  (\lambda_0,\lambda)+(a,\tilde{a})
  &:&k^2(c_2(r)-c_2(l))\\
  \hat\lambda,\check\lambda&:&(k^2+k)\left[\frac{F_R^2}{8}
  +\frac{c_2(r)}{2}+\frac{p_1(T_2)}{24}\right]\nonumber\\
  b,\tilde{b}&:&-(k^2-k)\left[\frac{F_R^2}{8}+\frac{c_2(l)}{2}
  +\frac{p_1(T_2)}{24}\right]\ .\nonumber
\end{eqnarray}
$R$ is the $U(1)$ Cartan of $SU(2)_R$.
Here and later, we shall often use expressions like
$c_2(r)$, $c_2(R)=\frac{F_R^2}{4}$ assuming symmetry enhancement,
but only the $U(1)_J$ part is to be kept in UV. Namely,
one first keeps the Cartan parts of the field strengths,
for $J_r$, $J_R$. Then they are all replaced by $J$
and its field strength $F_J$.
We present the results using $c_2(R)$ and $c_2(r)$
since this may suggest possible patterns of IR symmetry enhancement.
(See also \cite{Kim:2016foj}.)
The fields charged under $SU(4)\times U(2)$ contribute
to $I_4$ as follows:
\begin{eqnarray}
  q,\tilde{q},\phi&:&
  -3k\left[\frac{1}{4}{\rm Tr}(F_{SU(4)}^2)
  +4\cdot\frac{F_R^2}{8}+4\cdot\frac{p_1(T_2)}{24}\right]\nonumber\\
  \Psi_a,\tilde\Psi_a&:&
  2k\left[\frac{1}{2}{\rm tr}_{\bf 2}(F_{U(2)}^2)+2\cdot\frac{p_1(T_2)}{24}\right]
  \ .
\end{eqnarray}
Adding all, and using $p_1(T)=p_1(T_2)-2c_2(l)-2c_2(r)$, one obtains
\begin{equation}\label{2d-anomaly-gauge}
  I_4=\frac{3}{2}k^2(c_2(r)-c_2(l))-3k\left[\frac{1}{4}{\rm Tr}(F_{SU(4)}^2)
  +\frac{5}{3}\cdot\frac{F_R^2}{4}+\frac{p_1(T)}{12}
  -\frac{1}{3}{\rm tr}_{\bf 2}(F_{U(2)}^2)\right]\ .
\end{equation}
Here and below, we shall frequently use the fact that ${\rm Tr}(F^2)$
remains the same after restricting $F$ to a subalgebra if a long root of the
original algebra is kept, so that unit instanton charge $\frac{1}{4}\int {\rm Tr}(F^2)$ remains the same \cite{Ohmori:2014kda}. Here it applies to
$SU(4)\subset SO(7)$. As for $U(2)\subset Sp(2)$, or more generally
$U(n)\subset Sp(n)$, the embedding is such that
${\rm tr}_{\bf 2n}\rightarrow 2{\rm tr}_{\bf n}$.
Taking these into account, (\ref{2d-anomaly-gauge})
agrees with (\ref{2d-anomaly}) upon restricting (\ref{2d-anomaly}) to
$SU(4)$, $U(2)$, $SU(2)_r\times SU(2)_R\rightarrow J$, and using
$\chi(T_4)=c_2(l)-c_2(r)$. Their mixed anomalies with $U(k)$ also vanish.

One can study the elliptic genera $Z_k$ of $k$ instanton strings, whose
spatial direction wraps $S^1$. The definition is almost identical
to (\ref{index-definition}), except that there is another factor
$e^{2\pi i\tau P}$ inside the trace, where $P$ is
the left-moving momentum on $S^1$. The basic formula is given in
\cite{Benini:2013nda,Benini:2013xpa}. The result is obtained by
simply replacing all $2\sinh$ functions in (\ref{SO7-integral}),
(\ref{SO7-spinor-integrand}), (\ref{SO7-spinor-residue}) by
$2\sinh\frac{z}{2}\rightarrow\frac{i\theta_1(\tau|\frac{z}{2\pi i})}{\eta(\tau)}
\equiv\theta(z)$. For instance, at $k=1$, one obtains
\begin{equation}
  Z_1(\tau,\epsilon_{1,2},v_i,m_a)=\frac{1}{\theta(\epsilon_{1,2})}
  \sum_{i=1}^4\frac{\theta(4\epsilon_+-2v_i)\prod_{a=1}^2
  \theta(m_a\pm(v_i-\epsilon_+))}
  {\prod_{j(\neq i)}\theta(v_{ij})\theta(2\epsilon_+-v_{ij})
  \theta(2\epsilon_+-v_i-v_j)}
\end{equation}
where $v_{ij}\equiv v_i-v_j$.
Some tests of these formulae will be given in section 4.2.

\subsection{$G_2$ instantons and matters in ${\bf 7}$}

With a hypermultiplet in ${\bf 8}$,
one can Higgs $SO(7)$ to $G_2$. Decomposing the scalar to
${\bf 8}\rightarrow{\bf 7}\oplus{\bf 1}$ in $G_2$, ${\bf 1}$ is given
VEV and decouples in IR. ${\bf 7}$ is eaten up by the broken
part of the $SO(7)$ gauge fields, since
${\bf 21}\rightarrow{\bf 14}\oplus{\bf 7}$. The matter consists of
two half hypermultiplets,
forming a doublet of flavor symmetry $Sp(1)_F$.
The scalar can be written as
$[\Phi_{Aa}]_{(s_1,s_2,s_3)}$, where $A=1,2$ is the doublet index of $SU(2)_{R}$
R-symmetry, $a=1,2$ is that of $Sp(1)_F$, and $s_{1,2,3}=\pm\frac{1}{2}$ label the
components of ${\bf 8}$. It satisfies the reality condition
$(\Phi^\ast)^{Aa}_{(s_1,s_2,s_3)}=\epsilon^{AB}\epsilon^{ab}
(\Phi_{Bb})_{(s_1,s_2,s_3)}$.
Let us take $\Phi_{Aa}\equiv[\Phi_{Aa}]_{(+,+,+)}+[\Phi_{Aa}]_{(-,-,-)}$,
satisfying $(\Phi^\ast)^{Aa}=\epsilon^{AB}\epsilon^{ab}\Phi_{Bb}$.
One takes $\Phi_{Aa}=\epsilon_{Aa}\Phi$, with a pure imaginary VEV $\Phi$.
This preserves a diagonal subgroup of $SU(2)_R\times Sp(1)_F$, which is
the $SU(2)_R$ symmetry after Higgsing. At general $n_{\bf 8}$,
we give VEV to the last hypermultiplet scalar, $a=n_{\bf 8}$.
One should lock the chemical potentials as
\begin{equation}\label{higgs-chemical}
m_{\rm last}-\epsilon_{+}\pm v_4=0
\end{equation}
with both signs, not to rotate the scalar VEV.
So we should take $m_{\rm last}-\epsilon_+=0$, $v_4=0$.
The former condition turns off the $Sp(1)\subset Sp(n_{\bf 8})$
chemical potential $m_{\rm last}$, and the latter reduces the rank of gauge
group by $1$. As the index is invariant under the
RG flow triggered by the scalar VEV, one can get the IR $G_2$ index
by constraining the $SO(7)$ index.

In our $SU(4)$ formalism, the bulk scalars are written as
$Q_i,\tilde{Q}^i$, where $i=1,2,3,4$. Giving VEV to ${\bf 1}$ amounts to
turning on $Q_4=\tilde{Q}^4=M\neq 0$ (real), where we take
the unbroken $SU(3)\subset G_2$ to be labeled by $i=1,2,3$.
In 1d, the background fields couple to the 1d fields as
\begin{equation}
  J_{\Psi_{\rm last}}\sim Q_i\tilde{q}^i
  \ ,\ \ J_{\tilde{\Psi}_{\rm last}}\sim\tilde{Q}^iq_i\ .
\end{equation}
The second potential $|J_{\tilde\Psi_{\rm last}}|^2\sim M^2|q_4|^2$ gives mass to
$q_4$, while the first one gives mass to
$\tilde{q}^4$.\footnote{One may more generally take
$J_\Psi\sim \alpha Q_i\tilde{q}^i+\beta\tilde{Q}^i\phi_i$,
compatible with $U(k)\times SU(4)$.
However, with $SU(4)$ broken to $SU(3)$, $\tilde{q}^4$ and $\phi_4$ have
same charges in unbroken symmetries, and $\alpha,\beta$ does not
affect the IR physics.} The
$SU(4)$ ADHM fields reduce to
the $SU(3)$ ADHM fields at low energy. Among the extra fields, $\phi_i$ with
$i=1,2,3,4$ decomposes into $\phi_i$ with $i=1,2,3$ in $(\bar{\bf k},\bar{\bf 3})$,
and $\phi_4$ in $(\bar{\bf k},{\bf 1})$.
If $n_{\bf 8}\geq 2$, one still has $n_{\bf 7}=n_{\bf 8}-1$
pairs of Fermi multiplets $\Psi_a,\tilde\Psi_a$ left in
$({\bf k},{\bf 1})+(\bar{\bf k},{\bf 1})$, $a=1,\cdots,n_{\bf 7}$.
To summarize, one first has the $SU(3)$ ADHM fields,
\begin{eqnarray}\label{G2-matter-1}
  A_\mu,\lambda_0,\lambda&:&\mathcal{N}=(0,4)\ U(k)\ \textrm{vector multiplet }
  \nonumber\\
  q_i,\tilde{q}^i&:&({\bf k},\bar{\bf 3})+(\bar{\bf k},{\bf 3})\ \ \ (i=1,2,3)
  \nonumber\\
  a,\tilde{a}&:&({\bf adj},{\bf 1})\ .
\end{eqnarray}
In addition, one has
\begin{eqnarray}\label{G2-matter-2}
  \phi_i,\phi_4&:&\textrm{chiral in }(\bar{\bf k},\bar{\bf 3})_{J=\frac{1}{2}}
  +(\bar{\bf k},{\bf 1})_{J=\frac{1}{2}}\nonumber\\
  b,\tilde{b}&:&\textrm{chiral in }(\overline{\bf anti}_2,{\bf 1})_{J=\frac{1}{2}}
  \nonumber\\
  \hat\lambda&:&\textrm{Fermi in }({\bf sym}_2,{\bf 1})_{J=0}\nonumber\\
  \check\lambda&:&\textrm{Fermi in }({\bf sym}_2,{\bf 1})_{J=-1}\ .
\end{eqnarray}
For $n_{\bf 7}\leq 3$ hypermultiplet matters in representation ${\bf 7}$,
there are extra Fermi multiplets:
\begin{equation}\label{G2-matter-3}
  \Psi_a\ ,\ \tilde\Psi_a\ :\ ({\bf k},{\bf 1})+(\bar{\bf k},{\bf 1})\ \ ,\ \ \ \
  a=1,\cdots,n_{\bf 7}\ .
\end{equation}
The $\mathcal{N}=(0,1)$ action follows from a construction similar to
$SO(7)$ in section 2.1.

The index $Z_k$ for $k$ $G_2$ instantons
can either be obtained from the Witten
index of the above gauge theory, or by taking the Higgsing condition
of the $SO(7)$ index, $m_{n_{\bf 8}}=\epsilon_+$, $v_4=0$. It may
be more illustrative to write both the contour integral expression and
the residue sum. The contour integral expression for the index is given
by
\begin{eqnarray}\label{G2-integral}
  Z_{k}&=&\frac{1}{k!}\oint\prod_{I=1}^k\frac{d\phi_I}{2\pi i}\cdot
  \frac{\prod_{I\neq J}2\sinh\frac{\phi_{IJ}}{2}\cdot
  \prod_{I,J}2\sinh\frac{2\epsilon_+-\phi_{IJ}}{2}}
  {\prod_{I=1}^k\prod_{i=1}^3 2\sinh\frac{\epsilon_+\pm(\phi_I-v_i)}{2}
  \cdot\prod_{I,J}2\sinh\frac{\epsilon_{1,2}+\phi_{IJ}}{2}}\\
  &&\times\frac{\prod_{I\leq J}\left(2\sinh\frac{\phi_I+\phi_J}{2}
  \cdot 2\sinh\frac{\phi_I+\phi_J-2\epsilon_+}{2}\right)}
  {\prod_I\left(\prod_{i=1}^3 2\sinh\frac{\epsilon_+-\phi_I-v_i}{2}
  \cdot 2\sinh\frac{\epsilon_+-\phi_I}{2}\right)
  \cdot\prod_{I<J}2\sinh\frac{\epsilon_{1,2}-\phi_I-\phi_J}{2}}
  \cdot \prod_{I=1}^k\prod_{a=1}^{n_{\bf 7}}
  2\sinh\frac{m_a\pm\phi_I}{2}
  \nonumber
\end{eqnarray}
The residue sum, labeled by $SU(3)$ colored Young diagrams,
is given by\footnote{The factors on the first line of
(\ref{G2-residue}) containing $E_{ij}$ are the residues for pure $SU(3)$ theory.
In this type of expression, one finds an overall $(-1)^{Nk}$ factor
for pure $SU(N)$ instantons. This is why we have $(-1)^k$ in (\ref{G2-residue}).}
\begin{eqnarray}\label{G2-residue}
  Z_k\!&\!=\!&\!(-1)^k\sum_{\vec{Y};|\vec{Y}|=k}\prod_{i=1}^3\prod_{s\in Y_i}
  \frac{2\sinh(\phi(s))\cdot 2\sinh(\epsilon_+-\phi(s))}
  {\prod_{j=1}^3\left(2\sinh\frac{E_{ij}(s)}{2}\cdot
  2\sinh\frac{E_{ij}(s)-2\epsilon_+}{2}\cdot 2\sinh\frac{\epsilon_+-\phi(s)-v_j}{2}
  \right)\cdot 2\sinh\frac{\epsilon_+-\phi(s)}{2}}\nonumber\\
  \hspace*{-1cm}&&\cdot\prod_{i\leq j}^3\prod_{s_{i,j}\in Y_{i,j};s_i<s_j}
  \frac{2\sinh\frac{\phi(s_i)+\phi(s_j)}{2}\cdot
  2\sinh\frac{\phi(s_i)+\phi(s_j)-2\epsilon_+}{2}}
  {2\sinh\frac{\epsilon_{1,2}-\phi(s_i)-\phi(s_j)}{2}}
  \cdot\prod_{i=1}^3\prod_{s\in Y_i}\prod_{a=1}^{n_{\bf 7}}
  2\sinh\frac{m_a\pm\phi(s)}{2}
\end{eqnarray}
where $\phi(s)$ and $E_{ij}(s)$ are defined in (\ref{phi(s)}), (\ref{Eij}).

We first study the case with $n_{\bf 7}=0$.
We can test the results against known $G_2$ instanton partition functions of \cite{Cremonesi:2014xha}. We tested (\ref{G2-residue}) till $k\leq 3$.
Firstly, at $k=1$, it will be illustrative to make a basic presentation, directly
from the contour integral. (\ref{G2-integral}) at $k=1$ is given by
\begin{equation}
  \left(2\sinh\frac{\epsilon_{1,2}}{2}\right)Z_1=
  \oint d\phi\frac{2\sinh\epsilon_+\cdot 2\sinh\phi\cdot 2\sinh(\phi-\epsilon_+)}
  {\prod_{i=1}^3\left(2\sinh\frac{\epsilon_+\pm(\phi-v_i)}{2}
  \cdot 2\sinh\frac{\epsilon_+-\phi-v_i}{2}\right)
  \cdot 2\sinh\frac{\epsilon_+-\phi}{2}}\ .
\end{equation}
At $\eta>0$, the poles are chosen at $\phi=v_i-\epsilon_+$, $i=1,2,3$.
So one obtains
\begin{equation}
  \left(2\sinh\frac{\epsilon_{1,2}}{2}\right)Z_1=
  \sum_{i=1}^3\frac{2\sinh(v_i-2\epsilon_+)}
  {\prod_{j(\neq i)}\left(2\sinh\frac{v_{ij}}{2}\cdot
  2\sinh\frac{2\epsilon_+-v_{ij}}{2}\cdot 2\sinh\frac{2\epsilon_++v_j}{2}\right)
  \cdot 2\sinh\frac{v_i-2\epsilon_+}{2}}
\end{equation}
where we used $v_1+v_2+v_3=0$. Each residue only exhibits Weyl symmetry
of $SU(3)$, given by $3!$ permutations of $v_1,v_2,v_3$. However,
the sum of three residues exhibits enhanced Weyl symmetry of $G_2$,
the dihedral group $D_6$ of order $12$. The extra transformation
generating full $D_6$ is $v_i\rightarrow -v_i$ for all $i=1,2,3$,
$SU(3)$ charge conjugation.
One can show that $Z_1$ is given by
\begin{equation}\label{G2-k=1}
  \left(2\sinh\frac{\epsilon_{1,2}}{2}\right)Z_1=
  \frac{t^{3}(1+t^2)(1+t^2\chi_{\bf 7}^{G_2}(v)+t^4)}
  {\prod_{i<j}(1-t^2e^{v_{ij}})(1-t^2e^{-v_{ij}})}=
  t^{3}\sum_{n=0}^\infty
  \chi_{(0,n)}^{G_2}(v)t^{2n}\ ,
\end{equation}
where $t\equiv e^{-\epsilon_+}$.
$\chi_{\bf 7}^{G_2}=1\!+\!\chi_{\bf 3}^{SU(3)}\!+\!\chi_{\bar{\bf 3}}^{SU(3)}
=1+\sum_{i=1}^3(e^{v_i}+e^{-v_i})$
is the character of ${\bf 7}$. $\chi_{(0,n)}^{G_2}$ is the character of
the irrep $(0,n)$ of $G_2$, which is the $n$'th symmetric product of the
adjoint representation ${\bf 14}$. (\ref{G2-k=1}) is known as the correct
$G_2$ instanton partition function at $k=1$ \cite{Cremonesi:2014xha}.

At $k=2$, $Z_2$ can be rearranged into (where $t=e^{-\epsilon_+}$,
$u=e^{-\epsilon_-}$)
\begin{equation}\label{G2-k=2}
  \Big(\prod_{n=1}^2 2\sinh\frac{n\epsilon_{1,2}}{2}\Big)Z_2=
  \frac{t^{24}}{\prod_{i<j} (1-t^2e^{\pm v_{ij}})
  (1-t^{3}u^{\pm \frac{1}{2}}e^{\pm v_{ij}})}
  \Big[\chi_{\bf 20}^{SU(2)}+\sum_{n=1}^{18}
  \chi_{\bf n}^{SU(2)}g_n(v_i,u)\Big]\ ,
\end{equation}
where $SU(2)$ is still ${\rm diag}[SU(2)_r\times SU(2)_R]$, and
$g_n(v_i,u)$'s are given by
\begin{eqnarray}
  g_{18}&=&\chi_{\bf 7}^{G_2}+1\ ,\ \
  g_{17}=\chi_{\bf 2}^{SU(2)_l}(\chi_{\bf 7}^{G_2}+1)\ ,\ \
  g_{16}=\chi_{\bf 7}^{G_2}+\chi_{\bf 27}^{G_2}+1\ ,\ \
  g_{15}=\chi_{\bf 2}^{SU(2)_l} (3\chi_{\bf 7}^{G_2}+1)\ ,\nonumber\\
  g_{14}&=&\chi_{\bf 3}^{SU(2)_l}( \chi_{\bf 7}^{G_2}+ 1)
  +2 \chi_{\bf 7}^{G_2}+\chi_{\bf 27}^{G_2}+1\ ,\ \
  g_{13}=\chi_{\bf 2}^{SU(2)_l} (
  \chi_{\bf 7}^{G_2}+\chi_{\bf 14}^{G_2}
  + \chi_{\bf 27}^{G_2}- \chi_{\bf 64}^{G_2}+2 )\ ,\nonumber\\
  g_{12}&=&\chi_{\bf 3}^{SU(2)_l} ( \chi_{\bf 7}^{G_2}
  +\chi_{\bf 14}^{G_2}-\chi_{\bf 64}^{G_2} +2 )
  +2 \chi_{\bf 7}^{G_2}  +2 \chi_{\bf 14}^{G_2} -2 \chi_{\bf 64}^{G_2} +2
  \nonumber\\
  g_{11}&=&\chi_{\bf 4}^{SU(2)_l}
  + \chi_{\bf 2}^{SU(2)_l}
  ( 2 \chi_{\bf 7}^{G_2}- \chi_{\bf 64}^{G_2}- \chi_{\bf 189}^{G_2}+1)\nonumber\\
  g_{10}&=&\chi_{\bf 3}^{SU(2)_l}(\chi_{\bf 7}^{G_2}
  - \chi_{\bf 14}^{G_2} - \chi_{\bf 77}^{G_2})
   +3 \chi_{\bf 7}^{G_2} -\chi_{\bf 14}^{G_2} -\chi_{\bf 64}^{G_2}
  -2 \chi_{\bf 77}^{G_2} +\chi_{\bf 182}^{G_2} -\chi_{\bf 189}^{G_2} +1\nonumber\\
  g_9&=&\chi_{\bf 4}^{SU(2)_l} ( \chi_{\bf 7}^{G_2} - \chi_{\bf 14}^{G_2} )
  +\chi_{\bf 2}^{SU(2)_l}
  ( 3 \chi_{\bf 7}^{G_2} - \chi_{\bf 14}^{G_2}
  - \chi_{\bf 64}^{G_2}-3 \chi_{\bf 77}^{G_2}+ \chi_{\bf 182}^{G_2} + 1 )
  \nonumber\\
  g_8&=&
  \chi_{\bf 3}^{SU(2)_l} ( \chi_{\bf 77}^{G_2} - \chi_{\bf 14}^{G_2})
  + 2 \chi_{\bf 7}^{G_2}
  -\chi_{\bf 64}^{G_2}
  -\chi_{\bf 77}^{G_2}+\chi_{\bf 182}^{G_2}-\chi_{\bf 189}^{G_2}+\chi_{\bf 378}^{G_2}+1\nonumber\\
  g_7&=&
  \chi_{\bf 4}^{SU(2)_l} ( \chi_{\bf 77}^{G_2} - \chi_{\bf 7}^{G_2} )
  + \chi_{\bf 2}^{SU(2)_l} (2 \chi_{\bf 7}^{G_2} - \chi_{\bf 64}^{G_2}
  +2 \chi_{\bf 182}^{G_2}- \chi_{\bf 286}^{G_2}+ \chi_{\bf 448}^{G_2}+ 1)\nonumber\\
  g_6&=&
  \chi_{\bf 3}^{SU(2)_l} ( \chi_{\bf 77}^{G_2}\!+ \!\chi_{\bf 182}^{G_2}\!
  - \!\chi_{\bf 64}^{G_2}\! + \!\chi_{\bf 189}^{G_2}\!+\! 1)\!
  +\chi_{\bf 7}^{G_2}\!+\!\chi_{\bf 14}^{G_2}\!+\!\chi_{\bf 27}^{G_2}\!
  -\!\chi_{\bf 64}^{G_2}\! +\!\chi_{\bf 182}^{G_2}\!-\!\chi_{\bf 286}^{G_2}\!
  +\!\chi_{\bf 378}^{G_2}\! +\!\chi_{\bf 448}^{G_2}\!+\!1\nonumber\\
  g_5&=&
  \chi_{\bf 4}^{SU(2)_l} ( \chi_{\bf 77}^{G_2}
  -2 \chi_{\bf 7}^{G_2} - \chi_{\bf 14}^{G_2} - 1 )\nonumber\\
  &&+\chi_{\bf 2}^{SU(2)_l} ( - \chi_{\bf 7}^{G_2}\!+\!2 \chi_{\bf 14}^{G_2}
  \!+\! \chi_{\bf 27}^{G_2}\! -\!2 \chi_{\bf 64}^{G_2} \!+\!2 \chi_{\bf 77}^{G_2}
  \!+\! \chi_{\bf 182}^{G_2}\!+\! \chi_{\bf 189}^{G_2}\!-\! \chi_{\bf 286}^{G_2}
  \!+\! \chi_{\bf 378}^{G_2}\!+\!2 )\ \ (\textit{continued})\nonumber
\end{eqnarray}
\begin{eqnarray}
  g_4&=&
  -\chi_{\bf 5}^{SU(2)_l} \chi_{\bf 7}^{G_2}
  + \chi_{\bf 3}^{SU(2)_l} (2 \chi_{\bf 77}^{G_2}
  - \chi_{\bf 7}^{G_2}+ \chi_{\bf 14}^{G_2}- \chi_{\bf 27}^{G_2}
  -2 \chi_{\bf 64}^{G_2}+ \chi_{\bf 182}^{G_2}
  - \chi_{\bf 286}^{G_2} + 1)
  \nn\\&&
  +2 \chi_{\bf 14}^{G_2}+\chi_{\bf 27}^{G_2} -\chi_{\bf 64}^{G_2}
  +\chi_{\bf 77}^{G_2}+\chi_{\bf 182}^{G_2}-\chi_{\bf 273}^{G_2}
  -2 \chi_{\bf 286}^{G_2}+\chi_{\bf 448}^{G_2}+1
  \nonumber\\
  g_3&=&
  -\chi_{\bf 4}^{SU(2)_l} (\chi_{\bf 7}^{G_2}
  + \chi_{\bf 27}^{G_2} + \chi_{\bf 64}^{G_2})
  +\chi_{\bf 2}^{SU(2)_l} ( \chi_{\bf 7}^{G_2}- \chi_{\bf 14}^{G_2}
  + \chi_{\bf 182}^{G_2} - \chi_{\bf 189}^{G_2}
  - \chi_{\bf 286}^{G_2}- \chi_{\bf 729}^{G_2} )
  \nonumber\\
  g_2&=&-\chi_{\bf 5}^{SU(2)_l} ( \chi_{\bf 7}^{G_2} + \chi_{\bf 14}^{G_2} + 1 )
  +\chi_{\bf 3}^{SU(2)_l}
  ( \chi_{\bf 64}^{G_2}
  -4  \chi_{\bf 14}^{G_2}
  -2 \chi_{\bf 77}^{G_2}+ \chi_{\bf 182}^{G_2}- \chi_{\bf 189}^{G_2}
  - \chi_{\bf 448}^{G_2}-2 )
  \nn\\&&
  +2 \chi_{\bf 7}^{G_2}-2 \chi_{\bf 14}^{G_2}+2 \chi_{\bf 64}^{G_2}
  -3 \chi_{\bf 77}^{G_2}-\chi_{\bf 273}^{G_2}
  -\chi_{\bf 729}^{G_2}-1
  \nonumber\\
  g_1&=&
  \chi_{\bf 4}^{SU(2)_l} ( 2 \chi_{\bf 7}^{G_2}
  -2  \chi_{\bf 14}^{G_2}
  - \chi_{\bf 27}^{G_2}
  + \chi_{\bf 64}^{G_2}
  -2 \chi_{\bf 77}^{G_2}
  - 1 )\nonumber\\
  &&+ \chi_{\bf 2}^{SU(2)_l} (2 \chi_{\bf 7}^{G_2}-3 \chi_{\bf 14}^{G_2}
  +2  \chi_{\bf 64}^{G_2} -4  \chi_{\bf 77}^{G_2}+  \chi_{\bf 182}^{G_2}
  - \chi_{\bf 273}^{G_2}- \chi_{\bf 448}^{G_2}  -1)\ .
\end{eqnarray}
As the numerator is
manifestly arranged into $G_2$ characters, it shows enhanced $G_2$ Weyl symmetry.
The denominator is also invariant under the extra generator
$v_i\rightarrow-v_i$ of $D_6$, being invariant under $G_2$ Weyl symmetry.
One can also check the
agreement with the known $G_2$ partition function at $k=2$. For the simplicity of
comparison, let us turn off all $v_i=0$ and $\epsilon_-=0$.
Then, (\ref{G2-k=2}) becomes
\begin{eqnarray}\label{G2-k=2-unrefined}
  \left(2\sinh\frac{\epsilon_{1,2}}{2}\right)Z_2&=&
  \frac{t^7}{(1-t)^{14}(1+t)^8(1+t+t^2)^7}\left[1+t+10t^2+31t^3+75t^4+180t^5
  +385t^6\right.\nonumber\\
  &&\left.+637t^7+ 975t^8+ 1360t^9 +1614 t^{10} +1666 t^{11}+1614t^{12}+\cdots
  +t^{22}\right]
\end{eqnarray}
where the omitted terms $\cdots$ can be restored by
the $t\rightarrow t^{-1}$ Weyl symmetry of $SU(2)$
(i.e. the coefficients of $t^p$ and $t^{22-p}$ are same on the numerator).
The overall $t^7$ factor is like a zero point energy factor, and is needed
to have this Weyl symmetry. Apart from
this factor, (\ref{G2-k=2-unrefined}) agrees with eqn.(9.5) of
\cite{Hanany:2012dm} after correcting a typo there, as noted
in \cite{Cremonesi:2014xha}.

At $k=3$, we only show the simplified form of (\ref{G2-residue}) at
$v_i=0$, $\epsilon_-=0$, which is
\begin{eqnarray}
  \left(2\sinh\frac{\epsilon_{1,2}}{2}\right)Z_3&\!=\!&\!
  \frac{t^{11}}{(1-t)^{22} (1 + t)^{14} (1 + t^2)^7 (1 + t + t^2)^9}
 \left[1 + t +
   11 t^2 + 34 t^3 + 124 t^4+ 352 t^5\right.\nonumber\\
   &&\left.  + 1055 t^6 + 2657 t^7 +
   6584 t^8 + 14635 t^9 + 31194 t^{10} + 61229 t^{11}+ 114367 t^{12}
   \right.\nonumber\\
   &&\left. + 198932 t^{13} + 329172 t^{14} +
   511194 t^{15} + 755093 t^{16} + 1051845 t^17 + 1394817 t^{18} \right.\nonumber\\
   &&\left.+ 1749632 t^{19} + 2091341 t^{20} +
   2368619 t^{21} + 2557449 t^{22} + 2619060 t^{23}\right.\nonumber\\
   &&\left. + 2557449 t^{24} + \cdots + t^{46}\right]\ ,
\end{eqnarray}
where $\cdots$ can again be restored by noting that coefficients of
$t^p$ and $t^{46-p}$ are same on the numerator. Apart from the overall
$t^{11}$ factor which guarantees Weyl symmetry, this again agrees with
eqn.(4.16) of \cite{Cremonesi:2014xha}. Although we did comparisons till $k=3$,
one can in principle continue to test for higher $k$'s whether our
(\ref{G2-residue}) agrees with the results of \cite{Cremonesi:2014xha}.

Now as for the indices at $n_{\bf 7}\geq 1$, these observables
have not been computed or studied in the literature, to the best of our knowledge.
Here we simply note that, making expansions of the indices in $t=e^{-\epsilon_+}$,
one observes that the coefficients are characters of $G_2\times Sp(n_{\bf 7})$.
At least at $k=1$, this does not need independent
calculations, since we already illustrated the symmetry enhancement of
$SO(7)\times Sp(n_{\bf 8})$ in the previous subsection. Also, whenever
we provide concrete tests of some $SO(7)$ results in section 3
and section 4, this implies similar tests of the $G_2$ results at
$n_{\bf 7}=n_{\bf 8}-1$ by Higgsing.

At $n_{\bf 7}=1$, 6d SCFT exists with $G_2$ gauge group. This can be obtained from
6d $SO(7)$ theory at $n_{\bf 8}=2$ by Higgsing. Our 2d gauge theories
on $G_2$ instantons can also be uplifted to 2d gauge theories. As in the previous
subsection, this gauge theory is free of $U(k)$ gauge anomaly.
The 2d anomaly of $G_2\times Sp(n_{\bf 7})\times SU(2)_R\times SU(2)_r\times SU(2)_l$ global symmetries, computed from anomaly inflow, is also compatible with
the $SU(3)\times U(n_{\bf 7})\times U(1)_J\times SU(2)_l$ anomalies of our 2d gauge
theories. To see this, one first restricts $F_{SO(7)}\rightarrow F_{G_2}$
which leaves ${\rm Tr}(F^2)$ invariant,
${\rm tr}_{\bf 4}(F^2_{Sp(2)})
\rightarrow{\rm tr}_{\bf 2}(F^2_{Sp(1)})+{\rm tr}_{\bf 2}(F^2_{Sp(1)^\prime})$,
and note that $c_2(R)=\frac{1}{4}{\rm Tr}(F^2_R)
=\frac{1}{2}{\rm tr}_{\rm fund}(F^2_R)$ \cite{Ohmori:2014kda}.
Since we lock $Sp(1)^\prime$ and $SU(2)_R$ during Higgsing,
one identifies $F_{Sp(1)^\prime}=F_R$. Then, both anomaly 4-forms
(\ref{2d-anomaly}), (\ref{2d-anomaly-gauge}) reduce to
\begin{equation}\label{G2-2d-anomaly}
  I_4=-\frac{3}{2}k^2\chi(T_4)-3k\left[\frac{1}{4}{\rm Tr}(F^2_{G_2})
  +\frac{4}{3}c_2(R)+\frac{p_1(T)}{12}-\frac{1}{6}{\rm tr}_{\bf 2}(F_{Sp(1)}^2)
  \right]\ ,
\end{equation}
with restrictions to UV symmetry understood for gauge theory anomalies.
So the inflow anomaly and 2d gauge theory anomaly continue to agree with each other.

The elliptic genera for the strings can be computed similarly.
One takes the formulae (\ref{G2-integral}) or (\ref{G2-residue}), and replace
$2\sinh\frac{z}{2}\rightarrow
\frac{i\theta_1(\tau|\frac{z}{2\pi i})}{\eta(\tau)}\equiv\theta(z)$
for all $2\sinh$ functions. The $G_2$ symmetry of this
elliptic genus at $k=1$ is systematically discussed in \cite{klp}.

At $n_{\bf 7}=1$, one has a pair $\Psi,\tilde\Psi$ of Fermi
multiplets. One can again investigate the effect of bulk Higgsing
$G_2\rightarrow SU(3)$.
In the bulk, one decomposes ${\bf 7}\rightarrow{\bf 3}+\bar{\bf 3}+{\bf 1}$,
where scalar in ${\bf 1}$ assumes VEV and breaks $G_2$ into $SU(3)$. The other
hypermultiplet fields are eaten up by vector multiplets for the broken
symmetry. The constant VEV of the bulk scalar $\equiv Q$ in ${\bf 1}$ will
behave as a background field in 1d/2d ADHM-like models. With foresight on
the $SU(3)$ instantons studied in \cite{Kim:2016foj},
we propose that the coupling of the background bulk field $Q$ to
the $G_2$ ADHM-like gauge theory is given by
\begin{equation}
  J_{\Psi}\sim Q\phi\ ,\ \ J_{\tilde\Psi}\sim Q\epsilon^{ijk}q_i\phi_jq^\dag_k\ ,
\end{equation}
where $\phi\equiv \phi_4$.
The $\mathcal{N}=(0,1)$ superpotential $J_{\tilde\Psi}$ is compatible with symmetries,
but at this stage it may not be obvious why we should turn it on in this way.
$\Psi$ and the chiral multiplet $\phi$ become massive due to $J_\Psi$,
and decouple at low energy. However, $\tilde\Psi$ does not
decouple at low energy, since it does not acquire mass.
In fact, the remaining system (including $\tilde\Psi$, which was called
$\zeta$ in \cite{Kim:2016foj}) with the above cubic
superpotential was studied in \cite{Kim:2016foj}, which showed various
nontrivial physics of the $SU(3)$ instanton strings.
In 1d, this provides a novel alternative ADHM-like description for $SU(3)$
instanton particles. In 2d, this is (by now) the uniquely known $SU(3)$ ADHM
construction of instanton strings without matters.
All models presented so far in this paper, for $SO(7)$ and $G_2$ instantons,
were initially constructed by guessing the un-Higgsing procedures from
$SU(3)$. See \cite{Kim:2016foj}  for further discussions on
the last $SU(3)$ model.

\section{Exceptional instantons from D-branes}

\subsection{Brane setup and quantum mechanics}
\label{sbsec:brane}

\begin{figure}[t!]
    \centering\vspace{0.3cm}
    \subcaptionbox{}{
	\includegraphics[width=6cm]{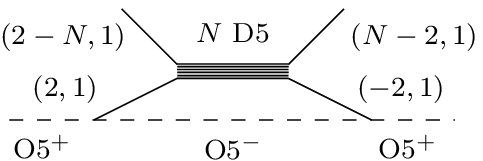}}
	\quad
    \subcaptionbox{}{
	\includegraphics[width=6cm]{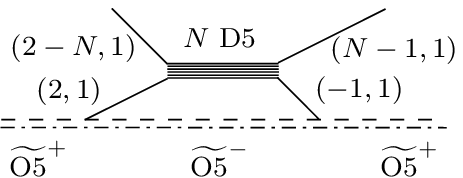}}
	\caption{Brane realizations of (a) $SO(2N)$ and (b) $SO(2N+1)$ gauge theories}
	\label{fig:puregauge}
\end{figure}
\begin{figure}[t!]
    \centering\vspace{0.3cm}
    \subcaptionbox{}{
	 \includegraphics[width=6cm]{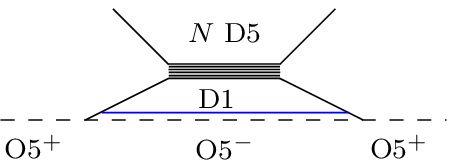}}
	\quad
    \subcaptionbox{}{
	\includegraphics[width=6cm]{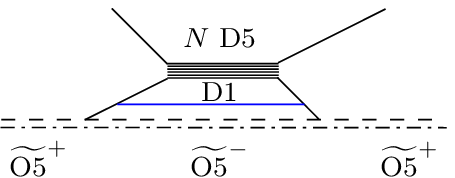}}
	\caption{Instantons of (a) $SO(2N)$ and (b) $SO(2N+1)$ theories}
	\label{fig:D1inst}
\end{figure}

\begin{figure}[t!]
    \centering\vspace{0.3cm}
    \subcaptionbox{}{
	\includegraphics[width=8cm]{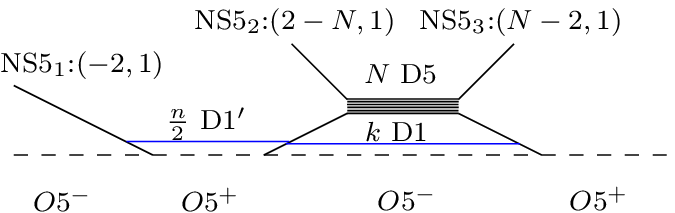}}
	\quad
    \subcaptionbox{}{
	\includegraphics[width=8cm]{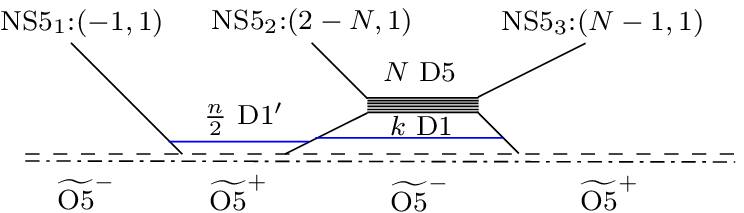}}
	\caption{Hypermultiplet in the spinor representation
of (a) $SO(2N)$ and (b) $SO(2N+1)$. }
	\label{fig:D1spin}
\end{figure}

\begin{figure}[t!]
    \centering\vspace{0.3cm}
    \subcaptionbox{}{
	\includegraphics[width=8cm]{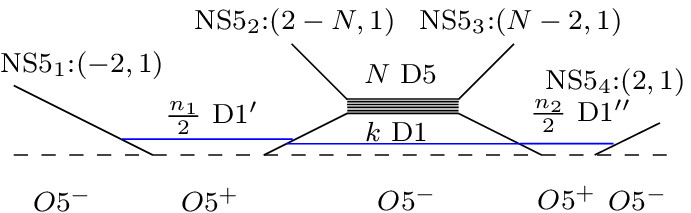}}
	\quad
    \subcaptionbox{}{
	\includegraphics[width=8cm]{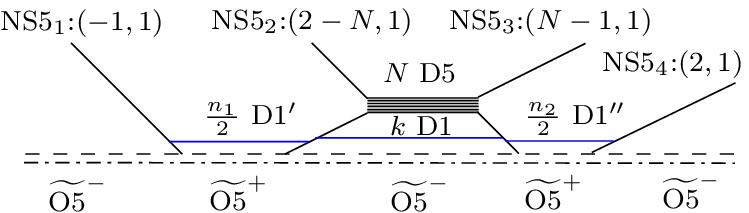}}
	\caption{Two hypermultipets in the spinor representations
of (a) $SO(2N)$ and (b) $SO(2N+1)$.}
	\label{fig:D1spin2}
\end{figure}

In this section, we test some indices of the
previous section, using 5-brane webs for the 5d
$\mathcal{N}=1$ gauge theories with $SO(N)$ gauge groups and
matters in spinor representations \cite{Zafrir:2015ftn}.
A type IIB 5-brane web on $x^5,x^6$ plane consists of
$(p,q)$ 5-branes stretched along lines with slope $q/p$:
e.g. D5-branes $(1,0)$ along $x^5$ and NS5-branes $(0,1)$ along $x^6$
directions. They occupy $x^0,\cdots,x^4$ directions for the 5d QFT.
$SO(N)$ gauge theories are realized by 5-brane webs
with orientifold 5-planes. An NS5-brane crossing the O5-plane bends to a
suitable $(p,1)$-brane, and changes the types of
O5 across NS5. An $SO(2N)$ theory is engineered by
suspending $N$ D5-branes between two NS5-branes, also with
an $O5^-$, as shown in Fig.~\ref{fig:puregauge}(a). $SO(2N\!+\!1)$ theory
is realized by $N$ D5-branes and an $\widetilde{O5}^-$
plane, which is an $O5^-$ with a half D5. See Fig.~\ref{fig:puregauge}(b).
Dashed-dotted line is a monodromy cut, to have
$(p,q)$ 5-branes at right angles with properly quantized charges
\cite{Zafrir:2015ftn}. In these constructions, instanton particles
are D1-branes stretched between two NS5-branes,
as shown in Fig. \ref{fig:D1inst}.
In this setting, a 5d hypermultiplet in the spinor representation is introduced
as follows \cite{Zafrir:2015ftn}. One introduces another NS5-brane as
shown in Fig.~\ref{fig:D1spin}.
D1$'$-branes suspended between NS5$_{1}$ and NS5$_{2}$
are the particles obtained by quantizing the hypermultiplet in
the $SO(N)$ spinor representation. (See \cite{Zafrir:2015ftn} for
the chirality of the $SO(2N)$ spinor.)
The mass of this field is proportional to the distance between
NS5$_{1}$ and NS5$_{2}$. To introduce two hypermultiplets in the
spinor representation, one puts another NS5-brane on the right side, as shown in
Fig.~\ref{fig:D1spin2}.
Note that for $SO(N)$ gauge theory with $N\leq6$, NS5$_{1}$ and NS5$_{2}$-branes do
not intersect. For $N=7,8$, NS5$_{1}$ and NS5$_{2}$ are parallel to
each other. In the last cases, there are extra continua of D1$'$ branes, 
orthogonally suspended between these parallel NS5-branes, which 
can escape to infinity and do not belong to 5d QFT. In section \ref{sec:Partition_function} we discuss
this extra sector in more detail. When $N\geq 9$, NS5$_{1}$ and NS5$_{2}$ meet at
a certain point. In this case, we do not know how to use this
setting to study the 5d QFT. So in the rest of this paper, we focus on $SO(N)$
QFTs with $N\leq 8$.


We discuss the quantum mechanical gauge theory, with given numbers
of $SO(N)$ instantons $k$ and hypermultiplet particles $n_a$. Their Witten indices
will be used to test some results of section 2.
In section 2, we did not fix the numbers of hypermultiplet particles, but instead
had chemical potentials $m_a$ for $Sp(n_{\bf 8})$. Expanding
the indices of section 2 in $e^{-m_a}$, the coefficients
will be the indices with fixed $k,n_a$, studied in this section.

We start from the case with $1$ hypermultiplet, and consider
the quantum mechanics of the D1 and D1$^\prime$ branes.
We first explain the symmetries. There is $SO(4)\sim SU(2)_{l}\times SU(2)_{r}$ rotating $x^1,\cdots,x^4$, and $SO(3)\sim SU(2)_{R}$ rotating $x^7,x^8,x^9$. The quantum mechanics preserves 4 real SUSY $\bar{Q}^{\dot{\alpha}A}$, where $\dot\alpha$ and $A$ are doublet indices of $SU(2)_{r}$ and $SU(2)_{R}$. It can be regarded as the
1d reduction of 2d $\mathcal{N}=(0,4)$ SUSY.
There are symmetries associated with D-branes and orientifolds.
For $r$ D1's and $N$ D5's on various O5-planes, the
symmetries are given as follows:
\begin{table}[h!]
\centering
\vspace{-0.5cm}
\begin{tabular}{c|cccc}
\hline branes  & $O5^{+}$ & $O5^{-}$ & $\widetilde{O5}^{+}$ & $\widetilde{O5}^{-}$ \\ \hline
$N$ D5  & $Sp(N)$ & $SO(2N)$ & $Sp(N)$ & $SO(2N+1)$ \\
$r$ D1 & $O(2r)$ & $Sp(r)$ & $O(2r)$ & $Sp(r)$ \\ \hline
\end{tabular}
\end{table}\vspace{-0.3cm}\\
Here $r$ is a half-integer $r=n/2$ for $O5^{+}$, $\widetilde{O5}^{+}$.
So D1 and D1$^\prime$ in Fig.~\ref{fig:D1spin} have
$Sp(k)\times O(n)$ gauge symmetry, while D5's induce
$SO(2N)$ or $SO(2N+1)$ global symmetry.

\begin{figure}[t!]
	\centering\vspace{0.3cm}
	\subcaptionbox{}{
	\includegraphics[height=5cm]{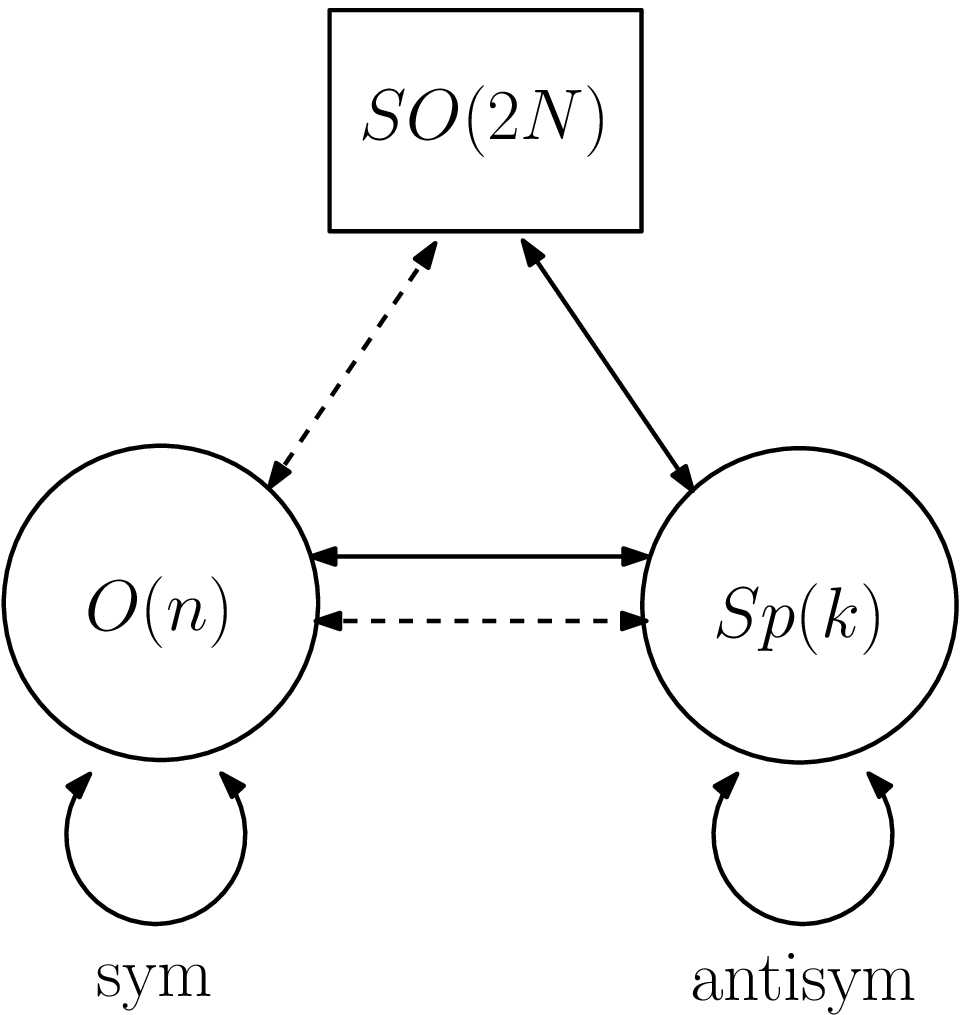}
	}\hfill
	\subcaptionbox{}{
	\begin{tabular}{c|c|c|ccc}
		Mode & Field & Type & $Sp(k)$ & $O(n)$ & $SO(2N)$ \\
		\hline
		D1$-$D1 & $(A_{t},\varphi, \lambda^{\dot{\alpha}A})$ & vector & \textbf{sym} & $-$ & $-$ \\
		& $(a_{\alpha\dot{\beta}}, \chi_{\alpha}^{A})$ & hyper & \textbf{anti} & $-$ & $-$ \\
		D1$-$D5 & $(q_{\dot{\alpha}}, \psi^{A})$ & hyper & \textbf{2k} & $-$ & \textbf{2N} \\
		D1$'-$D1$'$ & $(\hat{A}_{t},\hat{\varphi},\hat{\lambda}^{\dot{\alpha}A})$ & vector & $-$ & \textbf{anti} & $-$ \\
		& $(\hat{a}_{\alpha\dot{\beta}}, \hat{\chi}_{\alpha}^{A})$ & hyper & $-$ & \textbf{sym} & $-$ \\
		D1$'-$D5 & $\Lambda_{l}$ & Fermi & $-$ & \textbf{n} & \textbf{2N} \\
		D1$-$D1$'$ & $(\Phi_{A}, \Psi^{\dot{\alpha}})$ & twisted hyper & \textbf{2k} & \textbf{n} & $-$ \\
		& $\Xi_{\alpha}$ & Fermi & \textbf{2k} & \textbf{n} & $-$ \\
	\end{tabular}
	}
	
	\caption{(a) 1d quiver and (b) matters for $SO(2N)$. (bold/dashed lines
for hyper/Fermi)}
	\label{fig:matter_even}
\end{figure}
\begin{figure}[t!]
	\centering\vspace{0.2cm}
	\subcaptionbox{}{
	\includegraphics[height=4.9cm]{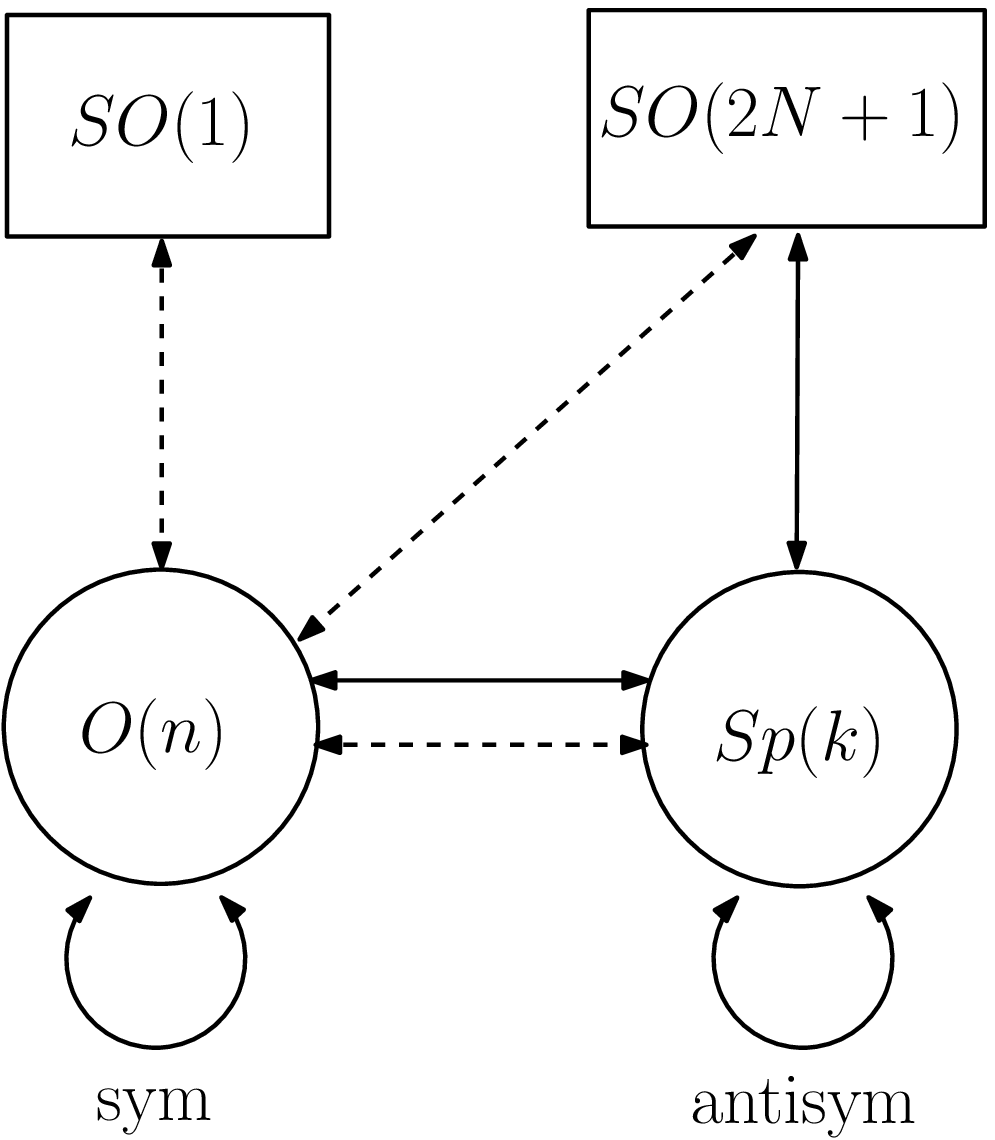}
	}\hfill
	\subcaptionbox{}{
	\begin{tabular}{c|c|c|ccc}
		Mode & Field & Type & $Sp(k)$ & $O(n)$ & $SO(2N+1)$ \\
		\hline
		D1$-$D1 & $(A_{t},\varphi, \lambda^{\dot{\alpha}A})$ & vector & \textbf{sym} & $-$ & $-$ \\
		& $(a_{\alpha\dot{\beta}}, \chi_{\alpha}^{A})$ & hyper & \textbf{anti} & $-$ & $-$ \\
		D1$-$D5 & $(q_{\dot{\alpha}}, \psi^{A})$ & hyper & \textbf{2k} & $-$ & \textbf{2N+1} \\
		D1$'-$D1$'$ & $(\hat{A}_{t},\hat{\varphi}, \hat{\lambda}^{\dot{\alpha}A})$ & vector & $-$ & \textbf{anti} & $-$ \\
		& $(\hat{a}_{\alpha\dot{\beta}}, \hat{\chi}_{\alpha}^{A})$ & hyper & $-$ & \textbf{sym} & $-$ \\
		D1$'-$D5 & $\Lambda_{l}$ & Fermi & $-$ & \textbf{n} & \textbf{2N+1} \\
		& $\Lambda$ & Fermi & $-$ & \textbf{n} & $-$  \\
		D1$-$D1$'$ & $(\Phi_{A}, \Psi^{\dot{\alpha}})$ & twisted hyper & \textbf{2k} & \textbf{n} & $-$ \\
		& $\Xi_{\alpha}$ & Fermi & \textbf{2k} & \textbf{n} & $-$ \\
	\end{tabular}
	}
	\caption{(a) 1d quiver and (b) matters for $SO(2N+1)$.}
	\label{fig:matter_odd}
\end{figure}
The quantum mechanical `fields' are derived from open strings.
They are shown in Fig.~\ref{fig:matter_even} and \ref{fig:matter_odd} for
$SO(2N)$ and $SO(2N+1)$. The formal `$SO(1)$' in Fig.~\ref{fig:matter_odd}(a)
comes from the half D5-brane on $\widetilde{\textrm{O5}}^{-}$, on the left side of NS5$_{1}$ in Fig.~\ref{fig:D1spin}.
The Lagrangian of this system preserving $\mathcal{N}=(0,4)$ supersymmetry
can be written down in a canonical manner. We focus on the bosonic part here.
Along the strategy of \cite{Tong:2014yna},
we first construct the Lagrangian in $\mathcal{N}=(0,2)$ formalism,
specifying the two possible types of superpotentials
$E$ and $J$ for each Fermi multiplet \cite{Witten:1993yc}.
Our $(0,4)$ multiplets decompose to $(0,2)$ multiplets as follows:
\begin{align}
\textrm{vector}\;(A_{t}, \varphi, \lambda^{\dot{\alpha}A})\quad &\longrightarrow\quad \textrm{vector}\;V\;(A_{t},\varphi,  \lambda^{\dot{1}2}, \lambda^{\dot{2}1})\;+\;\textrm{Fermi}\;\lambda\;(\lambda^{\dot{1}1}, \lambda^{\dot{2}2})\nonumber\\
\textrm{vector}\;(\hat{A}_{t},\hat{\varphi}, \hat{\lambda}^{\dot{\alpha}A})\quad &\longrightarrow\quad \textrm{vector}\;\hat{V}\;(\hat{A}_{t},\hat{\varphi}, \hat{\lambda}^{\dot{1}2}, \hat{\lambda}^{\dot{2}1})\;+\;\textrm{Fermi}\;\hat{\lambda}\;(\hat{\lambda}^{\dot{1}1}, \hat{\lambda}^{\dot{2}2})\nonumber\\
\textrm{hyper}\;(a_{\alpha\dot{\beta}}, \chi_{\alpha}^{A})\quad &\longrightarrow\quad\textrm{chiral}\;B\;(a_{c\dot{1}}, \chi_{c}^{2})\;+\;\textrm{chiral}\;\tilde{B}^{\dagger}\;(a_{c\dot{2}}, \chi_{c}^{1})\nonumber\\
\textrm{hyper}\;(\hat{a}_{\alpha\dot{\beta}}, \hat{\chi}_{\alpha}^{A})\quad &\longrightarrow\quad\textrm{chiral}\;C\;(\varphi_{c\dot{1}}, \xi_{c}^{2})\;+\;\textrm{chiral}\;\tilde{C}^{\dagger}\;(\varphi_{c\dot{2}}, \xi_{c}^{1})\nonumber\\
\textrm{hyper}\;(q_{\dot{\alpha}}, \psi^{A})\quad&\longrightarrow\quad\textrm{chiral}\;q\;(q_{\dot{1}},\psi^{2})\;+\;\textrm{chiral}\;\tilde{q}^{\dagger}\;(q_{\dot{2}}, \psi^{1})\nonumber\\
\textrm{twisted hyper}\;(\Phi_{A}, \Psi^{\dot{\alpha}})\quad&\longrightarrow\quad\textrm{chiral}\;\Phi\;(\Phi_{1}, \Psi^{2})\;+\;\textrm{chiral}\;\tilde{\Phi}^{\dagger}\;(\Phi_{2}, \Psi^{1})\nonumber\\
\textrm{Fermi}\;(\Lambda_{l}),\;(\Lambda),\;(\Xi_{\alpha})\quad&
\longrightarrow\quad\textrm{Fermi}\;(\Lambda_{l}),\;(\Lambda),\;(\Xi_{\alpha})\ .
\end{align}
The scalars in rank $2$ symmetric or antisymmetric
representations are real. It decomposes to two $(0,2)$ chiral multiplets whose scalars are complexified as $a_{c\dot{\beta}}=a_{1\dot{\beta}}+ia_{2\dot{\beta}}$ and likewise $\hat{a}_{c\dot{\beta}}$.

In $(0,2)$ theories, one can turn on two types of holomorphic `superpotentials'
for each Fermi multiplet $\Psi$, $J_\Psi$ and $E_\Psi$.
$(0,2)$ supersymmetry demands the superpotentials to satisfy
\begin{align}\label{E.J}
\sum_{\nu\in\textrm{Fermi}}E_{\nu}J_{\nu}=0\ .
\end{align}
We first consider the $SO(2N)$ theory, in which case
$E$ and $J$ for Fermi multiplets are given by
\begin{align}
J_{\lambda}&=\sqrt{2}(q\tilde{q}+[B, \tilde{B}]) &
E_{\lambda}&=-\sqrt{2}\widetilde{\Phi}\Phi\nonumber\\
J_{\hat{\lambda}}&=\sqrt{2}[C, \widetilde{C}] &
E_{\hat{\lambda}}&=\sqrt{2}\Phi\widetilde{\Phi}\nonumber\\
J_{\Xi_{1}}&=\sqrt{2}(\widetilde{\Phi}\widetilde{C}-\widetilde{B}\widetilde{\Phi}) &
E_{\Xi_{1}}&=\sqrt{2}(C\Phi-\Phi B)\nonumber\\
J_{\Xi_{2}}&=-\sqrt{2}(\widetilde{\Phi}C-B\widetilde{\Phi}) &
E_{\Xi_{2}}&=\sqrt{2}(\widetilde{C}\Phi-\Phi\widetilde{B})\nonumber\\
J_{\Lambda_{l}}&=\sqrt{2}\widetilde{q}\widetilde{\Phi} &
E_{\Lambda_{l}}&=\sqrt{2}\Phi q\ .
\label{eq:EJ}
\end{align}
The first two lines, for $(0,4)$ gauginos of $Sp(k)\times O(n)$, are
required by demanding $(0,4)$ SUSY enhancement \cite{Tong:2014yna}. Namely,
gaugino fields' $J$ and $E$ acquire contributions only from hypermultiplets and twisted hypermultiplets, respectively. But with the first two lines only,
(\ref{E.J}) is not met. The next three lines are fixed (up to sign choices)
by demanding (\ref{E.J}) to hold, as illustrated
in \cite{Tong:2014yna} in different models. D-terms are given by
\begin{eqnarray}
  D_{Sp(k)}&=&qq^{\dagger}-\widetilde{q}^{\dagger}\widetilde{q}
  +[B,B^{\dagger}]-[\widetilde{B}^{\dagger},\widetilde{B}]-\Phi^{\dagger}\Phi
  +\widetilde{\Phi}\widetilde{\Phi}^{\dagger}\nonumber\\
  D_{O(n)}&=&[C, C^{\dagger}]-[\widetilde{C}^{\dagger}, \widetilde{C}]+\Phi\Phi^{\dagger}-\widetilde{\Phi}^{\dagger}\widetilde{\Phi}.
\label{eq:D}
\end{eqnarray}
With these superpotentials and D-terms,
the bosonic potential energy is given by
\cite{Tong:2014yna,Witten:1993yc},
\begin{align}
V=\sum_{G\in\textrm{gauge}}\frac{1}{2}D_{G}^{2}
+\sum_{\nu\in\textrm{Fermi}}(|E_{\nu}|^{2}+|J_{\nu}|^{2})\ .
\label{eq:2dpot}
\end{align}
One can show that \eqref{eq:2dpot} exhibits enhanced
$SO(4)=SU(2)_r\times SU(2)_R$ R-symmetry,
\begin{eqnarray}\label{eq:2Npot}
  V&=&
  \frac{1}{2}\left(q_{\dot{\alpha}}(\sigma^{m})^{\dot{\alpha}}{}_{\dot{\beta}}
  {q^{\dagger}}^{\dot{\beta}}+(\sigma^{m})^{\dot{\alpha}}{}_{\dot{\beta}}
  \left[a_{\alpha\dot{\alpha}}, {a^{\dagger}}^{\alpha\dot{\beta}}\right]\right)^{2}
  +\frac{1}{2}\left((\sigma^{m})^{\dot{\alpha}}{}_{\dot{\beta}}
  \left[\hat{a}_{\alpha\dot{\alpha}}, {\hat{a}^{\dagger}}{}^{\alpha\dot{\beta}}\right]\right)^{2}\nonumber\\
  &&+\frac{1}{2}\left(\Phi_{A}(\sigma^{m})^{A}{}_{B}{\Phi^{\dagger}}^{B}\right)^{2}
    +\frac{1}{2}\left({\Phi^{\dagger}}^{A}(\bar{\sigma}^{m})_{A}{}^{B}\Phi_{B}\right)^{2}
    +|\Phi_{A}q_{\dot{\alpha}}|^{2}\nonumber\\
  &&+|\hat{a}_{\alpha\dot{\beta}}\Phi_{A}-\Phi_{A}a_{\alpha\dot{\beta}}|^{2}
    +|{\Phi^{\dagger}}^{A}\hat{a}_{\alpha\dot{\beta}}
    -a_{\alpha\dot{\beta}}{\Phi^{\dagger}}^{A}|^{2}.
\end{eqnarray}
Since $SO(4)$ is the $\mathcal{N}=(0,4)$ R-symmetry,
this is a strong indication that the classical action indeed has
$(0,4)$ SUSY.
We content ourselves with this observation, rather than checking $(0,4)$ SUSY
of the full action.
The fields in the last expression satisfy the pseudo-reality condition of
$Sp(k)$, $\widetilde{q}^{T}=\Lambda q$,
$\widetilde{\Phi}^{T}=\Phi(\Lambda^{-1})^{T}$,
where $\Lambda$ is the $Sp(k)$ skew-symmetric matrix.

One can repeat the analysis for the $SO(2N+1)$ quiver. One point to note
here is that there is no superpotential
for the Fermi multiplet $\Lambda$. So despite the presence of
$2N+2$ $O(n)$ fundamental Fermi multiplets $\Lambda_l$, $\Lambda$, their
flavor symmetry is $SO(2N+1)$, as we expect from 5d bulk.

\begin{figure}[t!]
	\centering\vspace{0.2cm}
	\subcaptionbox{}{
	\includegraphics[height=4.9cm]{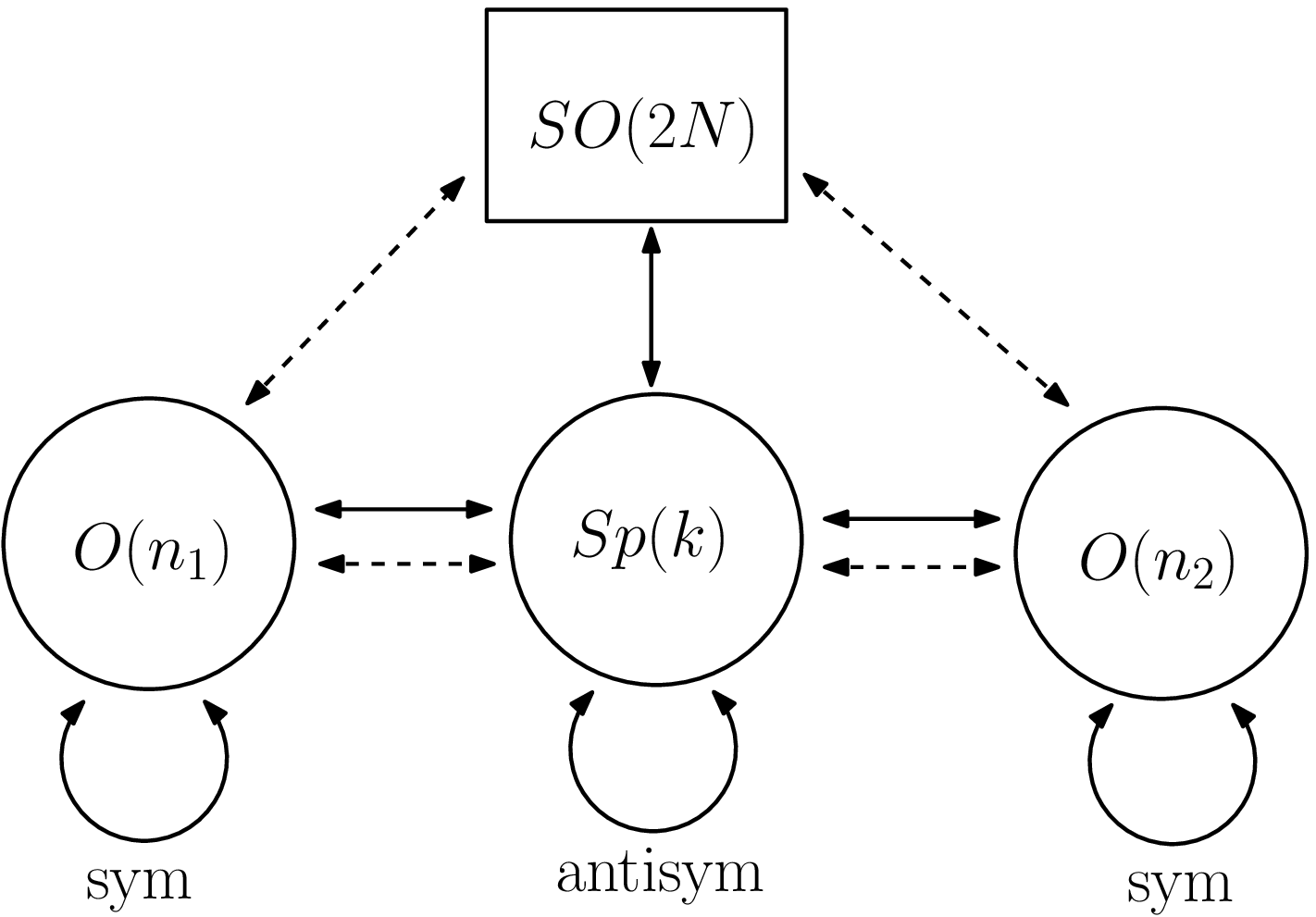}
	}\hfill
	\subcaptionbox{}{
	\begin{tabular}{c|c|c|cccc}
		Mode & Field & Type & $Sp(k)$ & $O(n_{1})$ & $O(n_{2})$ & $SO(2N)$ \\
		\hline
		D1$-$D1 & $(A_{t},\varphi, \lambda^{\dot{\alpha}A})$ & vector & \textbf{sym} & $-$ & $-$ & $-$ \\
		& $(a_{\alpha\dot{\beta}}, \chi_{\alpha}^{A})$ & hyper & \textbf{anti} & $-$ & $-$ & $-$ \\
		D1$-$D5 & $(q_{\dot{\alpha}}, \psi^{A})$ & hyper & \textbf{2k} & $-$ & $-$ & \textbf{2N} \\
		D1$'-$D1$'$ & $(\hat{A}_{t},\hat{\varphi}, \hat{\lambda}^{\dot{\alpha}A})$ & vector & $-$ & \textbf{anti} & $-$ & $-$ \\
		& $(\hat{a}_{\alpha\dot{\beta}}, \hat{\chi}_{\alpha}^{A})$ & hyper & $-$ & \textbf{sym} & $-$ & $-$ \\
		D1$'-$D5 & $\Lambda_{l}$ & Fermi & $-$ & $\boldsymbol{n_{1}}$ & $-$ & \textbf{2N} \\
		D1$-$D1$'$ & $(\Phi_{A}, \Psi^{\dot{\alpha}})$ & twisted hyper & \textbf{2k} & $\boldsymbol{n_{1}}$ & $-$ & $-$ \\
		& $\Xi_{\alpha}$ & Fermi & \textbf{2k} & $\boldsymbol{n_{1}}$ & $-$ & $-$ \\
		D1$''-$D1$''$ & $(\hat{A}_{t},\hat{\varphi}, \hat{\lambda}^{\dot{\alpha}A})$ & vector & $-$ & $-$ & \textbf{anti} & $-$ \\
		& $(\hat{a}_{\alpha\dot{\beta}}, \hat{\chi}_{\alpha}^{A})$ & hyper & $-$ & $-$ & \textbf{sym} & $-$ \\
		D1$''-$D5 & $\Lambda_{l}$ & Fermi & $-$ & $-$ & $\boldsymbol{n_{2}}$ & \textbf{2N} \\
		D1$-$D1$''$ & $(\Phi_{A}, \Psi^{\dot{\alpha}})$ & twisted hyper & \textbf{2k} & $-$ & $\boldsymbol{n_{2}}$ & $-$ \\
		& $\Xi_{\alpha}$ & Fermi & \textbf{2k} & $-$ & $\boldsymbol{n_{2}}$ & $-$ \\
	\end{tabular}
	}
	\caption{The 1d quiver (a) and matters (b) for 5d $SO(2N)$ theory with two  hypermultiplets.}
	\label{fig:matter_two_even}
\end{figure}

\begin{figure}[t!]
	\centering\vspace{0.2cm}
	\subcaptionbox{}{
	\includegraphics[height=4.9cm]{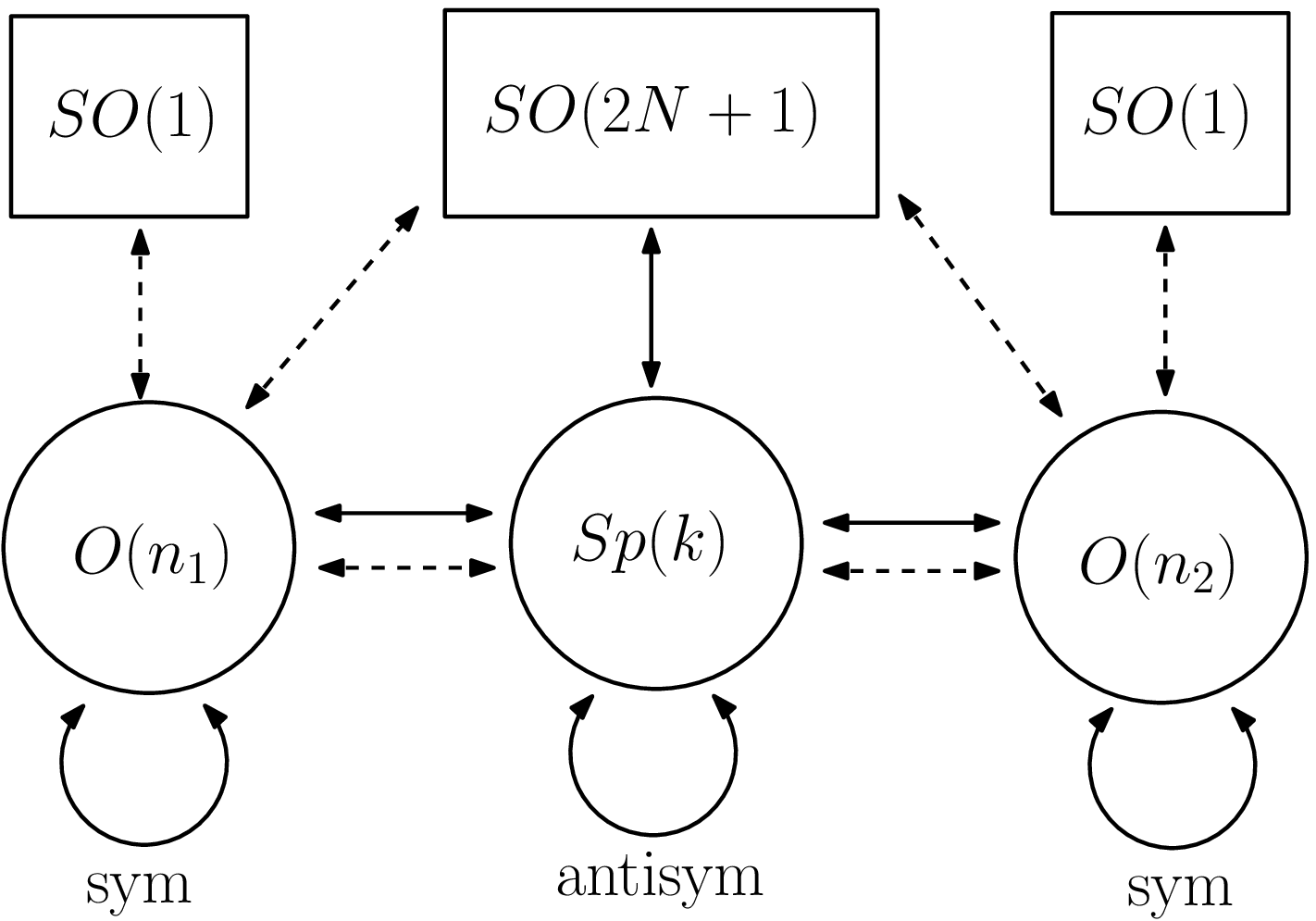}
	}\hfill
	\subcaptionbox{}{
	\begin{tabular}{c|c|c|cccc}
		Mode & Field & Type & $Sp(k)$ & $O(n_{1})$ & $O(n_{2})$ & $SO(2N+1)$ \\
		\hline
		D1$-$D1 & $(A_{t},\varphi, \lambda^{\dot{\alpha}A})$ & vector & \textbf{sym} & $-$ & $-$ & $-$ \\
		& $(a_{\alpha\dot{\beta}}, \chi_{\alpha}^{A})$ & hyper & \textbf{anti} & $-$ & $-$ & $-$ \\
		D1$-$D5 & $(q_{\dot{\alpha}}, \psi^{A})$ & hyper & \textbf{2k} & $-$ & $-$ & \textbf{2N+1} \\
		D1$'-$D1$'$ & $(\hat{A}_{t},\hat{\varphi}, \hat{\lambda}^{\dot{\alpha}A})$ & vector & $-$ & \textbf{anti} & $-$ & $-$ \\
		& $(\hat{a}_{\alpha\dot{\beta}}, \hat{\chi}_{\alpha}^{A})$ & hyper & $-$ & \textbf{sym} & $-$ & $-$ \\
		D1$'-$D5 & $\Lambda_{l}$ & Fermi & $-$ & $\boldsymbol{n_{1}}$ & $-$ & \textbf{2N+1} \\
		& $\Lambda$ & Fermi & $-$ & $\boldsymbol{n_{1}}$ & $-$ & $-$  \\
		D1$-$D1$'$ & $(\Phi_{A}, \Psi^{\dot{\alpha}})$ & twisted hyper & \textbf{2k} & $\boldsymbol{n_{1}}$ & $-$ & $-$ \\
		& $\Xi_{\alpha}$ & Fermi & \textbf{2k} & $\boldsymbol{n_{1}}$ & $-$ & $-$ \\
		D1$''-$D1$''$ & $(\hat{A}'_{t},\hat{\varphi}', \hat{\lambda}'^{\dot{\alpha}A})$ & vector & $-$ & $-$ & \textbf{anti} & $-$ \\
		& $(\hat{a}'_{\alpha\dot{\beta}}, \hat{\chi}'^{A}_{\alpha})$ & hyper & $-$ & $-$ & \textbf{sym} & $-$ \\
		D1$''-$D5 & $\Lambda'_{l}$ & Fermi & $-$ & $-$ & $\boldsymbol{n_{2}}$ & \textbf{2N+1} \\
		& $\Lambda'$ & Fermi & $-$ & $-$ & $\boldsymbol{n_{2}}$ & $-$ \\
		D1$-$D1$''$ & $(\Phi_{A}', \Psi'^{\dot{\alpha}})$ & twisted hyper & \textbf{2k} & $-$ & $\boldsymbol{n_{2}}$ & $-$ \\
		& $\Xi'_{\alpha}$ & Fermi & \textbf{2k} & $-$ & $\boldsymbol{n_{2}}$ & $-$ \\
	\end{tabular}
	}
	\caption{The 1d quiver (a) and matters (b)
for 5d $SO(2N+1)$ theory with two hypermultiplets.}
	\label{fig:matter_two_odd}
\end{figure}
When there are two 5d hypermultiplets in the spinor representation of
$SO(N)$, we can consider a sector with $n_{1}$ and $n_{2}$ particles 
and $k$ instantons. The 1d quivers and fields are
shown in Figs.~\ref{fig:matter_two_even}, \ref{fig:matter_two_odd}. The
Lagrangians can be constructed by following the completely same procedures,
which we do not present here.

\subsection{The instanton partition functions}
\label{sec:Partition_function}
We shall compute the Witten indices of
the quantum mechanics presented in the previous subsection. They count
BPS states preserving $Q=-\bar{Q}^{1\dot{2}}$ and $Q^{\dagger}=\bar{Q}^{2\dot{1}}$,
and is defined by
\begin{align}
Z_{QM}=\textrm{Tr}\left[(-1)^{F}e^{-\beta\{Q, Q^{\dagger}\}}
e^{-2\epsilon_{+}(J_{r}+J_{R})}e^{-2\epsilon_{-}J_{l}}e^{-v_{i}q_{i}}e^{-z\cdot F}\right].
\label{eq:Witten}
\end{align}
$J_{l}$, $J_{r}$ and $J_{R}$ are Cartans of $SO(4)=SU(2)_{l}\times SU(2)_{r}$  and $SU(2)_{R}$ respectively, while $q_{i}$ are the $SO(N)$ electric charges.
$F$, $z$ denote other charges and their chemical potentials.

We compute \eqref{eq:Witten} using the contour integral formula of \cite{Hwang:2014uwa,Kim:2014dza,Hori:2014tda}.
The zero modes in the path integral appear
as the contour integral variables. They are the eigenvalues of the scalar $\varphi$ and $A_{\tau}$ in the vector multiplet. For $O(n)$,
the flat connections on $S^1$ have two disconnected sectors $O(n)_\pm$.
$U=e^{R\phi}\equiv e^{R(\varphi+iA_\tau)}$, where $R$ is the radius of
the temporal circle, is given by
\begin{eqnarray}
  &&U^{+}_{O(2n)}=\textrm{diag}(e^{i\phi_{1}\sigma_{2}},e^{i\phi_{2}\sigma_{2}},
  \cdots,e^{i\phi_{n}\sigma_{2}})\ \ ,\ \
  U^{-}_{O(2n)}=\textrm{diag}(e^{i\phi_{1}\sigma_{2}},e^{i\phi_{2}\sigma_{2}},
  \cdots,e^{i\phi_{n-1}\sigma_{2}},\sigma_{3})\nonumber\\
  &&U^{+}_{O(2n+1)}=\textrm{diag}(e^{i\phi_{1}\sigma_{2}},e^{i\phi_{2}\sigma_{2}},
  \cdots,e^{i\phi_{n}\sigma_{2}},1)\ \ ,\ \
  U^{-}_{O(2n+1)}=\textrm{diag}(e^{i\phi_{1}\sigma_{2}},e^{i\phi_{2}\sigma_{2}},
  \cdots,e^{i\phi_{n}\sigma_{2}},-1)\nonumber\\
  &&U_{Sp(k)}=\textrm{diag}(e^{i\phi_{1}\sigma_{2}},e^{i\phi_{2}\sigma_{2}},
  \cdots,e^{i\phi_{k}\sigma_{2}})\ .
\label{eq:hol}
\end{eqnarray}
$\sigma_{i}$ are Pauli matrices, `diag' mean block-diagonal matrices, and
$\det(U^\pm)=\pm 1$.
The integrand acquires contributions from various multiplets.
A chiral multiplet $\Phi$ and a Fermi multiplet $\Psi$ contribute as
\begin{equation}
  Z_{\Phi}=\prod_{\rho\in R_{\Phi}}\frac{1}{2\,\textrm{sinh}(\frac{\rho(\phi)+2J \epsilon_{+}+Fz}{2})}\ \ ,\ \
  Z_{\Psi}=\prod_{\rho\in R_{\Psi}}2\,\textrm{sinh}(\frac{\rho(\phi)+2J \epsilon_{+}+Fz}{2})
\label{eq:intch}
\end{equation}
respectively. $\rho$ runs over the weights of $Sp(k)$, $O(n)$ in the
representation $R_\Phi$, $R_\Psi$, and $J$ is defined by $J=J_{r}+J_{R}$.
$F$ collectively denotes the remaining charges.
A $(0,2)$ vector multiplet $V$ contributes similarly as
$Z_{V}=\prod_{\alpha\in\textrm{root}}2\,\textrm{sinh}\frac{\alpha(\phi)}{2}$,
where we used the formula for a Fermi multiplet at $J=0$, $F=0$.
Collecting all, the Witten index is given by
\begin{align}
  Z=\frac{1}{|W|}\oint\frac{d\phi}{2\pi i}Z_{\textrm{1-loop}}
  \ \ ,\ \ \ Z_{\textrm{1-loop}}\equiv
  \prod_{V}Z_{V}\prod_{\Phi}Z_{\Phi}\prod_{\Psi}Z_{\Psi}\ .
\end{align}
The $O(n)$ holonomy has two discrete sectors.
The Witten index is given by \cite{Kim:2012gu}
\begin{align}
Z=\frac{Z^{+}+Z^{-}}{2}\ .
\end{align}
The Weyl factors $|W|$ of $O(2n)_\pm$, $O(2n+1)_\pm$, $Sp(k)$
are given by \cite{Kim:2012gu}
\begin{equation}
  |W_{O(2n)_{+}}|=\frac{1}{2^{n-1}n!}\ ,\ |W_{O(2n)_{-}}|=\frac{1}{2^{n-1}(n-1)!}
  \ ,\ |W_{O(2n+1)_{+}}|=|W_{O(2n+1)_{-}}|=\frac{1}{2^{n}n!}\ ,\
  |W_{Sp(k)}|=\frac{1}{2^{k}k!}\ .
\end{equation}
For $SO(N)$ with odd $N$, one can show that
$Z_{\textrm{1-loop}}=0$ in $O(2n)_{-}$ and $O(2n+1)_{+}$ sectors, since the
fermionic zero modes from $\Lambda$ (in Table \ref{fig:matter_odd}(b))
provide factors of $0$'s.

Let us call $Z_{k,n}$ the index of the $Sp(k)\times O(n)$
quiver. Being a multi-particle index, it acquires
contribution from $n$ hypermultiplet particles either bound or unbound to $k$
instantons. Also, as we shall explain in more detail below, $Z_{k,n}$ for
$n\geq 2$ also contains a spurious contribution from
particles not belonging to the 5d QFT. To explain these structures
clearly, we first discuss the indices $Z_{0,n}$ before considering the
instanton partition functions at $k\neq 0$.
At $n=1$, $k=0$, the $O(1)$ indices do not contain
integrals. The results are given by
\begin{align}
&Z^{SO(2N)}_{0,1}=\frac{1}{2}\left(\frac{\prod_{l=1}^{N}2\,\textrm{sinh}\frac{v_{l}}{2}}{2\,\textrm{sinh}\frac{\epsilon_{1}}{2}\cdot 2\,\textrm{sinh}\frac{\epsilon_{2}}{2}}+\frac{\prod_{l=1}^{N}2\,\textrm{cosh}\frac{v_{l}}{2}}{2\,\textrm{sinh}\frac{\epsilon_{1}}{2}\cdot 2\,\textrm{sinh}\frac{\epsilon_{2}}{2}}\right)\nonumber\\
&Z^{SO(2N+1)}_{0,1}=\frac{\prod_{l=1}^{N}2\,\textrm{cosh}\frac{v_{l}}{2}}{2\,\textrm{sinh}\frac{\epsilon_{1}}{2}\cdot 2\,\textrm{sinh}\frac{\epsilon_{2}}{2}}\ .
\label{eq:Z01}
\end{align}
The overall factors  $(2\,\textrm{sinh}\frac{\epsilon_{1}}{2}\cdot 2\,\textrm{sinh}\frac{\epsilon_{2}}{2})^{-1}$ in \eqref{eq:Z01} come
from the center-of-mass motion on $\mathbb{R}^4$. The remaining factor is
the character of the $SO(2N)$ chiral spinor
$\frac{\prod_{l=1}^N2\sinh\frac{v_l}{2}+\prod_{l=1}^N2\cosh\frac{v_l}{2}}{2}$,
and that of $SO(2N\!+\!1)$ spinor $\prod_{l=1}^N2\cosh\frac{v_l}{2}$,
respectively. They are the perturbative partition functions of
matters in $SO(N)$ in spinor representations.
Next, $Z_{0,2}$ is given by
\begin{align}\label{Z02-integral}
&Z^{SO(2N)}_{0,2}=\frac{1}{2}\left[\oint\frac{d\chi}{2\pi i}\frac{2\,\textrm{sinh}\,\epsilon_{+}\cdot\prod_{l=1}^{N}2\,
\textrm{sinh}(\frac{v_{l}\pm\chi}{2})}{2\,\textrm{sinh}(\frac{\epsilon_{1,2}}{2})
\cdot 2\,
\textrm{sinh}(\frac{\epsilon_{1,2}\pm2\chi}{2})}+\frac{2\,\textrm{cosh}\,
\epsilon_{+}\cdot\prod_{l=1}^{N}2\,\textrm{sinh}\,v_{l}}{2\,
\textrm{cosh}\frac{\epsilon_{1,2}}{2}\cdot(2\,\textrm{sinh}\frac{\epsilon_{1,2}}{2}
)^{2}}\right]\nonumber\\
&Z^{SO(2N+1)}_{0,2}=\frac{1}{2}\oint\frac{d\chi}{2\pi i}\frac{2\,\textrm{sinh}\,\epsilon_{+}\cdot\prod_{l=1}^{N}2\,
\textrm{sinh}(\frac{v_{l}\pm\chi}{2})\cdot 2\,\textrm{sinh}(\pm\frac{\chi}{2})}{2\,\textrm{sinh}(\frac{\epsilon_{1,2}}{2})\cdot 2\,\textrm{sinh}(\frac{\epsilon_{1,2}\pm2\chi}{2})}.
\end{align}
For $SO(2N+1)$ and the first term of $SO(2N)$ index,
one should evaluate JK-Res. With the choice $\eta>0$,
one keeps the residues at $\chi=-\frac{\epsilon_{1,2}}{2}$ and $\chi=-\frac{\epsilon_{1,2}}{2}+\pi i$.\footnote{For $O(n)$ and
$Sp(k)$ gauge theories, the choice of $\eta$ does not affect the results
due to Weyl symmetry \cite{Hwang:2014uwa}.}
For $N\leq 6$, one obtains
\begin{align}
Z^{SO(N)}_{0,2}&=\frac{Z^{SO(N)}_{0,1}(\epsilon_{\pm},v_{l})^{2}
+Z^{SO(N)}_{0,1}(2\epsilon_{\pm},2v_{l})}{2}\ ,
\label{eq:Z02pert}
\end{align}
while for $N=7,8$ one obtains
\begin{align}
Z^{SO(N)}_{0,2}&=\frac{Z^{SO(N)}_{0,1}(\epsilon_{\pm},v_{l})^{2}
+Z^{SO(N)}_{0,1}(2\epsilon_{\pm},2v_{l})}{2}
-\frac{1}{2}\cdot\frac{2\,\textrm{cosh}\,\epsilon_{+}}{2\,
\textrm{sinh}\frac{\epsilon_{1,2}}{2}}\ .
\label{eq:Z02ext}
\end{align}
\eqref{eq:Z02pert} and the first term of \eqref{eq:Z02ext} are
the indices of two non-interacting identical particles, whose single particle
index is given by $Z_{0,1}^{SO(N)}$. There are no bound states
formed by these perturbative hypermultiplet particles, as expected.
The second term of \eqref{eq:Z02ext}
requires more explanations, which we now turn to.

\begin{figure}[t!]
    \centering\vspace{0.3cm}
        \includegraphics[width=6cm]{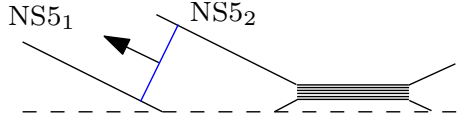}
	\caption{D1$'$-brane escaping the 5d QFT}
	\label{fig:extra}
\end{figure}
The second term of (\ref{eq:Z02ext}) comes from extra states in the brane
system that do not belong to the 5d QFT.
In particular, the fractional coefficient in the fugacity expansion
implies that it comes from a sector which has
a continuum unlifted by our massive deformations. In fact, following the
arguments presented between (\ref{SO7-spinor-integrand}) and
(\ref{SO7-spinor-residue}), one finds that
the linear 1-loop potential from (\ref{Z02-integral}) vanishes for
$N=7,8$, implying continua.
Physically, this comes from a D1$^\prime$-brane moving away
from 5d QFT, suspended between two parallel 5-branes as in Fig.~\ref{fig:extra}.
Although we are not aware of fully logical arguments, it has been
empirically observed that the last term
$-\frac{1}{2}\frac{2\cosh\epsilon_+}{2\sinh\frac{\epsilon_{1,2}}{2}}$ is
the contribution from the escaping particle for strings suspended between
parallel 5-branes: e.g. see eqn.(3.62) of \cite{Hwang:2014uwa}. See also \cite{Hayashi:2013qwa,Bao:2013pwa,Bergman:2013ala} for related results.
The suspended string of Fig.~\ref{fig:extra} carries the same spacetime and
R-symmetry quantum numbers as a 5d vector multiplet particles, since the
configuration of Fig.~\ref{fig:extra} is locally dual to a fundamental
string suspended between two D5-branes (a 5d vector W-boson). Indeed, the chemical
potential dependence
$\sim\frac{2\cosh\epsilon_+}{2\sinh\frac{\epsilon_{1,2}}{2}}$ is precisely that
of a 5d W-boson and its superpartners. Such extra states start to appear at
$n\geq 2$, since at $n=1$, one only has fractional D1$^\prime$ stuck to O5.

Collecting all, we expect that the partition function at $k=0$ is given by
\begin{align}
\sum_{n=0}^{\infty}e^{-nm}Z^{SO(N)}_{0,n}=Z_{\textrm{pert}}\equiv
\textrm{PE}\left[e^{-m}Z^{SO(N)}_{0,1}\right]
\end{align}
for $N\leq 6$, while for $N=7,8$ we expect that it is given by
\begin{align}
\sum_{n=0}^{\infty}e^{-nm}Z^{SO(N)}_{0,n}=Z_{\textrm{pert}} Z_{\textrm{extra}}=\textrm{PE}\left[e^{-m}Z^{SO(N)}_{0,1}\right]
\textrm{PE}\left[-\frac{1}{2}e^{-2m}\frac{2\textrm{cosh}\,\epsilon_{+}}{2\,
\textrm{sinh}\frac{\epsilon_{1,2}}{2}}\right]\ .
\label{eq:Zrest}
\end{align}
Here, $Z_{0,0}\equiv 1$ by definition,
and $PE\left[f(x,y,\cdots)\right]\equiv\exp\left[\sum_{n=1}^\infty\frac{1}{n}
f(nx,ny,\cdots)\right]$ is the multi-particle index for the single particle index
$f$.

The full partition function would factorize as
\begin{align}
\sum_{k,n=0}^\infty
q^ke^{-nm}Z^{SO(N)}_{k,n}=Z_{\textrm{inst}}(q,\epsilon_{1,2},m)
Z_{\textrm{pert}}(\epsilon_{1,2},m)Z_{\textrm{extra}}(\epsilon_{1,2},m).
\label{eq:Zfact}
\end{align}
Expanding $Z_{\rm inst}(q)=\sum_{k=0}^\infty Z_kq^k$ where $Z_0\equiv 1$,
\eqref{eq:Zfact} implies at given $q^k$ order that
\begin{align}\label{pert-extra}
\sum_{n=0}^{\infty}e^{-nm}Z_{k,n}=Z_{k}Z_{\textrm{pert}} Z_{\textrm{extra}}.
\end{align}
When there are two 5d hypermultiplets as in Fig. \ref{fig:D1spin2},
the full partition function is
\begin{align}
Z^{(2)}=\sum_{k,n_{1},n_{2}=0}^\infty
q^ke^{-n_{1}m_{1}-n_{2}m_{2}}Z_{k_{1},n_{1},n_{2}}
\end{align}
where $Z_{k,n_{1},n_{2}}$ is the index for $k$ D1, $n_{1}$ D1$'$ and $n_{2}$ D1$''$.  $m_{1,2}$ are the $Sp(2)$ flavor chemical potentials. The contributions from
perturbative and extra degrees of freedom in this case are
\begin{eqnarray}
Z^{(2)}_{\textrm{pert}}(\epsilon_{1,2},m_{1,2})&=&
Z_{\textrm{pert}}(\epsilon_{1,2},m_{1}) Z_{\textrm{pert}}(\epsilon_{1,2},m_{2})\nonumber\\
Z^{(2)}_{\textrm{extra}}(\epsilon_{1,2},m_{1,2})&=&
Z_{\textrm{extra}}(\epsilon_{1,2},m_{1})Z_{\textrm{extra}}(\epsilon_{1,2},m_{2})
\end{eqnarray}
where $Z_{\textrm{pert}}$ and $Z_{\textrm{extra}}$ take the same forms
as in (\ref{eq:Zrest}).

Although our methods apply well to both $SO(8)$ and $SO(7)$,
we only study the cases with $SO(7)$ in this paper.
We start from the case with one hypermultiplet field.
From the field contents of Fig.~\ref{fig:matter_odd}, $Z_{\textrm{1-loop}}$
for $k$ instantons and $2n$($=$even) hypermultiplet particles is given by
\begin{eqnarray}
\hspace*{-1cm}[Z_{\textrm{1-loop}}]^{SO(7)}_{k,2n}&=&
\frac{1}{2^{k}\,k!}\frac{(2\,\textrm{sinh}\,\epsilon_{+})^{k}\cdot
\prod_{i=1}^{k}2\,\textrm{sinh}(\epsilon_{+}\pm\phi_{i})\,2\,
\textrm{sinh}(\pm\phi_{i})\cdot\prod_{i>j}2\,
\textrm{sinh}(\frac{2\epsilon_{+}\pm\phi_{i}\pm\phi_{j}}{2})\,2\,
\textrm{sinh}(\frac{\pm\phi_{i}\pm\phi_{j}}{2})}
{(2\,
\textrm{sinh}(\frac{\epsilon_{1,2}}{2}))^{k}\cdot\prod_{i>j}2\,
\textrm{sinh}(\frac{\epsilon_{1,2}\pm\phi_{i}\pm\phi_{j}}{2})\cdot
\prod_{i=1}^{k}\prod_{l=1}^{3}2\,\textrm{sinh}(\frac{\epsilon_{+}\pm\phi_{i}\pm
v_{l}}{2})\cdot\prod_{i=1}^{k}2\,\textrm{sinh}(\frac{\epsilon_{+}\pm\phi_{i}}{2})}
\nonumber\\
&&\cdot\frac{1}{2^{n}\,n!}\frac{(2\,\textrm{sinh}\,\epsilon_{+})^{n}\cdot
\prod_{I>J}2\,\textrm{sinh}(\frac{2\epsilon_{+}\pm\chi_{I}\pm\chi_{J}}{2})
\cdot\prod_{I>J}2\,\textrm{sinh}(\frac{\pm\chi_{I}\pm\chi_{J}}{2})}{(2\,
\textrm{sinh}(\frac{\epsilon_{1,2}}{2}))^{n}\cdot\prod_{I=1}^{n}2\,
\textrm{sinh}(\frac{\epsilon_{1,2}\pm 2\chi_{I}}{2})\cdot\prod_{I>J}2\,
\textrm{sinh}(\frac{\epsilon_{1,2}\pm\chi_{I}\pm\chi_{J}}{2})}\nonumber\\
&&\cdot\prod_{I=1}^{n}\prod_{l=1}^{3}2\,\textrm{sinh}(\frac{\pm\chi_{I}+v_{l}}{2})
\cdot\prod_{I=1}^{n}2\,\textrm{sinh}(\frac{\chi_{I}}{2})\cdot\frac{\prod_{i=1}^{k}
\prod_{I=1}^{n}2\,\textrm{sinh}(\frac{\epsilon_{-}\pm\phi_{i}\pm\chi_{I}}{2})}
{\prod_{i=1}^{k}\prod_{I=1}^{n}2\,\textrm{sinh}(\frac{-\epsilon_{+}\pm
\phi_{i}\pm\chi_{I}}{2})}\ ,
\label{eq:SO7Z1loopeven}
\end{eqnarray}
while $Z_{\textrm{1-loop}}$ for $k$ instantons and $2n+1$ hypermultiplet particles is given by
\begin{eqnarray}
\hspace*{-1cm}[Z_{\textrm{1-loop}}]^{SO(7)}_{k,2n+1}&=&
\frac{1}{2^{k}\,k!}\frac{(2\,\textrm{sinh}\,\epsilon_{+})^{k}\cdot
\prod_{i=1}^{k}2\,\textrm{sinh}(\epsilon_{+}\pm\phi_{i})\,2\,
\textrm{sinh}(\pm\phi_{i})\cdot\prod_{i>j}2\,
\textrm{sinh}(\frac{2\epsilon_{+}\pm\phi_{i}\pm\phi_{j}}{2})\,2\,
\textrm{sinh}(\frac{\pm\phi_{i}\pm\phi_{j}}{2})}{(2\,
\textrm{sinh}(\frac{\epsilon_{1,2}}{2}))^{k}\cdot\prod_{i>j}2\,
\textrm{sinh}(\frac{\epsilon_{1,2}\pm\phi_{i}\pm\phi_{j}}{2})\cdot
\prod_{i=1}^{k}\prod_{l=1}^{3}2\,\textrm{sinh}(\frac{\epsilon_{+}\pm\phi_{i}\pm
v_{l}}{2})\cdot\prod_{i=1}^{k}2\,\textrm{sinh}(\frac{\epsilon_{+}\pm\phi_{i}}{2})}
\nonumber\\
&&\cdot\frac{1}{2^{n}\,n!}\frac{(2\,\textrm{sinh}\,\epsilon_{+})^{n}\cdot
\prod_{I>J}2\,\textrm{sinh}(\frac{2\epsilon_{+}\pm\chi_{I}\pm\chi_{J}}{2})\cdot
\prod_{I>J}2\,\textrm{sinh}(\frac{\pm\chi_{I}\pm\chi_{J}}{2})}{(2\,
\textrm{sinh}(\frac{\epsilon_{1,2}}{2}))^{n+1}\cdot\prod_{I=1}^{n}2\,
\textrm{sinh}(\frac{\epsilon_{1,2}\pm 2\chi_{I}}{2})\cdot\prod_{I>J}2\,
\textrm{sinh}(\frac{\epsilon_{1,2}\pm\chi_{I}\pm\chi_{J}}{2})\prod_{I=1}^{n}2\,
\textrm{cosh}(\frac{\epsilon_{1,2}\pm\chi_{I}}{2})}\nonumber\\
&&\cdot\prod_{I=1}^{n}\prod_{l=1}^{3}2\,\textrm{sinh}(\frac{\pm\chi_{I}+v_{l}}{2})
\cdot\prod_{I=1}^{n}2\,\textrm{sinh}(\frac{\chi_{I}}{2})\cdot\prod_{I=1}^{n}2\,
\textrm{cosh}(\frac{2\epsilon_{+}\pm\chi_{I}}{2})\cdot\prod_{I=1}^{n}2\,
\textrm{cosh}(\frac{\chi_{I}}{2})\prod_{l=1}^{3}2\,\textrm{cosh}(\frac{v_{l}}{2})
\nonumber\\
&&\cdot\frac{\prod_{i=1}^{k}\prod_{I=1}^{n}2\,
\textrm{sinh}(\frac{\epsilon_{-}\pm\phi_{i}\pm\chi_{I}}{2})\cdot\prod_{i=1}^{k}2\,
\textrm{cosh}(\frac{\epsilon_{-}\pm\phi_{i}}{2})}{\prod_{i=1}^{k}\prod_{I=1}^{n}2\,
\textrm{sinh}(\frac{-\epsilon_{+}\pm\phi_{i}\pm\chi_{I}}{2})\cdot\prod_{i=1}^{k}2\,
\textrm{cosh}(\frac{-\epsilon_{+}\pm\phi_{i}}{2})}\ .
\label{eq:SO7Z1loopodd}
\end{eqnarray}
$i,j=1,\cdots, k$ are $Sp(k)$ indices, $I,J=1,\cdots,n$ are $O(2n)$ or $O(2n+1)$ indices, and $l=1,2,3$ are $SO(7)$ indices. \eqref{eq:SO7Z1loopeven} and \eqref{eq:SO7Z1loopodd} are computed on either $O(n)_{+}$ or $O(n)_{-}$ sector,
where $\chi_I$ are eigenvalues of $\log U^\pm$ given by (\ref{eq:hol}).

The partition function at $k=1$, $n=0$ is given by
\begin{align}
Z^{SO(7)}_{1,0}=\oint\frac{d\phi}{2\pi i}\,\frac{1}{2}\cdot\frac{2\,\textrm{sinh}\,\epsilon_{+}\cdot2\,
\textrm{sinh}(\epsilon_{+}\pm\phi)\cdot 2\,\textrm{sinh}(\pm\phi)}{2\,\textrm{sinh}(\frac{\epsilon_{1,2}}{2})
\cdot\prod_{l=1}^{3}2\,\textrm{sinh}(\frac{\epsilon_{+}\pm\phi\pm v_{l}}{2})\cdot 2\,\textrm{sinh}(\frac{\epsilon_{+}\pm\phi}{2})}\ .
\end{align}
Poles chosen at $\eta>0$ are $\phi=-\epsilon_{+}$, $\phi=-\epsilon_{+}\pm v_{l}$,
but the residue from $\phi=-\epsilon_+$ vanishes. Collecting the residues,
one obtains
\begin{eqnarray}
  Z^{SO(7)}_{1,0}&=&\frac{t}{(1-tu)(1-t/u)}\prod_{i<j}
  \frac{t^4}{(1-t^2b_i^\pm b_j^\pm)}(\chi_{\boldsymbol{9}}^{SU(2)}
  +\chi_{\boldsymbol{7}}^{SU(2)}(\chi_{\boldsymbol{7}}^{SO(7)}+1)\nonumber\\
  &&+\chi_{\boldsymbol{5}}^{SU(2)}(-\chi_{\boldsymbol{35}}^{SO(7)}
  +\chi_{\boldsymbol{7}}^{SO(7)}+1)
  +\chi_{\boldsymbol{3}}^{SU(2)}(-\chi_{\boldsymbol{35}}^{SO(7)}
  +\chi_{\boldsymbol{27}}+1)+\chi_{\boldsymbol{105}}^{SO(7)}
  -\chi_{\boldsymbol{21}}^{SO(7)}+\chi_{\boldsymbol{7}}^{SO(7)})\nonumber\\
  &=&\frac{t}{(1-tu)(1-t/u)}\sum_{p=0}^{\infty}\chi_{(0,p,0)}(v_{l})\,t^{2p+4}
\label{eq:SO7Z10}
\end{eqnarray}
where $t=e^{-\epsilon_{+}}$ and $u=e^{-\epsilon_{-}}$. Here $\chi_{\bf R}$ is
the character of $SO(7)$ representation ${\bf R}$.
This is simply the well-known 1-instanton partition function of $SO(7)$ gauge theory. E.g. see \cite{Benvenuti:2010pq} for the above character expansion form.

Next, consider the sector at $k=1$, $n=1$. $Z_{1,1}^{SO(7)}$ is given by
\begin{align}
Z^{SO(7)}_{1,1}=\oint\frac{d\phi}{2\pi i}\,[Z_{\textrm{1-loop}}]^{SO(7)}_{1,0}
\frac{\prod_{l=1}^{3}2\,\textrm{cosh}(\frac{v_{l}}{2})}{2\,
\textrm{sinh}(\frac{\epsilon_{1,2}}{2})}\cdot\frac{2\,
\textrm{cosh}(\frac{\epsilon_{-}\pm\phi}{2})}{2\,
\textrm{cosh}(\frac{-\epsilon_{+}\pm\phi}{2})}.
\end{align}
Poles chosen at $\eta>0$ with nonzero residues are at $\phi=-\epsilon_{+}\pm v_{l}$.
As we explained around (\ref{pert-extra}),
$Z^{SO(7)}_{1,1}$ has contributions from $Z_{\textrm{pert}}$ at $n=1$.
Let us call the proper contribution to the instanton partition function
$\hat{Z}^{SO(7)}_{k,n}$. From (\ref{pert-extra}), one obtains
\begin{align}
\hat{Z}^{SO(7)}_{1,1}=Z^{SO(7)}_{1,1}-Z^{SO(7)}_{1,0}Z^{SO(7)}_{0,1}\ .
\end{align}
Hat denotes the instanton partition function at level $(k,n)$, while
$Z_{k,n}$ is simply the Witten index of our $Sp(k)\times O(n)$ quantum mechanics.
From this formula, one obtains
\begin{eqnarray}
  \hspace*{-0.5cm}\hat{Z}^{SO(7)}_{1,1}&=&\frac{t}{(1-tu^{\pm 1})}\prod_{i<j}
  \frac{t^4}{(1-t^2b_i^\pm b_j^\pm)}(-\chi_{\boldsymbol{8}}^{SU(2)}\chi_{\boldsymbol{8}}^{SO(7)}
  -\chi_{\boldsymbol{6}}^{SU(2)}\chi_{\boldsymbol{8}}^{SU(2)}
  +\chi_{\boldsymbol{4}}^{SU(2)}\chi_{\boldsymbol{112}}^{SO(7)}
  -\chi_{\boldsymbol{2}}^{SU(2)}\chi_{\boldsymbol{168}}^{SO(7)})\nonumber\\
  &=&-\frac{t}{(1-tu^{\pm 1})}\sum_{p=0}^{\infty}\chi_{(0,p,1)}\,t^{2p+5}
\label{eq:SO7Z11}
\end{eqnarray}
where $\sum_{p=0}^{\infty}\chi_{(0,p,1)}\,t^{2p+5}=\chi_{\boldsymbol{8}}(v_{l})
+\chi_{\boldsymbol{112}}(v_{l})\,t^{2}+\chi_{\boldsymbol{720}}(v_{l})\,t^{4}
+\cdots$. Then consider the sector at $k=1$, $n=2$. $Z_{1,2}^{SO(7)}$ is given by
the $Sp(1)\times O(2)$ contour integral
\begin{equation}
Z^{SO(7)}_{1,2}=\oint\frac{d\phi\,d\chi}{(2\pi i)^{2}}\,[Z_{\textrm{1-loop}}]^{SO(7)}_{1,0}\cdot\frac{1}{2}\cdot\frac{2\,\textrm{sinh}\,
\epsilon_{+}\cdot\prod_{l=1}^{3}2\,\textrm{sinh}(\frac{\pm\chi+v_{l}}{2})\cdot
2\,\textrm{sinh}(\frac{\pm\chi}{2})}{2\,\textrm{sinh}(\frac{\epsilon_{1,2}}{2})\cdot
2\,\textrm{sinh}(\frac{\epsilon_{1,2}\pm 2\chi}{2})}\cdot\frac{2\,
\textrm{sinh}(\frac{\epsilon_{-}\pm\phi\pm\chi}{2})}{2\,\textrm{sinh}
(\frac{-\epsilon_{+}\pm\phi\pm\chi}{2})}\ .
\end{equation}
Taking $\eta=(1,1+\epsilon)$  for small positive $\epsilon$
\cite{Hwang:2014uwa}, the poles at
$(\phi, \chi)=(-\epsilon_{+}\pm v_{l}, -\frac{\epsilon_{1,2}}{2}[+\pi i])$, $(-\epsilon_{+}\pm v_{l}, \pm v_{l})$, $(\epsilon_{+}\pm v_{l}, \mp v_{l})$, $(0[+\pi i], \epsilon_{+}+[\pi i])$, $(\frac{\epsilon_{1,2}}{2}[+\pi i],-\frac{\epsilon_{2,1}}{2}[+\pi i])$, $(\frac{3\epsilon_{+}\pm\epsilon_{-}}{2},-\frac{\epsilon_{1,2}}{2}[+\pi i])$ are
chosen. $[+\pi i]$ means that there are two cases with and without $+\pi i$ addition.
Subtracting the contribution from $Z_{\textrm{pert}}Z_{\textrm{extra}}$ in (\ref{pert-extra}),
the instanton partition function $\hat{Z}_{1,2}^{SO(7)}$ at this order
is given by $\hat{Z}_{1,2}=Z_{1,2}-\hat{Z}_{1,1}Z_{0,1}-Z_{1,0}Z_{0,2}$.
One finds after computations that
\begin{align}
\hat{Z}^{SO(7)}_{1,2}=Z^{SO(7)}_{1,0}.
\label{eq:SO7Z12}
\end{align}
For $n\geq 3$, we find $\hat{Z}_{1,n}=0$. We checked this exactly for
$n=3$. For $n=4$, to save time, we plugged in random numbers in
the chemical potentials and checked that $\hat{Z}^{SO(7)}_{1,4}$
is very small. (Below, we present an argument for this phenomenon.)

Collecting all the computations at $n=0,1,2$, one obtains
\begin{eqnarray}\label{SO7-Nf=1-D-brane}
  Z_{k=1}&=&
  e^m\left[Z_{1,0}+e^{-m}\hat{Z}_{1,1}+e^{-2m}\hat{Z}_{1,2}\right]\\
  &=&\frac{t}{(1-tu^{\pm 1})}\prod_{i<j}\frac{t^4}{(1-t^2b_i^{\pm}b_j^{\pm})}
  \left[-\chi_{\bf 8}^{SU(2)}\chi_{\bf 8}^{SO(7)}
  -\chi_{\boldsymbol{6}}^{SU(2)}\chi_{\boldsymbol{8}}^{SO(7)}
  +\chi_{\boldsymbol{4}}^{SU(2)}\chi_{\boldsymbol{112}}^{SO(7)}
  -\chi_{\boldsymbol{2}}^{SU(2)}\chi_{\boldsymbol{168}}^{SO(7)}\right.\nonumber\\
  &&+\left(\chi_{\boldsymbol{9}}^{SU(2)}
  +\chi_{\boldsymbol{7}}^{SU(2)}(\chi_{\boldsymbol{7}}^{SO(7)}+1)
  +\chi_{\boldsymbol{5}}^{SU(2)}(-\chi_{\boldsymbol{35}}^{SO(7)}
  +\chi_{\boldsymbol{7}}^{SO(7)}+1)\right.\nonumber\\
  &&\left.\left.
  +\chi_{\boldsymbol{3}}^{SU(2)}(-\chi_{\boldsymbol{35}}^{SO(7)}
  +\chi_{\boldsymbol{27}}+1)+\chi_{\boldsymbol{105}}^{SO(7)}
  -\chi_{\boldsymbol{21}}^{SO(7)}+\chi_{\boldsymbol{7}}^{SO(7)}\right)
  (e^m+e^{-m})\right]
  \nonumber
\end{eqnarray}
Here we multiplied an overall factor $e^m$, like the `zero point energy'
factor, to have the expected Weyl symmetry $m\rightarrow -m$ of the $Sp(1)$
flavor symmetry. Noting that $e^m+e^{-m}=\chi_{\bf 2}^{Sp(1)}$, (\ref{SO7-Nf=1-D-brane}) completely agrees with (\ref{SO7-Nf=1}), supporting
our ADHM-like proposals of section 2 at $n_{\bf 8}=1$.

Here we discuss more about the maximal value of $n$
with $\hat{Z}^{SO(7)}_{k,n}\neq 0$, at given $k$. Note that
\begin{equation}\label{weyl-Sp(1)}
  Z_{\rm inst}(q,\epsilon_{1,2},v,m)=e^{-\epsilon_0}
  \sum_{k=0}^\infty\sum_{n=0}^{\infty}
  q^ke^{-nm}\hat{Z}^{SO(7)}_{k,n}(\epsilon_{1,2},v)\ ,
\end{equation}
refining the previous definition by the zero point energy-like factor.
Note that $m$ is the flavor chemical potential for the 5d hypermultiplet.
Since a hypermultiplet only
adds fermion zero modes on the instanton moduli space, the rotation parameter
$m$ acts only on these fermions. So unlike the chemical potentials $v_i$,
$\epsilon_{1,2}$ which
act on noncompact zero modes, the coefficient $Z_k$ of $Z_{\rm inst}$ at given
$q^k$ order should not have any poles in $m$. Since $Z_k$ 
admits fugacity expansions, this implies that $Z_k$ is a
finite polynomial in $e^m$ and $e^{-m}$. So the sum over $n$
should truncate to $0\leq n\leq n_{\rm max}$ for some finite $n_{\rm max}$,
also with a suitable $m$ dependent $\epsilon_0$ to ensure the Weyl symmetry of
$Sp(1)$. One can also naturally infer the value of $n_{\rm max}$.
To see this, note that a 5d hypermultiplet in the spinor representation induces
$kD({\bf 8})=2kT({\bf 8})=2k$ complex fermion zero modes on the moduli
space, where we used $2T=2^{N-2}$ for $SO(2N\!+\!1)$ spinor representation.
Quantizing them into $2k$ pairs of fermionic harmonic oscillators, each oscillator
raises/lowers the particle number $n$ by $1$. This means that the charge difference
between the lowest and highest states is $2k$, implying $n_{\rm max}=2k$. Then
$Sp(1)$ Weyl symmetry implies $n\rightarrow -n$ symmetry, demanding
$\epsilon_0=-km$ and $\hat{Z}_{k,2k-n}=\hat{Z}_{k,n}$.
These completely agree with our empirical findings around (\ref{eq:SO7Z12}).
Below, we shall proceed with these properties assumed.

One can study the case with $k=2$ in the same manner. We computed it at $v_{l}=\epsilon_{-}=0$ due to computational complications.
We simply report the results:
\begin{align}
Z^{SO(7)}_{2,0}&=\frac{t^{10}}{(1-t)^{20}(1+t)^{10}(1+t+t^{2})^{9}}(1+t+15t^{2}+48t^{3}+152t^{4}+446t^{5}+1126t^{6}+2374t^{7}\nonumber\\
&+4674t^{8}+8184t^{9}+12680t^{10}+17816t^{11}+22957t^{12}+26449t^{13}+27622t^{14}+\cdots+t^{28})\nonumber\\
\hat{Z}^{SO(7)}_{2,1}&=-\frac{8\,t^{11}}{(1-t)^{20}(1+t)^{10}(1+t+t^{2})^{9}}(1+3t+17t^{2}+62t^{3}+183t^{4}+477t^{5}+1109t^{6}+2206t^{7}\nonumber\\
&+3921t^{8}+6285t^{9}+9004t^{10}+11543t^{11}+13459t^{12}+14194t^{13}+\cdots+t^{26})\nonumber\\
\hat{Z}^{SO(7)}_{2,2}&=\frac{t^{10}}{(1-t)^{20}(1+t)^{10}(1+t+t^{2})^{9}}
(1+3t+45t^{2}+176t^{3}+647t^{4}+2087t^{5}+5560t^{6}\!+12639t^{7}\nonumber\\
&+25923t^{8}\!+46880t^{9}\!+74843t^{10}\!+107589t^{11}\!+139877t^{12}\!+162758t^{13}
\!+170752t^{14}\!+\cdots+t^{28})
\label{eq:SO7Zk2}
\end{align}
Here, the omitted terms in $\cdots$ can be restored from the fact that
coefficients of $t^p$ and $t^{28-p}$ are same in the numerator of $Z_{2,0}$,
and also from similar reflection symmetries in $\hat{Z}_{2,1}$, $\hat{Z}_{2,2}$.
Assuming $\hat{Z}_{k,n}=0$ for $n>4$ and $\hat{Z}_{2,n}=\hat{Z}_{2,4-n}$,
as discussed in the previous paragraph, one can compute the full $2$ instanton
partition function for $SO(7)$ gauge theory at $n_{\bf 8}=1$,
\begin{equation}
  Z_{k=2}=e^{2m}\sum_{n=0}^4e^{-nm}\hat{Z}_{k=2,n}\ .
\end{equation}
We have checked that this completely agrees with our index of section 2.

Next we consider the instanton quantum mechanics of 5d $SO(7)$ gauge theory with two hypermultiplets. From Fig.~\ref{fig:matter_two_odd}, the contour integrand
$Z_{\textrm{1-loop}}$ of $k$ instantons with $n_{1}$ and $n_{2}$ hypermultiplet particles is given by
\begin{align}
[Z_{\textrm{1-loop}}]^{SO(7)}_{k,n_{1},n_{2}}(\phi_{i},\chi_{I},\chi'_{I'})
=\frac{[Z_{\textrm{1-loop}}]^{SO(7)}_{k,n_{1}}(\phi_{i},\chi_{I})\cdot [Z_{\textrm{1-loop}}]^{SO(7)}_{k,n_{2}}(\phi_{i},\chi'_{I'})}
{[Z_{\textrm{1-loop}}]^{SO(7)}_{k,0}(\phi_{i})}
\end{align}
where $[Z_{\textrm{1-loop}}]^{SO(7)}_{k,n}$ is given by
\eqref{eq:SO7Z1loopeven}, \eqref{eq:SO7Z1loopodd}. Here $i=1,\cdots,k$ is the
$Sp(k)$ index, $I=1,\cdots,n_{1}$ and $I'=1,\cdots,n_{2}$ are $O(n_{1})$ and
$O(n_{2})$ indices respectively. We summarize the results of our calculations:
\begin{align}\label{two-spinor-index}
  &Z_{1,0,0}^{SO(7)}=\hat{Z}^{SO(7)}_{1,0,2}=\hat{Z}^{SO(7)}_{1,2,0}=Z^{SO(7)}_{1,0}\\
  &\hat{Z}^{SO(7)}_{1,1,0}=\hat{Z}^{SO(7)}_{1,0,1}=\hat{Z}^{SO(7)}_{1,1,2}=
  \hat{Z}^{SO(7)}_{1,2,1}=\hat{Z}^{SO(7)}_{1,1}\nonumber\\
  &\hat{Z}^{SO(7)}_{1,1,1}=
\frac{t}{(1-tu^{\pm1})}\prod_{i<j}\frac{t^4}{(1-t^2b_i^{\pm}b_j^{\pm})}\bigr[
\chi^{SU(2)}_{\boldsymbol{9}}+\chi^{SU(2)}_{\boldsymbol{7}}
(\chi^{SO(7)}_{\boldsymbol{35}}+\chi^{SO(7)}_{\boldsymbol{7}}+1)
\nonumber\\
&+\chi^{SU(2)}_{\boldsymbol{5}}(-\chi^{SO(7)}_{\boldsymbol{105}}+1)
+\chi^{SU(2)}_{\boldsymbol{3}}(-\chi^{SO(7)}_{\boldsymbol{168'}}
+\chi^{SO(7)}_{\boldsymbol{77}}-\chi^{SO(7)}_{\boldsymbol{21}})
+\chi^{SO(7)}_{\boldsymbol{330}}+\chi^{SO(7)}_{\boldsymbol{189}}
+\chi^{SO(7)}_{\boldsymbol{27}}\bigr]\nonumber
\end{align}
where $Z^{SO(7)}_{1,0}$ and $Z^{SO(7)}_{1,1}$ are given by \eqref{eq:SO7Z10}, \eqref{eq:SO7Z11}.
With the data shown in (\ref{two-spinor-index}), one can compute
$Z_{k=1}$ for the $SO(7)$ at $n_{\bf 8}=2$, using the fermion zero mode structures
and $Sp(2)$ Weyl symmetry, extending the discussions for $n_{\bf 8}=1$ in the
paragraph containing (\ref{weyl-Sp(1)}). Namely, at $k$ instanton sector,
there are $2k$ fermion zero modes which rotate in $m_1$ and $m_2$, respectively.
This means that $(n_1)_{\rm max}=(n_2)_{\rm max}=2k$, with zero point energy
factor $e^{-\epsilon_0}=e^{k(m_1+m_2)}$ from Weyl symmetry.
Weyl symmetry also requires
$\hat{Z}_{k,n_1,n_2}=\hat{Z}_{k,2k-n_1,n_2}=\hat{Z}_{k,n_1,2k-n_2}$.
(Our calculus on the second line of (\ref{two-spinor-index}), relating
$\hat{Z}_{1,1,2}$, $\hat{Z}_{1,2,1}$ to other coefficients, partially reconfirms
this general argument.) With these structures and (\ref{two-spinor-index}),
one finds
\begin{eqnarray}
  Z_{1}&=&e^{m_1+m_2}\left[Z^{SO(7)}_{1,0}+(e^{-m_1}+e^{-m_2})\hat{Z}^{SO(7)}_{1,1}
  +(e^{-2m_1}+e^{-2m_2})Z^{SO(7)}_{1,0}+e^{-m_1-m_2}\hat{Z}^{SO(7)}_{1,1,1}\right.
  \nonumber\\
  &&\left.+(e^{-2m_1-m_2}+e^{-m_1-2m_2})\hat{Z}^{SO(7)}_{1,1}
  +e^{-2m_1-2m_2}Z^{SO(7)}_{1,0}
  \right]\nonumber\\
  &=&\chi_{\bf 4}^{Sp(2)}\hat{Z}_{1,1}^{SO(7)}+
  \chi_{\bf 5}^{Sp(2)}Z_{1,0}^{SO(7)}+\left(\hat{Z}_{1,1,1}^{SO(7)}
  -Z_{1,0}^{SO(7)}\right)
\end{eqnarray}
where $\chi_{\bf 4}^{Sp(2)}\!=\!\sum_{\pm}(e^{\pm m_1}\!+\!e^{\pm m_2})$,
$\chi_{\bf 5}^{Sp(2)}\!=\!1\!+\!\sum_{\pm,\pm}e^{\pm m_1\pm m_2}$.
This completely agrees with (\ref{SO7-Nf=2}).

As explained in section 2,
one can Higgs the $SO(7)$ gauge theory with a matter hypermultiplet in ${\bf 8}$, to pure $G_{2}$ Yang-Mills theory by giving VEV to the hypermultiplet. In the
index, this amounts to setting $m_{n_{\bf 8}}=\epsilon_+$, $v_4=0$. See section 2.2.
Since we have provided concrete tests of $SO(7)$ instanton partition functions
of section 2 using our D-brane-based methods, Higgsing both
sides do not yield any further significant information or tests. Namely,
calculations in this section at $n_{\bf 8}=1,2$ already tested our $G_2$
instanton calculus of section 2 at $n_{\bf 7}=0,1$. Therefore, we shall not
repeat the analysis of Higgsings to $G_2$ in our D-brane-based formalism.

\section{Strings of non-Higgsable 6d SCFTs}

In this section, we study the strings of non-Higgsable 6d SCFTs containing
$G_2$ theories or $SO(7)$ theories with matters in ${\bf 8}$. In particular,
we shall construct the 2d gauge theories for the strings of 6d atomic SCFTs
with $2$ and $3$ dimensional tensor branches \cite{Morrison:2012np}.

We first briefly review the `atomic classification'
\cite{Heckman:2013pva,Heckman:2015bfa,Morrison:2012np} of 6d $\mathcal{N}=(1,0)$
SCFTs. This is based on F-theory engineering of 6d SCFTs, on elliptic
Calabi-Yau 3-fold (CY$_3$). Elliptic CY$_3$ admits a $T^2$ fibration over a 4d
base $\mathcal{B}$, which is non-compact and singular. The singular point on
$\mathcal{B}$ hosts 6d degrees of freedom which decouple from 10d bulk at low energy.
In 6d QFT, resolving this singularity corresponds 
to going to the tensor branch. Namely, there is a 6d supermultiplet
called tensor multiplet, consisting of a self-dual 2-form potential $B_{\mu\nu}$
(whose field strength $H=dB+\cdots$ satisfies $H=\star_6 H$), a real scalar $\Phi$,
and fermions. Giving VEV to $\Phi$, one goes into the tensor branch.
Geometrically, the singularity of $\mathcal{B}$ is resolved into a collection of
intersecting 2-cycles $\mathbb{P}^1$.
Associated with the $i$'th $\mathbb{P}^1$, there is a
tensor multiplet $B^i$, $\Phi^i$, and sometimes a non-Abelian vector
multiplet $A^i$ with simple gauge group $G_i$.
The VEV of $\Phi^i$ is proportional to the volume of the $i$'th $\mathbb{P}^1$.
Depending on how the 2-cycles intersect, the vector multiplets form a sort of
`quiver' possibly with charged hypermultiplet matters.
Geometrically, the vector and hypermultiplets are determined by how the $T^2$
fiber degenerates on $\mathcal{B}$. Equivalently, they depend on the
7-branes wrapping $\mathcal{B}$.
With a given resolution of the singularity on $\mathcal{B}$,
there are families of theories related to others by Higgsings.
The classification of
\cite{Heckman:2013pva,Heckman:2015bfa,Morrison:2012np} proceeds by
first identifying possible non-Higgsable theories, and then considering
possible `un-Higgsings.'

\begin{table}[t!]
\begin{center}
\begin{tabular}{c||c|c|c|c|c|c|c|c|c
    }
	\hline
	$n$ & $1$ & $2$ & $3$ & $4$ & $5$ & $6$ & $7$ & $8$
    & $12$ \\
    \hline
    gauge symmetry & - & - & $SU(3)$ & $SO(8)$ & $F_4$ & $E_6$ & $E_7$ & $E_7$
    & $E_8$ \\
    \hline
    global symmetry &$ E_8$ & $SO(5)_R$ & - & - & - & - & - & - & 
    - \\
    \hline matter &&&-&-&-&-&$\frac{1}{2}{\bf 56}$&-&
    - \\
	\hline
\end{tabular}
\caption{Symmetries/matters of SCFTs with rank $1$
tensor branches}\label{minimal}
\vskip 0.5cm
\begin{tabular}{c||c|c|c}
	\hline
	base & $3,2$ & $3,2,2$ & $2,3,2$\\
    \hline
    gauge symmetry & $G_2\times SU(2)$ & $G_2\times SU(2)\times\{\ \}$ &
    $SU(2)\times SO(7)\times SU(2)$ \\
    \hline matter &$\frac{1}{2}({\bf 7}+{\bf 1},{\bf 2})$
    & $\frac{1}{2}({\bf 7}+{\bf 1},{\bf 2})$ &
    $\frac{1}{2}({\bf 2},{\bf 8},{\bf 1})+
    \frac{1}{2}({\bf 1},{\bf 8},{\bf 2})$ \\
	\hline
\end{tabular}
\caption{Non-Higgsable atomic SCFTs with higher rank
tensor branches}\label{other}
\end{center}
\end{table}
Non-Higgsable theories are constructed by first taking a finite set of `quiver
nodes' and connecting them with certain rules. Technically, the nodes are connected
by suitably gauging the E-string theory and identifying them with the
gauge groups of the quiver nodes. See \cite{Heckman:2013pva} for the detailed rules.
Roughly speaking, the possible `quiver nodes' are given in Tables~\ref{minimal} and
\ref{other}. More precisely, the SCFTs at $n=1$ and $n=2$ play
different roles: see \cite{Heckman:2013pva,Heckman:2015bfa} for
the precise ways of using the SCFTs in Table~\ref{minimal}, \ref{other}.
The SCFTs in Table \ref{minimal} are called `minimal SCFTs' in
\cite{Haghighat:2014vxa}.
Here, the numbers on the first rows denote the self-intersection numbers of
$\mathbb{P}^1$. Thus in Table~\ref{other},
there are two or three 2-cycles (tensor multiplets).

We are interested in the self-dual strings, which are charged under
$B_{\mu\nu}^i$ with equal
electric and magnetic charges. If a node has gauge symmetry, the string is identified
as an instanton string soliton. See, e.g. \cite{Kim:2016foj} and
references therein for a review. In this section,
we are interested in the strings of the SCFTs given in Table \ref{other}.
Since they involve $G_2$ gauge group with matters in ${\bf 7}$
or $SO(7)$ gauge group with matters in ${\bf 8}$, the gauge theories on
these strings will be constructed using our gauge theories of section 2 as
ingredients.

\subsection{$2,3,2$: $SU(2)\times SO(7)\times SU(2)$ gauge group}

Since this QFT has three factors of simple gauge groups, one can
assign three topological numbers $k_1,k_2,k_3$ for the instanton strings in
$SU(2)_1$, $SO(7)$, $SU(2)_2$. To construct the 2d quiver for these
strings, we proceed in steps. We first
consider the case in which two of the three gauge symmetries are ungauged in 6d,
when only one of $k_1,k_2,k_3$ is nonzero.
They are instanton strings of either $SU(2)$ or $SO(7)$ gauge
theory with certain matters. After identifying three ADHM(-like) gauge theories,
we then consider the case with all $k_1,k_2,k_3$ nonzero,
and form a quiver of the three ADHM(-like) theories.

We first consider the case with $k_1\!=\!k_3\!=\!0$, when $SU(2)_1\times SU(2)_2$
is ungauged. Then $SU(2)^2\sim Sp(1)^2$ becomes a flavor symmetry
rotating the hypermultiplets, which in the strict ungauging limit enlarges to $Sp(2)$.
This is because the matters in $\frac{1}{2}({\bf 2},{\bf 8},{\bf 1})
+\frac{1}{2}({\bf 1},{\bf 8},{\bf 2})$ will arrange into
$\frac{1}{2}({\bf 8},{\bf 4})$ of $SO(7)\times Sp(2)$ in the ungauging limit.
This theory was discussed in section 2.1, the 6d $SO(7)$ theory 
at $n_{\bf 8}=2$. So as the ADHM-like description,
we take this theory with $U(k_2)$ gauge symmetry and
reduced $SU(4)\times U(2)\subset SO(7)\times Sp(2)$ global symmetry.
Note that in section 2, our 2d gauge theory can have $U(4)$ global symmetry rotating
$4$ Fermi multiplets, but it reduced to $U(2)$ after coupling
to the 5d/6d background fields, especially the hypermultiplet scalar VEV.
So the relevant global symmetry of this
model (as describing higher dimensional QFT's soliton) depends on
the bulk information.
Here, since we shall use this model for the strings of the non-Higgsable
$2,3,2$ SCFT, with $SU(2)^2$ gauged, one cannot turn on such a background
hypermultiplet field. Instead, $SU(2)^2\subset U(4)$ global symmetry will remain
in 2d after 6d gauging. $4$ Fermi fields are divided into $2$ pairs,
and  we can rotate them only within a pair.

We also consider the limit in which $SO(7)\times SU(2)_2$ is
ungauged, and consider $k_1$ instanton strings in $SU(2)_1$.
The matter
$\frac{1}{2}({\bf 1},{\bf 8},{\bf 2})$ will not affect the ADHM construction
since it is neutral in $SU(2)_1$. $\frac{1}{2}({\bf 2},{\bf 8},{\bf 1})$
will reduce to four fundamental hypermultiplets in $SU(2)$.
Its ADHM construction is well known.
The 2d $(0,4)$ field contents are given as follows:
\begin{eqnarray}\label{SU2-Nf=4-ADHM}
  (A_\mu,\lambda_0,\lambda)&:&\textrm{vector mutiplet in }({\bf adj},{\bf 1})\\
  q_{\dot\alpha}=(q,\tilde{q}^\dag)&:&\textrm{hypermultiplet in }
  ({\bf k},\bar{\bf 2})\nonumber\\
  a_{\alpha\dot\beta}\sim(a,\tilde{a}^\dag)&:&
  \textrm{hypermultiplet in }({\bf adj},{\bf 1})\nonumber\\
  \Psi_a&:&\textrm{Fermi multiplet in }({\bf k},{\bf 1})\ ,\nonumber
\end{eqnarray}
where $a=1,\cdots\!,4$.
We showed the representations of $U(k_1)\times SU(2)$. As for the
hypermultiplets, we have only shown the scalar components. $\alpha,\dot\alpha=1,2$
are the doublet indices for $SU(2)_l$ and $SU(2)_r$. Although $\bar{\bf 2}\sim{\bf 2}$
for $SU(2)$, we put bar since the ADHM construction classically has $U(2)$ symmetry
as a default. This is the UV quiver description for the $SU(2)$ instanton string
at $n_{\bf 2}=4$. This quiver classically
has $U(k_1)$ gauge symmetry and $U(2)\times U(4)_F$ global symmetries.
$U(k_1)$ is anomaly-free \cite{Kim:2015fxa}.
The overall $U(1)_G\subset U(2)$ and $U(1)_F\subset U(4)_F$ has mixed anomaly with
$U(1)\subset U(k_1)$, and only $G+F$ is free of mixed anomaly \cite{Kim:2015fxa}.
Moreover, considering all fields in this ADHM quiver, $G+F$ can be eaten up by
$U(1)\subset U(k_1)$.
This implies that $U(k_1)$ gauge invariant observables will not see
$G,F$. So this system only has $SU(2)\times SU(4)_F$ symmetry \cite{Kim:2015fxa}.
In the IR, this enhances to $SU(2)\times SO(7)_F$. This is in contrast to the
$SU(2)$ theory at $n_{\bf 2}=4$ in lower dimensions, in which case $U(4)_F$
enhances to $SO(8)$. The $SO(7)_F$ symmetry of this model was noticed in
\cite{Heckman:2015bfa,Ohmori:2015pia}.
Replacing $k_1$ by $k_3$, one can also obtain the ADHM gauge theory
when $SU(2)_1\times SO(7)$ is ungauged in 6d.

\begin{figure}[t!]
  \centering
	\includegraphics[width=9cm]{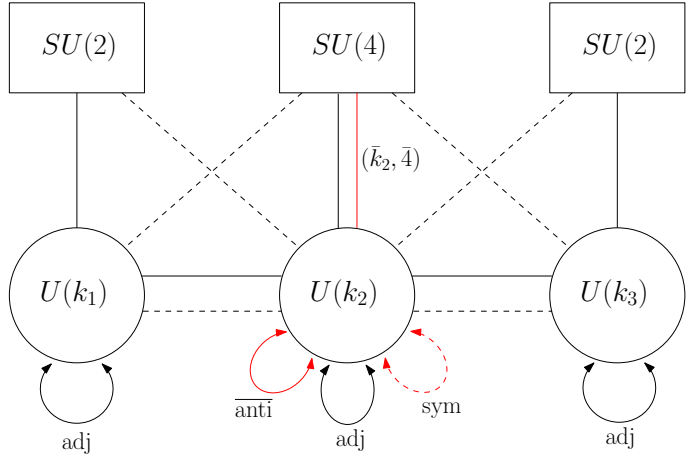}
\caption{2d quiver for the strings of
6d $2,3,2$ SCFT. Black lines are fields taking the form of
$\mathcal{N}=(0,4)$ multiplets, being either
hypermultiplet/twisted hypermultiplet (bold line) or Fermi multiplet (dashed).
Red lines are $\mathcal{N}=(0,2)$ chiral(bold)/Fermi(dashed) multiplets.}
	\label{232-quiver}
\end{figure}
Now when all $k_1,k_2,k_3$ are nonzero, one can form a quiver of the above
three ADHM(-like) theories.
We shall add more 2d matters to account for the zero modes coming from
6d hypermultiplets, and introduce extra potentials.
Between adjacent $SU(2)_1\times SO(7)$ or $SO(7)\times SU(2)_2$ pair of nodes,
one has bi-fundamental hypermultiplet in $\frac{1}{2}({\bf 8},{\bf 2})$.
Since we seek for a 2d UV description seeing $SU(2)\times SU(4)$ subgroup
only, this hypermultiplet is in $({\bf 2},\bar{\bf 4})$ bi-fundamental
representation of the latter. Usually in D-brane models with bi-fundamental
matters, the induced $(0,4)$ matters on the $U(k_1)\times U(k_2)$ ADHM construction
of instantons are
\begin{eqnarray}\label{bif-ADHM-1}
  \Phi_A=(\Phi,\tilde\Phi^\dag)&:&\textrm{twisted hypermultiplet in }
  ({\bf k}_1,\overline{\bf k}_2)\\
  \Psi_\alpha=(\Psi_1,\Psi_2)&:&\textrm{two Fermi multiplet fields in }
  ({\bf k}_1,\overline{\bf k}_2)\nonumber
\end{eqnarray}
and
\begin{eqnarray}\label{bif-ADHM-2}
  \Psi_a&:&\textrm{Fermi multiplets in }({\bf k}_1,\overline{\bf 4})
  \textrm{ of }U(k_1)\times SU(4)
  \ \ \ (a=1,\cdots,4)\nonumber\\
  \Psi_i&:&\textrm{Fermi multiplets in }(\overline{\bf k}_2,{\bf 2})
  \textrm{ of }U(k_2)\times SU(2)\ \ \ (i=1,2)\ .
\end{eqnarray}
See, e.g. \cite{Kim:2015fxa,Gadde:2015tra} for the details. Although our construction
is not guided by D-brane models, we advocate the same field contents as our natural
ansatz. The fields $\Psi_a$ with $a=1,\cdots,4$ are not new, but come from
the last line of (\ref{SU2-Nf=4-ADHM}). This is natural because
the 6d $SU(4)\subset SO(7)$ gauge symmetry is obtained by gauging the global
symmetry in the setting of (\ref{SU2-Nf=4-ADHM}).
Also, $\Psi_i$ with $i=1,2$ can also be found in the ADHM-like quiver in
section 2. Namely, in section 2, we had four Fermi multiplet fields in
${\bf k_2}$ representation of $U(k_2)$, at $n_{\bf 8}=2$. $\Psi_i$ of
(\ref{bif-ADHM-2}) is obtained by taking two of these four. (The other two will be
associated with the $SO(7)\times SU(2)_2$ pair.)
The bi-fundamental fields in (\ref{bif-ADHM-1}) are new, and link
the two ADHM(-like) gauge nodes.
Similarly, between the second and third nodes, bi-fundamental fields of the form of (\ref{bif-ADHM-1}), replacing $k_1\rightarrow k_3$, are added. The remaining
Fermi fields in the second and third nodes take the form of (\ref{bif-ADHM-2}),
with $k_1\rightarrow k_3$. The flavor symmetry of these Fermi multiplets in
an ADHM node is locked with the 6d gauge symmetry of the adjacent ADHM node.
The resulting quiver is shown schematically in Fig.~\ref{232-quiver}.

In the previous paragraph, and in Fig.~\ref{232-quiver},
we locked some 6d flavor symmetries of an ADHM theory with
6d gauge symmetries of adjacent
ADHM theories. This has to be justified by writing down the interactions
which lock the symmetries as claimed. Now
we explain such superpotentials. In the $(0,2)$ off-shell description
\cite{Tong:2014yna} of $(0,4)$ theories, one can introduce interactions by
two kinds of superpotentials $J_\Psi$, $E_\Psi$ given for each Fermi multiplet.
There are some constraints on $J_\Psi$'s and $E_\Psi$'s to be met, either for
$(0,2)$ SUSY or for $(0,4)$ enhancement of the classical action. These conditions
are all mentioned in section 3, when we discussed models with manifest $(0,4)$ SUSY.
In our current ADHM-like models, some part of the matters and interactions inevitably
break manifest $(0,4)$ SUSY. However, most of the fields still take the form of
$(0,4)$ multiplets, so that we find it convenient to turn on classical interactions
in two steps. We first turn on manifestly $(0,4)$ supersymmetric classical
interactions for the fields shown in  Fig.~\ref{232-quiver} with black lines/nodes.
Then we rephrase these interactions in $\mathcal{N}=(0,1)$ language, after which we
turn on further $(0,1)$ interactions for the fields shown as red lines in
Fig.~\ref{232-quiver}. We find that securing the partial $(0,4)$ SUSY structure
plays important roles for the correct physics, e.g.
yielding the right multi-particle structures of the elliptic genus, etc.

In $(0,4)$ gauge theories, one has two types of
hypermultiplets: hypermultiplet whose scalars form a doublet of $SU(2)_r$,
and twisted hypermultiplet whose scalars form a doublet of $SU(2)_R$.
These two multiplets contribute differently to the $J,E$ superpotentials
for the fermions in the $(0,4)$ vector multiplet. Namely, in the $(0,2)$ formalism
of \cite{Tong:2014yna}, a $(0,4)$ vector multiplet decomposes into a $(0,2)$
vector multiplet $A_\mu,\lambda_0$ and an adjoint Fermi multiplet $\lambda$
(plus auxiliary field). A hypermultiplet field
$(\Phi_{\dot\alpha})_{\bf R}=(\Phi,\tilde\Phi^\dag)_{\bf R}$ in the representation
${\bf R}$ of the gauge group contributes $J_{\lambda^a}=
\Phi_{\bf R}[T^a_{\bf R}]\tilde\Phi_{\bf R}$. A twisted hypermultiplet
$(\Phi_A)_{\bf R}=(\Phi,\tilde\Phi^\dag)_{\bf R}$ contributes to
$E_{\lambda^a}=\Phi_{\bf R}[T^a_{\bf R}]\tilde\Phi_{\bf R}$. This is
the requirement of $(0,4)$ supersymmetry. (In our normalization of section 3,
one has $\sqrt{2}$ factors multiplied.) However, from
the $(0,2)$ SUSY, they should satisfy $\sum_\Psi J_\Psi E_\Psi=0$. 
To meet this condition, one has to turn on extra
potentials for the Fermi multiplets shown as black lines in Fig.~\ref{232-quiver}.
This is in complete parallel with the results shown in section 3.
Let us name the fields in Fig. \ref{232-quiver}
with black lines/nodes as follows. The ADHM fields within an ADHM node 
are named as
\begin{eqnarray}
  \textrm{node 1}&:&q_1,\tilde{q}_1\in({\bf k}_1,\bar{\bf 2}_1)+
  (\bar{\bf k}_1,{\bf 2}_1)\ ,\ \ a,\tilde{a}\in{\bf adj}_1\nonumber\\
  \textrm{node 2}&:&q_2,\tilde{q}_2\in({\bf k}_2,\bar{\bf 4})+
  (\bar{\bf k}_2,{\bf 4})\ ,\ \ a,\tilde{a}\in{\bf adj}_2\nonumber\\
  \textrm{node 3}&:&q_3,\tilde{q}_3\in({\bf k}_3,\bar{\bf 2}_3)+
  (\bar{\bf k}_3,{\bf 2}_3)\ ,\ \ a,\tilde{a}\in{\bf adj}_3\ ,
\end{eqnarray}
while the fields linking the adjacent nodes are named as 
\begin{eqnarray}
  \textrm{link 1-2}&:&\Phi_{12},\tilde\Phi_{12}
  \in({\bf k}_1,\bar{\bf k}_2)+(\bar{\bf k}_1,{\bf k}_2)\ ,\ \
  \Psi_{12},\tilde\Psi_{12}
  \in({\bf k}_1,\bar{\bf k}_2)+(\bar{\bf k}_1,{\bf k}_2)\nonumber\\
  &&\psi_{12},\tilde{\psi}_{12}\in({\bf k}_1,\bar{\bf 4})+(\bar{\bf k}_2,{\bf 2}_1)
  \nonumber\\
  \textrm{link 2-3}&:&\Phi_{23},\tilde\Phi_{23}
  \in({\bf k}_2,\bar{\bf k}_3)+(\bar{\bf k}_2,{\bf k}_3)\ ,\ \
  \Psi_{23},\tilde\Psi_{23}
  \in({\bf k}_2,\bar{\bf k}_3)+(\bar{\bf k}_2,{\bf k}_3)\nonumber\\
  &&\psi_{23},\tilde{\psi}_{23}\in({\bf k}_2,\bar{\bf 2}_3)
  +(\bar{\bf k}_3,{\bf 4})\ .
\end{eqnarray}
Here, notations like ${\bf 2}_1$, ${\bf 2}_3$ mean representations
of $SU(2)$ on the first (leftmost) and the third (rightmost) nodes, respectively.
Then, using the results of \cite{Kim:2015gha}, eqns. (3.3) and (3.4),
we find the following superpotentials after mapping our fields with
those in Table 4 of \cite{Kim:2015gha}:
\begin{eqnarray}\label{232-potential-1}
  \textrm{nodes}&:&J_{\lambda_i}=\sqrt{2}(q_i\tilde{q}_i+[a_i,\tilde{a}_i])
  \ \ \ (\textrm{for }i=1,2,3)\ ,\ \
  E_{\lambda_1}=\sqrt{2}\Phi_{12}\tilde{\Phi}_{12}\ ,\nonumber\\
  &&E_{\lambda_2}=\sqrt{2}(\Phi_{23}\tilde{\Phi}_{23}-\tilde{\Phi}_{12}\Phi_{12})
  \ ,\ \ E_{\lambda_3}=-\sqrt{2}\tilde{\Phi}_{23}\Phi_{23}\nonumber\\
  \textrm{links}&:&E_{\Psi_{i-1,i}}=
  \sqrt{2}(\Phi_{i-1,i}a_i-a_{i-1}\Phi_{i-1,i})\ ,\ \
  J_{\Psi_{i-1,i}}=
  \sqrt{2}(\tilde{a}_i\tilde{\Phi}_{i-1,i}-\tilde{\Phi}_{i-1,i}\tilde{a}_{i-1})
  \ ,\nonumber\\
  &&E_{\tilde{\Psi}_{i-1,i}}=
  \sqrt{2}(\tilde{a}_{i-1}\Phi_{i-1,i}-\Phi_{i-1,i}\tilde{a}_{i})\ ,\ \
  J_{\tilde{\Psi}_{i-1,i}}=\sqrt{2}
  (a_i\tilde{\Phi}_{i-1,i}-\tilde{\Phi}_{i-1,i}a_{i-1})\ ,\nonumber\\
  &&E_{\psi_{i-1,1}}=\sqrt{2}\Phi_{i-1,i}q_i\ ,\ \
  J_{\psi_{i-1,i}}=\sqrt{2}\tilde{q}_i\tilde{\Phi}_{i-1,i}\nonumber\\
  &&E_{\tilde{\psi}_{i-1,1}}=\sqrt{2}\tilde{q}_{i-1}\Phi_{i-1,i}\ ,\ \
  J_{\tilde{\psi}_{i-1,i}}=-\sqrt{2}
  \tilde{\Phi}_{i-1,i}q_{i-1}\ \ \ (\textrm{for }i=2,3)\ .
\end{eqnarray}
(We correct overall normalization of \cite{Kim:2015gha} by $\sqrt{2}$ factors.)
These are part of the interactions, and we shall add more
interactions later preserving less SUSY. Only with the
interactions shown above, one can check the $(0,4)$ SUSY of the classical action,
for instance in the bosonic potential \cite{Tong:2014yna,Kim:2015gha}.
The rearrangement of the potential energy with
$SU(2)_r\times SU(2)_R$ symmetry can be made similar to
eqn.(3.6) of \cite{Kim:2015gha}.
In particular, the flavor symmetries which rotate Fermi multiplets
are locked by these interactions as shown in Fig.~\ref{232-quiver}.

We now proceed to write down all the interactions preserving only $(0,1)$ symmetry,
for the red fields associated with the middle `$3$' node. This will basically be
the same as the interactions explained in section 2.1, for $SO(7)$ instanton strings
at $n_{\bf 8}\neq 0$. However, before doing that, we should 
rephrase the previous $(0,4)$ interactions in the $(0,1)$ superfield language.
In $(0,2)$ superfield, one has a pair of complex superspace coordinates
$\theta,\bar\theta$. $E_\Psi$ appears as the top component
$\sim\theta\bar\theta E_\Psi(\Phi)$ of the Fermi multiplet \cite{Witten:1993yc}.
On the other hand, $J_\Psi$ appears as a term in the Lagrangian, of the form
$\int d\theta \Psi J_\Psi+h.c.$.
However, since $(0,1)$ supersymmetry only has one real superspace coordinate
$\theta$, there is no separate notion of $E_\Psi$. There can be 
superpotentials $\int d\theta(\Psi J^{(0,1)}_\Psi-h.c.)$, where
$J^{(0,1)}_\Psi$ can be any non-holomorphic function of the scalars.
To realize $J_\Psi$ and $E_\Psi$ in the previous paragraph, one writes
\begin{equation}
  {\sum}_\Psi\int d\theta\left[\Psi(J_\Psi(\Phi)+\bar{E}_\Psi(\bar\Phi))-
  {\rm h.c.}\right]\ .
\end{equation}
One finds the correct bosonic potential
$\sum_{\Psi}\left|J_\Psi+\bar{E}_\Psi\right|^2=\sum_{\Psi}
\left(|J_\Psi|^2+|E_\Psi|^2\right)$,
using $\sum_\Psi J_\Psi E_\Psi=0$ of (\ref{232-potential-1}).
The Yukawa couplings associated with $J_\Psi$ and $E_\Psi$ $\sim\sum_\Psi\Psi(\frac{\partial J_\Psi}{\partial\phi^i}\psi^i
+\frac{\partial\bar{E}}{\partial\bar\phi_i}\psi_i)$ are also correctly
reproduced. Now with (\ref{232-potential-1}) rewritten as
$J_\Psi^{(0,1)}=J_\Psi+\bar{E}_\Psi$, we add further interactions for
$\hat\lambda$, $\check\lambda$ on the middle node, as
given by (\ref{SO7-(0,1)}).

With these potentials, one can show that the moduli space
is that of each ADHM-like quiver, at $\Phi_{i-1,i}=0$, $\tilde{\Phi}_{i-1,i}=0$.
In particular, no extra branch is formed by $\Phi_{i-1,i},
\tilde{\Phi}_{i-1,i}$.

One can compute the 2d anomalies from our
gauge theory, and compare with the result known from anomaly inflow. The
6d 1-loop anomaly 8-form in the tensor branch is given by
\begin{eqnarray}
  I_{\textrm{1-loop}}&=&-\frac{3}{32}\left[{\rm Tr}(F^2_{SO(7)})\right]^2
  -\frac{1}{16}\left[{\rm Tr}(F^2_{SU(2)_1})\right]^2
  -\frac{1}{16}\left[{\rm Tr}(F^2_{SU(2)_2})\right]^2\\
  &&+\frac{1}{16}{\rm Tr}(F_{SO(7)}^2)\left[{\rm Tr}(F^2_{SU(2)_1})
  +{\rm Tr}(F^2_{SU(2)_2})\right]-\frac{1}{16}p_1(T){\rm Tr}(F^2_{SO(7)})\nonumber\\
  &&-\frac{1}{4}c_2(R)\left[5{\rm Tr}(F_{SO(7)}^2)+
  2{\rm Tr}(F^2_{SU(2)_1})+2{\rm Tr}(F^2_{SU(2)_2})\right]+\cdots\ .\nonumber
\end{eqnarray}
We only showed the terms containing $SU(2)_1\times SO(7)\times SU(2)_2$
gauge fields. This can be written as
$I_{\textrm{1-loop}}=-\frac{1}{2}\Omega^{ij}I_iI_j+\cdots$,
with $i,j=1,2,3$, where
\begin{equation}\label{232-GS-anomaly}
  \Omega^{ij}=
  \left(\begin{array}{ccc}
    2&-1&0\\-1&3&-1\\0&-1&2
  \end{array}\right)\ \ ,\ \ \
  I_i=\left(\begin{array}{c}
    \frac{1}{4}{\rm Tr}(F_{SU(2)_1}^2)+\frac{11}{4}c_2(R)+\frac{1}{16}p_1(T)\\
    \frac{1}{4}{\rm Tr}(F_{SO(7)}^2)+\frac{7}{2}c_2(R)+\frac{1}{8}p_1(T)\\
    \frac{1}{4}{\rm Tr}(F_{SU(2)_2}^2)+\frac{11}{4}c_2(R)+\frac{1}{16}p_1(T)
  \end{array}\right)\ .
\end{equation}
Using (\ref{inflow-anomaly}), one finds the following anomaly 4-form $I_4$
\begin{equation}\label{232-anomaly-inflow}
  I_4
  =\left(k_1k_2+k_2k_3-k_1^2-\frac{3}{2}k_2^2-k_3^2\right)\chi(T_4)
  +k_2(I_1+I_3-3I_2)+k_1(I_2-2I_1)+k_3(I_2-2I_3)\ .
\end{equation}
on the instanton strings with string numbers $k_i=(k_1,k_2,k_3)$.

We now compute the anomaly from our gauge theory.
We first compute the anomalies of three ADHM
quivers $I_4^{(i)}$ ($i=1,2,3$), restricting them according to the
symmetry locking rules. We then compute the anomalies $I^{\rm bif}_4$ of 
matters $\Phi_{i-1,i},\tilde{\Phi}_{i-1,i}$. The net anomaly is
$I_4=\sum_{i=1}^3I^{(i)}_4+I^{\rm bif}_4$. Using
(\ref{2d-anomaly}), one first finds
\begin{equation}
  I^{(2)}_4=-\frac{3}{2}k_2^2\chi(T_4)-k_2\left[\frac{3}{4}{\rm Tr}(F^2_{SO(7)})+
  5c_2(R)+\frac{p_1(T)}{4}
  -\frac{1}{4}({\rm Tr}(F_{SU(2)_1}^2)+{\rm Tr}(F_{SU(2)_2}^2))\right]
\end{equation}
where we replaced ${\rm tr}_{\bf 4}(F_{Sp(2)}^2)\rightarrow
{\rm tr}_{\bf 2}(F_{SU(2)_1}^2)+{\rm tr}_{\bf 2}(F_{SU(2)_2}^2)
=\frac{1}{2}[{\rm Tr}(F_{SU(2)_1}^2)+{\rm Tr}(F_{SU(2)_2}^2)]$.
As in section 2.1, $F_{SO(7)}$ is restricted to $SU(4)$ in our UV gauge theory,
and fields in $c_2(R)$, $c_2(r)$ are also restricted to $F_J$.
$I^{(1)}_4$ and $I^{(3)}_4$ can be computed from the known anomaly polynomial for
the instanton strings of 6d $SU(2)$ theory at $n_{\bf 2}=4$. The result is
eqn.(5.19) of \cite{Kim:2016foj} at $N=2$, with $k$ replaced by $k_1$ or $k_3$:
\begin{equation}\label{SU2-Nf=4-anomaly}
  I^{(1)}_4=-k_1^2\chi(T_4)-\frac{k_1}{2}{\rm Tr}(F_{SU(2)_1}^2)
  +\frac{k_1}{4}{\rm Tr}(F_{SO(7)}^2)-2k_1c_2(R)\ ,\ \
  I^{(3)}_4=\left(k_1,SU(2)_1\rightarrow k_3,SU(2)_2\right)\ .
\end{equation}
Here we replaced $F=SU(4)$ of \cite{Kim:2016foj} by $SO(7)$, assuming symmetry
enhancement. Finally, $I^{\rm bif}_4$ is also computed in \cite{Kim:2016foj},
eqn.(3.58), which for our model is
\begin{equation}
  I^{\rm bif}_4=(k_1k_2+k_2k_3)\chi(T_4)\ .
\end{equation}
One finds that $I_4=\sum_{i=1}^3I^{(i)}_4+I^{\rm bif}_4$ agrees with
(\ref{232-anomaly-inflow}), providing a check of our gauge theory.

The elliptic genus of this gauge theory is given by 
(note again the definition
$\theta(z)\equiv\frac{i\theta_1(\tau|\frac{z}{2\pi i})}{\eta(\tau)}$)
\begin{eqnarray}\label{232-contour}
  Z_{k_1,k_2,k_3}&=&\oint\prod_{I_1=1}^{k_1}\frac{\prod_{i=1}^4\theta(v_i-u_{I_1})}
  {\theta(\epsilon_+\pm u_{I_1}\pm \nu)}\cdot
  \frac{\prod_{I_1\neq J_1}\theta(u_{I_1J_1})
  \prod_{I_1,J_1}\theta(2\epsilon_++u_{I_1J_1})}
  {\prod_{I_1,J_1=1}^{k_1}\theta(\epsilon_{1,2}+u_{I_1J_1})}
  \cdot\left(\!\frac{}{}\!1\rightarrow 3,
  \nu\rightarrow\tilde{\nu}\right)\nonumber\\
  &&\cdot\prod_{I_2=1}^{k_2}
  \frac{\theta(v\pm u_{I_2})\theta(\tilde{v}\pm u_{I_2})}
  {\prod_{i=1}^4\theta(\epsilon_+\pm(u_{I_2}-v_i))
  \theta(\epsilon_+\!-\!u_{I_2}\!-\!v_i)}
  \cdot\frac{\prod_{I_2\neq J_2}\theta(u_{I_2J_2})
  \prod_{I_2,J_2}\theta(2\epsilon_++u_{I_2J_2})}
  {\prod_{I_2,J_2=1}^{k_2}\theta(\epsilon_{1,2}+u_{I_2J_2})}
  \nonumber\\
  &&\cdot\frac{\prod_{I_2\leq J_2}
  \theta(u_{I_2}\!+\!u_{J_2})\theta(u_{I_2}\!+\!u_{J_2}\!-\!2\epsilon_+)}
  {\prod_{I_2<J_2}\theta(\epsilon_{1,2}\!-\!u_{I_2}\!-\!u_{J_2})}
  \nonumber\\
  &&\cdot\prod_{I_1=1}^{k_1}\prod_{I_2=1}^{k_2}
  \frac{\theta(\epsilon_-\pm(u_{I_1}-u_{I_2}))}
  {\theta(-\epsilon_+\pm(u_{I_1}-u_{I_2}))}\cdot
  \prod_{I_2=1}^{k_2}\prod_{I_3=1}^{k_3}
  \frac{\theta(\epsilon_-\pm(u_{I_2}-u_{I_3}))}
  {\theta(-\epsilon_+\pm(u_{I_2}-u_{I_3}))}\ .
\end{eqnarray}
$v_i$ (with $\sum_iv_i=0$) is the $SU(4)\subset SO(7)$
chemical potential, $\pm \nu$ and $\pm\tilde{\nu}$ are the chemical potentials for
6d $SU(2)^2$. The contour integral is given with suitable
weight \cite{Benini:2013xpa}, including the $U(k_1)\times U(k_2)\times U(k_3)$
Weyl factor. The contour integral is again given by the JK residues
\cite{Benini:2013xpa}. We again choose
$\eta_1=(1,\cdots,1)$, $\eta_2=(1,\cdots,1)$, $\eta_3=(1,\cdots,1)$.
Then, similar to the residue choices made in section 2, one can show that
the residues are labeled by three sets of colored Young diagrams,
$(Y^{(1)}_1,Y^{(1)}_2)$ with $k_1$ boxes for $u_{I_1}$,
$(Y^{(2)}_1,\cdots,Y^{(2)}_4)$ with $k_2$ boxes for $u_{I_2}$,
and $(Y^{(3)}_1,Y^{(3)}_2)$ with $k_3$ boxes for $u_{I_3}$.
The residues all come from the poles at
\begin{eqnarray}\label{232-pole-Young}
  u_{I_1}&:&\epsilon_++u_{I_1}\pm \nu=0\ ,\ \ \epsilon_{1,2}+u_{I_1J_1}=0\nonumber\\
  u_{I_2}&:&\epsilon_++u_{I_2}-v_i=0\ ,\ \ \epsilon_{1,2}+u_{I_2J_2}=0\nonumber\\
  u_{I_3}&:&\epsilon_++u_{I_3}\pm\tilde{\nu}=0\ ,\ \
  \epsilon_{1,2}+u_{I_3J_3}=0\ ,
\end{eqnarray} 
coming from the first, second and third line of (\ref{232-contour}), respectively.
The residue sum is given by
\begin{align}\label{232-residue}
	Z_{k_1,k_2,k_3} =& \sum_{\substack{Y_{(1,2,3)}\\|Y_{(a)}| = k_a}} \prod_{i = 1}^2 \bigg(\prod_{s_1 \in Y_{(1)i}} \frac{\prod_{l=1}^4 \theta (v_l - \phi(s_1))}{\prod_{j=1}^2 \theta(E_{ij} (s_1))\theta(E_{ij}(s_1)-2\epsilon_+)} \bigg) \times  \prod_{i=1}^2 \bigg( 1 \rightarrow 3 \bigg)\\
	 & \times \prod_{i = 1}^4\bigg(  \prod_{s_2 \in Y_{(2)i}} \frac{\theta (2\phi(s_2)) \theta(2\phi(s_2)-2\epsilon_+)\cdot \theta(v \pm \phi(s_2))\theta(\tilde{v} \pm \phi(s_2))}{\prod_{j=1}^4 \theta(E_{ij} (s_2))\theta(E_{ij}(s_2)-2\epsilon_+) \theta(\epsilon_+-\phi(s_2)-v_j)}  \times\nonumber \\
	  &
	\qquad\quad \prod_{s_2 \in Y_{(2)i}} \prod_{j \geq i}^4 \prod_{\substack{\tilde{s}_{2} \in  Y_{(2)j}\nonumber \\
	s_{2} < \tilde{s}_2 }} \frac {\theta (\phi(s_2) + \phi(\tilde{s}_2))\theta (\phi(s_2) + \phi(\tilde{s}_2) - 2\epsilon_+)}{\theta(\epsilon_{1,2} - \phi(s_2) - \phi(\tilde{s}_2))} \times\nonumber\\
	&\qquad\quad\ \ \prod_{j=1}^2 \prod_{s_1 \in Y_{(1)j}} \frac{\theta(-\epsilon_- \pm (\phi(s_1) - \phi(s_2)))}{\theta(-\epsilon_+ \pm (\phi(s_1) - \phi(s_2)))} \times \prod_{j=1}^2 \prod_{s_3 \in Y_{(3)j}} \frac{\theta(-\epsilon_- \pm (\phi(s_2) - \phi(s_3)))}{\theta(-\epsilon_+ \pm (\phi(s_2) - \phi(s_3)))} \bigg) \nonumber
\end{align}
where $s_a$ (for $a=1,2,3$) labels the $k_a$ boxes in the $a$'th
colored Young diagram, and $(v_{(1)})_{1,2}=\pm \nu$, 
$(v_{(3)})_{1,2}=\pm\tilde{\nu}$.
$\phi(s_a)$ and $E_{ij}(s_a)$ are defined as
\begin{eqnarray}
  E_{ij}(s_a)&=& v_{(a)i}-v_{(a)j} -\epsilon_1 h_i(s_a)+\epsilon_2 (v_j(s_a)+1)\\
  \phi(s_a) &=& v_{(a)i} - \epsilon_+ - (n_a - 1)\epsilon_1 - (m_a - 1)\epsilon_2
\end{eqnarray}
for $s_a = (m_a, n_a) \in Y^{(a)}_{i}$.

It is important to note that
$\Phi_{i-1,i},\tilde{\Phi}_{i-1,i}$ do not provide extra JK-Res, for
the following reason. For instance, suppose that we take the `pole' from
$\theta(-\epsilon_++u_{I_1}-u_{I_2})^{-1}$ on the fourth line,
at $-\epsilon_++u_{I_1}-u_{I_2}=0$ to determine $u_{I_1}$, with
$u_{I_2}$ determined from (\ref{232-pole-Young}). Suppose that
$u_{I_2}$ is determined by $\epsilon_++u_{I_2}-v_i=0$. Then on the first
line of (\ref{232-contour}), a Fermi multiplet contribution
$\theta(v_i-u_{I_1})$ vanishes at the pole, because
$u_{I_1}-v_i=(-\epsilon_++u_{I_1}-u_{I_2})+(\epsilon_++u_{I_2}-v_i)=0$.
On the other hand, suppose that $u_{I_2}$ is determined by one of
$\epsilon_{1,2}+u_{I_2}-u_{J_2}=0$,
with $u_{J_2}$ determined by other equations. Then,
from $\Psi_{12},\tilde{\Psi}_{12}$'s contributions
$\theta(\epsilon_-\pm(u_{I_1}-u_{J_2}))$ on the fourth line, one again finds that
one of the two $\theta$ factors vanishes at the pole location.
Therefore, one finds that the residue vanishes due to the vanishing determinant
from certain Fermi multiplet. This idea turns out to hold most generally, so that
one can show that the fourth line of (\ref{232-contour}) never provides a pole with
nonzero JK residue. Based on these observations, one can make a recursive proof of
this statement, similar to that made for the 5d $\mathcal{N}=1^\ast$ instanton
partition function in \cite{Hwang:2014uwa}. Note that the symmetry locking
provided by the $(0,4)$ potentials (\ref{232-potential-1}) played crucial roles
for the vanishing of these residues.

\subsection{Tests from 5d descriptions}

In this subsection, we test the elliptic genera of section 4.1,
using a recently proposed 5d description for the 6d $2,3,2$ SCFT compactified
on $S^1$ \cite{Hayashi:2017jze}.
The description is available when the elliptic CY$_3$ in 
F-theory admits an orbifold description, of the form
$[\mathcal{B}\times T^2]/\Gamma$ with a discrete group $\Gamma$.
One can dualize F-theory to M-theory
on same CY$_3$. The small $S^1$ limit (together with suitably scaling other
massive parameters) on the F-theory side corresponds to
the large $T^2$ limit on the M-theory dual.
There may be fixed points of $\Gamma$ on $T^2$, as it decompactifies into
$\mathbb{R}^2$. Near each fixed point, there exists an interacting 5d SCFTs. So
in this 5d limit, one obtains factors of decoupled 5d SCFTs. The 6d KK momentum
degrees of freedom can be restored by locking certain global symmetries of these
5d SCFTs and gauging it, so that the instanton quantum number of this 5d gauge
theory provides the 6d KK momentum. See
\cite{DelZotto:2015rca,hck,DelZotto:2017pti,Hayashi:2017jze}
for the details.

\begin{figure}[t!]
    \centering
    \subcaptionbox{}{
	\includegraphics[width=10.5cm]{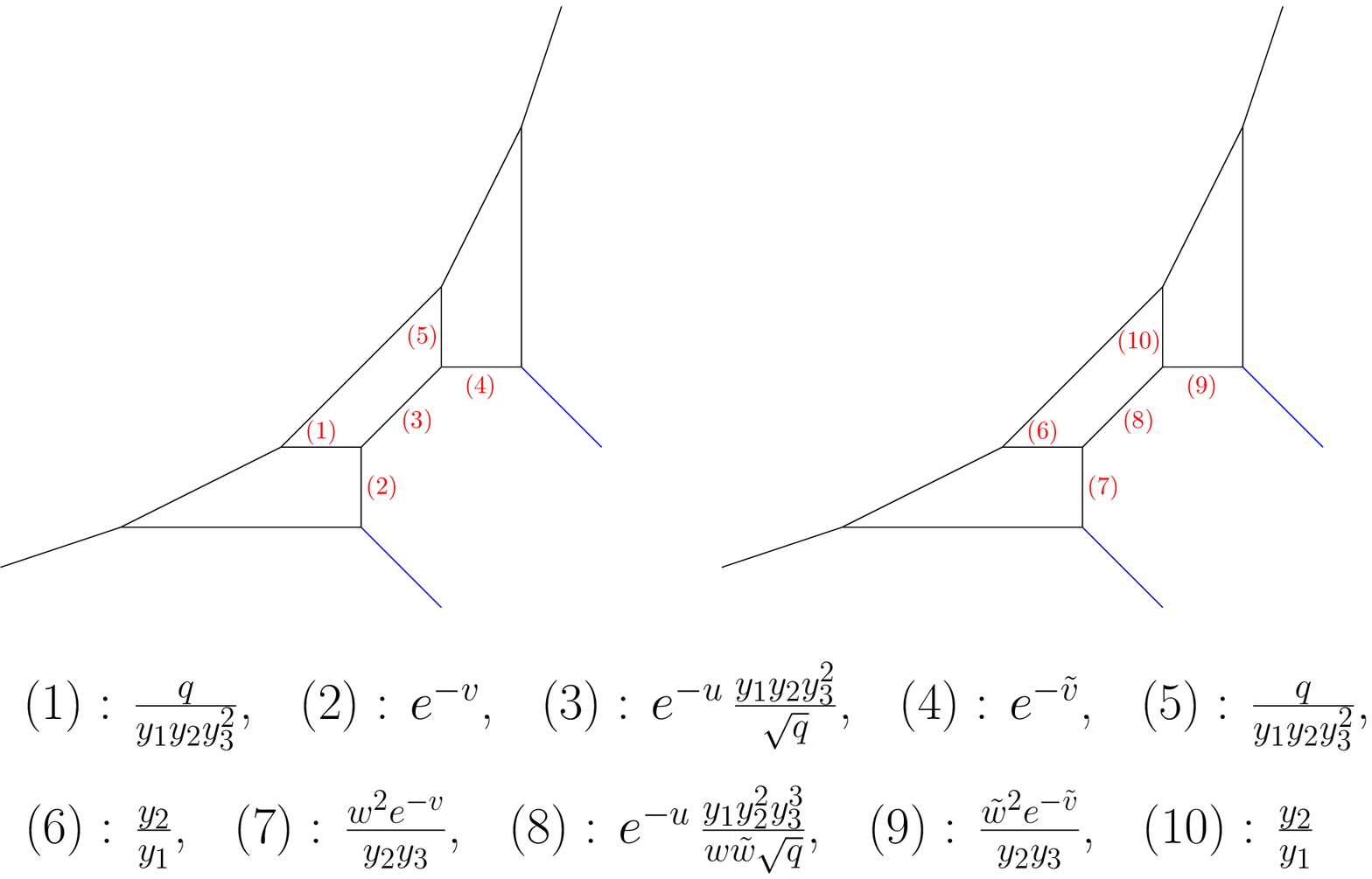}}
	\quad
    \subcaptionbox{}{
	\includegraphics[width=5.5cm]{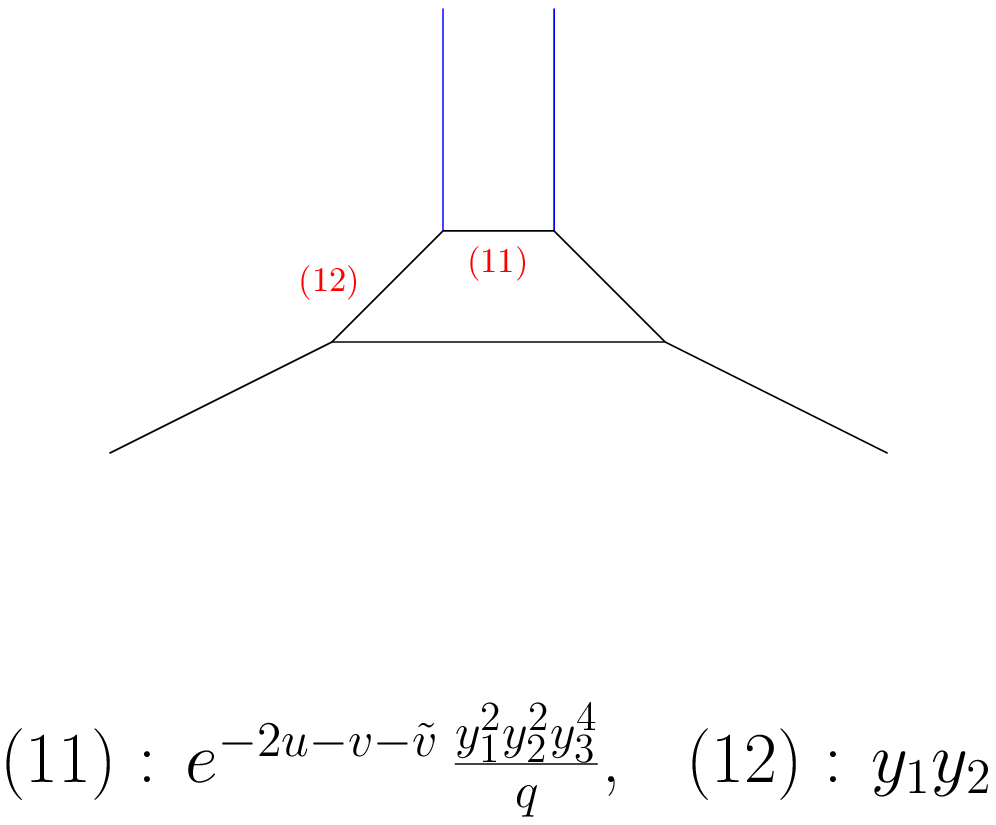}}
	\caption{5-brane webs for the 6d $2,3,2$ SCFT in a 5d limit.
(1)$\sim$(12) are the Kahler parameters in terms
of our fugacities. $v,u,\tilde{v}$ are tensor VEV's for
$SU(2)_1\times SO(7)\times SU(2)_2$.}
	\label{232-5d}
\end{figure}

If a 6d SCFT admits a 6d gauge theory description, an obvious 5d limit
is given by the 5d gauge theory with same gauge group.
This is obtained by a scaling limit with 6d tensor multiplet
scalar VEV, $v=\langle\Phi\rangle\rightarrow\infty$. Namely,
$v=g_{\rm 6d}^{-2}$ and $Rv=g_{\rm 5d}^{-2}$ are the 6d and 5d inverse gauge
couplings, respectively, where $R$ is the circle radius. If one
takes $R\rightarrow 0$, $v\rightarrow\infty$ with $g_{\rm 5d}^{-2}$ kept fixed,
one often gets a 5d SCFT with a relevant deformation made by $g_{\rm 5d}^{-2}\neq 0$
\cite{DelZotto:2017pti}. The 5d factorization limit described in the previous
paragraph takes different scaling limit of massive
parameters when taking $R\rightarrow 0$. The latter 5d limit scales
other massive parameters like the holonomies of gauge fields on $S^1$.
From the viewpoint of former 5d limit, the latter
5d limit keeps a different slice of 5d states, which contains states with nonzero
KK momenta from the former viewpoint. For our $2,3,2$ SCFT,
the new 5d limit consists of three 5d SCFTs. The three 5d SCFTs admit IIB
5-brane web engineerings, given by Fig.~\ref{232-5d} \cite{Hayashi:2017jze}.
Each factor in Fig.~\ref{232-5d}(a) is a non-Lagrangian theory, in that
it does not admit a relevant deformations to 5d Yang-Mills theory.
(Fig.~\ref{232-5d}(a) is related to that in \cite{Hayashi:2017jze} by a flop
transition.) To have states with general KK momenta, one locks the three $SU(2)_g$
flavor
symmetries associated with gauge symmetries on the blue-colored parallel 5-branes
of Fig.~\ref{232-5d}, and gauge it. The relations between 6d parameters
and the Kahler parameters of 5-brane web are shown below Fig.~\ref{232-5d},
which will be (empirically) justified. \cite{Hayashi:2017jze} also
discusses the gauging of $SU(2)_g$ in the brane web context as trivalent
gluing, with some prescriptions for computations.
But here we shall only discuss computations in the factorization limit.

We want to test our elliptic genera (\ref{232-contour}), (\ref{232-residue})
using this 5d description. The test will be made in the 5d factorization limit
in which $SU(2)_g$ is ungauged, as in Fig.~\ref{232-5d} with semi-infinite
blue lines. In some sectors with special values of
$k_1,k_2,k_3$, the BPS spectrum of the brane configuration is well known,
so our elliptic genera in these sectors will be tested against known results.
More generically, we shall do topological vertex calculus.
Technically, identifying the parameters of 6d gauge theory
(and our elliptic genus) and those in the 5-brane web is not straightforward.
The relation between the two sets of parameters are often determined empirically
in the literature. We follow the strategy of \cite{Hayashi:2017jze} which
studied the 5d description of 6d gauge theories. \cite{Hayashi:2017jze} used the
guidance from 6d affine gauge symmetry structure to partly determine the relations
between 5d/6d parameters, and then empirically fixed the rest. In our problem, we
shall use the affine $SO(7)$ symmetry to partly determine the relation, and then
focus on well-known subsectors to fix the rest.

\begin{figure}[t!]
  \centering
	\includegraphics[width=7cm]{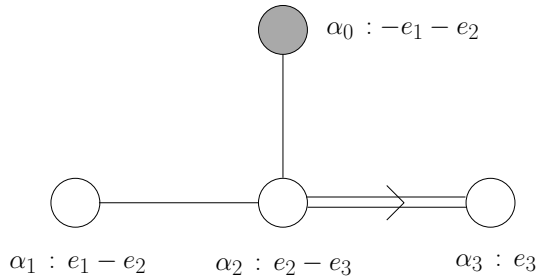}
	\caption{Affine Dynkin diagram of $SO(7)$}
	\label{SO7-affine}
\end{figure}
We first determine the parameter relation that can be inferred from
$SO(7)$ group theory. To this end, we focus on the part of web diagrams
of Fig.~\ref{232-5d} associated with the Kahler parameters
(1)$=$(5), (6)$=$(10), (12), and the blue 5-branes. Considering how the
associated four faces are connected to others (after $SU(2)_g$ gauging),
it is natural to conceive that the four Kahler parameters
are fugacities for
the affine $SO(7)$ symmetry. This is somewhat similar to the identifications of
6d $SU(3)$, $SO(8)$, $E_{6,7,8}$ fugacities in \cite{Hayashi:2017jze}, using
their affine Dynkin diagrams. For $SO(7)$, the affine Dynkin diagram is given
by Fig.~\ref{SO7-affine}, where $e_1,e_2,e_3$ are orthonormal vectors.
We call the fugacities
corresponding to the simple roots as $(t_1,t_2,t_3,t_4)\leftrightarrow
(\alpha_1,\alpha_0,\alpha_3,\alpha_2)$. From the expressions of the roots in
Fig. \ref{SO7-affine}, one obtains
\begin{equation}
  t_1=\frac{z_1^2}{z_2^2}\ ,\ \ t_2=\frac{q}{z_1^2z_2^2}\ ,\ \
  t_3=z_3^2\ ,\ \ t_4=\frac{z_2^2}{z_3^2}
\end{equation}
where we used the fact that the KK momentum fugacity $q\equiv e^{2\pi i\tau}$ 
is associated with
$\alpha_0$ in the affine Lie algebra. The root relation
$\alpha_1+2\alpha_2+2\alpha_3+\alpha_0=0$ is reflected in the above parameterization
as $t_1t_2t_3^2t_4^2=q$. $z_1,z_2,z_3$ are the fugacities of $SO(7)$
rotating three orthogonal 2-planes. More precisely, the characters
of ${\bf 7}$ and ${\bf 8}$ are given in these parameters by
\begin{eqnarray}
  \chi_{\bf 8}&=&z_1z_2z_3+\frac{z_1z_2}{z_3}+\frac{z_2z_3}{z_1}+\frac{z_3z_1}{z_2}
  +(\textrm{inverse of all four terms})\nonumber\\
  \chi_{\bf 7}&=&1+z_1^2+z_2^2+z_3^2+(\textrm{inverse of all three terms})\ .
\end{eqnarray}
The $SU(4)$ fugacity basis $y_i=e^{-v_i}$ ($i=1,2,3$)
that we have been using is related to
$z_{1,2,3}$ by $z_1^2=y_2y_3$, $z_2^2=y_3y_1$, $z_3^2=y_1y_2$, so that
the characters are given by
\begin{eqnarray}
  \chi_{\bf 8}&=&y_1+y_2+y_3+\frac{1}{y_1y_2y_3}
  +(\textrm{inverse of all four terms})\nonumber\\
  \chi_{\bf 7}&=&1+y_1y_2+y_2y_3+y_3y_1+(\textrm{inverse of all three terms})\ .
\end{eqnarray}
$t_{1,2,3,4}$ are given in terms of $y_{1,2,3}$, $q$ by
\begin{equation}
  t_1=\frac{y_2}{y_1}=(6)=(10)\ ,\ \ t_2=\frac{q}{y_1y_2y_3^2}=(1)=(5)\ ,\ \
  t_3=y_1y_2=(12)\ ,\ \ t_4=\frac{y_3}{y_2}\ .
\end{equation}
$t_4$ is roughly the Kahler parameter for the blue line in Fig.~\ref{232-5d},
which is sent to zero in the factorization limit,
with $t_{1,2,3}$ fixed. This limit requires
$q\sim y_3^2\rightarrow 0$ with fixed $y_1$, $y_2$.
To fully specify this 5d limit, we still have to specify the scaling of other
parameters in $q\rightarrow 0$. The remaining parameters are: two $SU(2)$
inverse gauge couplings (or tensor VEV's) which we call $e^{-v}$, $e^{-\tilde{v}}$ 
in this subsection, two
$SU(2)$ fugacities $w,\tilde{w}$ (related to $\nu,\tilde{\nu}$ of 
section 4.1 by $w=e^{-\nu}$, $\tilde{w}=e^{-\tilde{\nu}}$), 
$SO(7)$ inverse gauge coupling
$e^{-u}$. All the scaling rules except that of $e^{-u}$ will be determined
below by considering an $SU(2)$ subsector. The scaling of $e^{-u}$ will then
be determined next by considering the $SO(7)$ subsector, at which stage we shall
already make some tests of our elliptic genera. Then we consider more general
sectors for further tests.

\begin{figure}[t!]
    \centering
    \subcaptionbox{}{
	\includegraphics[width=9.5cm]{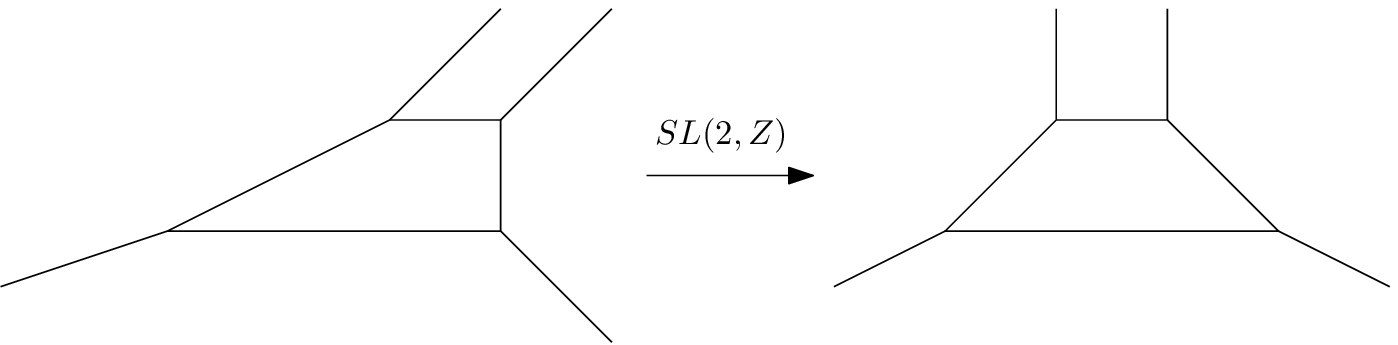}}
	\quad
    \subcaptionbox{}{
	\includegraphics[width=6.5cm]{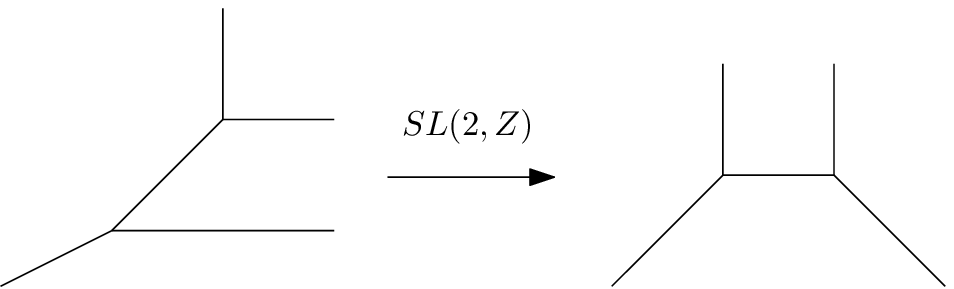}}
	\caption{Ingredients of the 5d description of 6d $SU(2)$ theory at $n_{\bf 2}=4$}
	\label{web-SU2-Nf=4}
\end{figure}
\hspace*{-0.65cm}{\bf \underline{$SU(2)$ subsector}:}
We first study the limit in which $SO(7)$ is ungauged, or
equivalently, when $k_2=0$.
The limit $u\rightarrow\infty$ should yield two 6d $SU(2)$ theories at
$n_{\bf 2}=4$, decoupled to each other. So in this limit, the 
brane web of Fig.~\ref{232-5d} (with $SU(2)_g$ gauged) 
should factorize into two. The natural identification
of $u\rightarrow\infty$ in the web is to take the distance between the parallel
blue lines to infinity.
(Assuming the identification of Kahler parameters in Fig.~\ref{232-5d},
the distance between two blue lines is proportional to
$(11)=(2)(3)^2(4)=(7)(8)^2(9)\propto e^{-2u}$.) The string
suspended between the two parallel blue lines is infinitely heavy in this limit.
So the 5d description suggests that the 6d $SU(2)$ theory at $n_{\bf 2}=4$ is
given by $U(1)_g(\subset SU(2)_g)$
gauging of three factors,
where two of them take the form of Fig.~\ref{web-SU2-Nf=4}(a), and one takes
the form of Fig.~\ref{web-SU2-Nf=4}(b). Upon a suitable $SL(2,\mathbb{Z})$
transformation, Fig.~\ref{web-SU2-Nf=4}(a) is the standard
5-brane web for the 5d $\mathcal{N}=1$ pure $SU(2)$ theory. Similarly,
Fig.~\ref{web-SU2-Nf=4}(b) describes the 5d `$SU(1)$ theory.'
The $SU(1)$ theory simply refers to the brane
configuration of Fig.~\ref{web-SU2-Nf=4}(b), not containing
an interacting 5d SCFT. This sector will be void.
So we shall take a suitable 5d scaling limit of the elliptic genera
of 6d $SU(2)$ theory at $n_{\bf 2}=4$, and find the parameter map
which exhibits two copies of 5d pure $SU(2)$ theories.

\begin{table}[t!]
$$
\begin{array}{c||c|c|c|c|c|c|c}
  \hline k\setminus n&1&2&3&4&5&6&7\\
  \hline\hline 0& -2&0&0&0&0&0&0\\
  \hline 1& -2&-4&-6&-8&-10&-12&-14\\
  \hline 2& 0&0&-6&-32&-110&-288&-644\\
  \hline 3& 0&0&0&-8&-110&-756&-3556\\
  \hline 4& 0&0&0&0&-10&-288&-3556\\
  \hline
\end{array}
$$
\caption{BPS spectrum of 5d $\mathcal{N}=1$ pure $SU(2)$ theory}\label{SU2}
\end{table}
The 5d $SU(2)$ theory's BPS spectrum can be computed from its
instanton partition function. It contains two fugacities, $Q$ for the instanton
number, and $W$ for the $SU(2)$ electric charge in the Coulomb branch.
It also contains Omega deformation parameters
$\epsilon_{1,2}$. Here, we only consider the unrefined single particle spectrum,
defined as follows. The partition function $Z^{SU(2)}(Q,W,\epsilon_{1,2})$ is
written as
$Z^{SU(2)}=\exp\left[\sum_{n=1}^\infty\frac{1}{n}f(Q^n,W^n,n\epsilon_{1,2})\right]$,
where $f$ is the single particle index. Then one considers the limit
\begin{equation}\label{single-index}
  \lim_{\epsilon_{1,2}\rightarrow 0}\left[
  \left(2\sinh\frac{\epsilon_{1,2}}{2}\right)f(Q,W,\epsilon_{1,2})
  \right]\equiv f_{\rm rel}(Q,W)=\sum_{k=0}^\infty
  \sum_{n=0}^\infty Q^kW^{2n}N_{k,n}\ .
\end{equation}
The subscript `rel' denotes the relative degrees of freedom of the bound
states, as we divided the contribution
$\frac{1}{4\sinh\frac{\epsilon_1}{2}\sinh\frac{\epsilon_2}{2}}$
from the center-of-mass degrees of freedom.
We list some known results of $N_{k,n}$ in Table \ref{SU2}. 
The states at $k=0$, $n=1$ come from the perturbative
partition function, from a massive 5d vector multiplet of W-boson.
We would like to identify two copies of Table~\ref{SU2}, by taking a 5d scaling limit
of the elliptic genus for the instanton strings of 6d $SU(2)$ theory at $n_{\bf 2}=4$.
The elliptic genus can be obtained as a special case of (\ref{232-residue}) at
$k_2=k_3=0$.

After some trial-and-errors, we find it useful to expand the
6d index as
\begin{equation}
  f_{\rm rel}(v,q,w,y_{1,2,3})=\sum_{n=0}^\infty e^{-nv}f_n(q,w,y_{1,2,3})=
  \sum_{n=0}^\infty \sum_{p=0}^\infty\sum_{m=0}^\infty
  e^{-nv}\left(\frac{q}{w^2y_3}\right)^p\left(wy_3^{-\frac{1}{2}}\right)^m
  N_{n,p,m}(y_{1,2,3})\ ,
\end{equation}
where $f_{\rm rel}$ is defined in the completely same manner as
(\ref{single-index}). $w$ is exponential of the $SU(2)$ Coulomb VEV,
and $y_i\equiv e^{-v_i}$. We take the scaling limit
$q\sim y_3^2\sim w^4\rightarrow 0$, with $v,y_1,y_2$ fixed.
Note that $q\sim y_3^2$ is compatible with the scaling rules we already found,
based on affine $SO(7)$ structure. The
nonzero terms in this limit are listed in Tables 6,7,8,9, for
$n\leq 4$. All terms except $N_{1,0,1}$ are finite in this limit.
The two terms in $N_{1,0,1}\sim y_3^{-\frac{1}{2}}$ are divergent in
the scaling limit. This implies the following situation.
Suppose that we reduce $q\sim y_3^2\sim w^4$, maintaining their ratios
finite. Reducing $q$ physically means reducing the radius $R$ of $S^1$. When
$e^{-v}wy_3^{-1}=1$ or $e^{-v}w(y_1y_2y_3)^{-1}=1$,
the two terms in $N_{1,0,1}$ becomes $1$, respectively.
This means that the two states labeled by these terms become massless, causing
a phase transition. Each term contributes $+1$ to the index, implying that
$N_{1,0,1}$ comes from two hypermultiplets. Massless hypermultiplets cause
flop phase transitions. Since the hypermultiplet's central charge changes
sign after the transition, one should get
$e^{v}w^{-1}y_3(1+y_1y_2)$ after the two phase transitions.
As we further reduce $q\sim y_3^2\sim w^4$ to zero after the phase transitions,
these two terms vanish, and we are left with the
remaining finite numbers in the tables. One can then show that the remaining
numbers in the tables are two copies of Table \ref{SU2}. Namely, one finds
\begin{eqnarray}
  f_{\rm rel}&\rightarrow&-2e^{-v}
  \left[1+\frac{\mathcal{Q}\mathcal{W}^2}{y_1y_2}\right]
  -4e^{-2v}\frac{\mathcal{Q}\mathcal{W}^2}{y_1y_2}-6
  e^{-3v}\left[\frac{\mathcal{Q}\mathcal{W}^2}{y_1y_2}
  +\left(\frac{\mathcal{Q}\mathcal{W}^2}{y_1y_2}\right)^2\right]\\
  &&-e^{-4v}\left[8\frac{\mathcal{Q}\mathcal{W}^2}{y_1y_2}
  +32\left(\frac{\mathcal{Q}\mathcal{W}^2}{y_1y_2}\right)^2
  +8\left(\frac{\mathcal{Q}\mathcal{W}^2}{y_1y_2}\right)^3\right]
  -\cdots\nonumber\\
  \hspace*{-2cm}&&-2\frac{\mathcal{W}^2e^{-v}}{y_2}\left[1+\frac{y_2}{y_1}\right]
  -4\left(\frac{\mathcal{W}^2e^{-v}}{y_2}\right)^2\cdot\frac{y_2}{y_1}
  -6\left(\frac{\mathcal{W}^2e^{-v}}{y_2}\right)^3
  \left[\frac{y_2}{y_1}+\frac{y_2^2}{y_1^2}\right]\nonumber\\
  &&-\left(\frac{\mathcal{W}^2e^{-v}}{y_2}\right)^4
  \left[8\frac{y_2}{y_1}+32\frac{y_2^2}{y_1^2}+8\frac{y_2^3}{y_1^3}\right]
  -\cdots\ ,\nonumber
\end{eqnarray}
where $\mathcal{Q}\equiv\frac{q}{w^2y_3}$, $\mathcal{W}\equiv wy_3^{-\frac{1}{2}}$.
The first two lines yield a 5d pure $SU(2)$ index, with the identification
of Kahler parameters
\begin{equation}
  Q_1=\frac{\mathcal{Q}\mathcal{W}^2}{y_1y_2}=\frac{q}{y_1y_2y_3^2}
  (\equiv t_2)\ ,\ \
  W_1^2=e^{-v}\ .
\end{equation}
The last two lines yield another copy of 5d $SU(2)$ index, with parameters
\begin{equation}
  Q_2=\frac{y_2}{y_1}(\equiv t_1)\ ,\ \
  W_2^2=\frac{\mathcal{W}^2e^{-v}}{y_2}=\frac{w^2e^{-v}}{y_2y_3}\ .
\end{equation}
Note that the identifications of $Q_1,Q_2$ are consistent with our previous
findings based on affine $SO(7)$ structure. This identifies
the parameters (2), (7) of Fig.~\ref{232-5d}, and similarly (4), (9).

\begin{table}[t!]
\parbox{0.49\textwidth}
{
$$
\begin{array}{c||c|c|c|c}
  \hline p\setminus m&0&1&2&3\\
  \hline\hline 0&-2&
  y_3^{-\frac{1}{2}}(1\!+\!\frac{1}{y_1y_2})&-2(\frac{1}{y_1}\!+\!\frac{1}{y_2})
  &0\\
  \hline 1& 0&0&-\frac{2}{y_1y_2}&0\\
  \hline 2& 0&0&0&0\\
  \hline 3& 0&0&0&0\\
  \hline
\end{array}
$$
\caption{$N_{1,p,m}$ in the scaling limit}\label{p=1}
}
\parbox{0.55\textwidth}
{
$$
\begin{array}{c||c|c|c|c|c|c}
  \hline p\setminus m&0&1&2&3&4&5\\
  \hline\hline 0& 0&0&0&0&-\frac{4}{y_1y_2}&0\\
  \hline 1& 0&0&-\frac{4}{y_1y_2}&0&0&0\\
  \hline 2& 0&0&0&0&0&0\\
  \hline 3& 0&0&0&0&0&0\\
  \hline
\end{array}
$$
\caption{$N_{2,p,m}$ in the limit}\label{p=2}
}
$$
\begin{array}{c||c|c|c|c|c|c|c}
  \hline p\setminus m&0&1&2&3&4&5&6\\
  \hline\hline 0& 0&0&0&0&0&0&
  -\frac{6}{y_1y_2}(\frac{1}{y_1}\!+\!\frac{1}{y_2})\\
  \hline 1& 0&0&-\frac{6}{y_1y_2}&0&0&0&0\\
  \hline 2& 0&0&0&0&-\frac{6}{y_1^2y_2^2}&0&0\\
  \hline 3& 0&0&0&0&0&0&0\\
  \hline
\end{array}
$$
\caption{$N_{3,p,m}$ in the limit}\label{p=3}
\end{table}
\begin{table}[t!]
$$
\begin{array}{c||c|c|c|c|c|c|c|c|c}
  \hline p\setminus m&0&1&2&3&4&5&6&7&8\\
  \hline\hline 0& 0&0&0&0&0&0&0&0&-\frac{1}{y_1y_2}
  (\frac{8}{y_1^2}\!+\!\frac{8}{y_2^2}\!+\!\frac{32}{y_1y_2})\\
  \hline 1& 0&0&-\frac{8}{y_1y_2}&0&0&0&0&0&0\\
  \hline 2& 0&0&0&0&-\frac{32}{y_1^2y_2^2}&0&0&0&0\\
  \hline 3& 0&0&0&0&0&0&-\frac{8}{y_1^3y_2^3}&0&0\\
  \hline
\end{array}
$$
\caption{$N_{4,p,m}$ in the limit}\label{p=4}
\end{table}

\begin{figure}[t!]
    \centering
	\includegraphics[width=10cm]{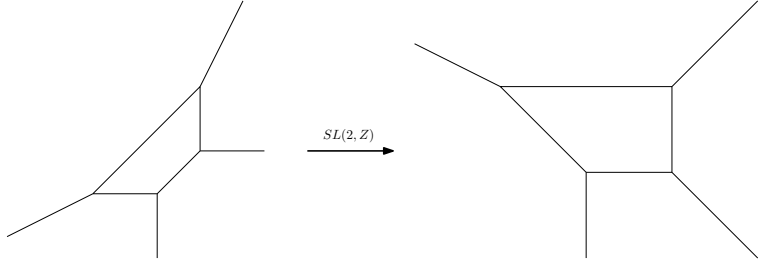}
	\caption{Brane web for the 5d $\tilde{E}_1$ SCFT}
	\label{tilde-E1}
\end{figure}

\hspace*{-0.65cm}{\bf \underline{$SO(7)$ subsector}:}
We now consider another subsector with
$k_1=k_3=0$, $k_2\neq 0$. We start from the elliptic genus of the $SO(7)$
instanton strings at $n_{\bf 8}=2$, studied in section 2.
In the 5d scaling limit, e.g. at $k_2=1$,
we found the following exact factorization,
\begin{eqnarray}\label{SO7-test-tilde}
  f_{\rm rel}&=&
  -U\left[\frac{1+Q_2}{W_2\tilde{W}_2(1-Q_2)^2}
  +(2\rightarrow 1)\right]\\
  &&-2U^2\left[\frac{3Q_2^2+4Q_2^3+3Q_2^4}{(W_2\tilde{W}_2)^2(1-Q_2)^4(1-Q_2^2)^2}
  +(2\rightarrow 1)\right]
  \nonumber\\
  &&-U^3
  \left[\frac{Q_2^3(27+70Q_2+119Q_2^2+119Q_2^3+70Q_2^4+27Q_2^5)}
  {(W_2\tilde{W}_2)^3(1-Q_2)^{8}(1-Q_2^3)^2}
  +(2\rightarrow 1)\right]+\cdots
  \nonumber
\end{eqnarray}
where $U\equiv e^{-u-\frac{v+\tilde{v}}{2}}q^{-\frac{1}{2}}y_1y_2y_3^2$.
So $f_{\rm rel}$ decomposes into two factors. To have such a factorization,
one should scale $e^{-u}\rightarrow\infty$ so that
$e^{-u}q^{-1/2}y_3^2\sim e^{-u}q^{1/2}$ is finite,
which guarantees that $U$ is finite. Here, one can show that each
factor takes the form of the instanton partition function for the 5d
$\tilde{E}_1$ SCFT, upon identifying
$U\left(Q_i^{1/2}W_i\tilde{W}_i\right)^{-1}$ as the instanton number fugacity
and $Q_i$ as the electric charge fugacity (Coulomb VEV), for $i=1,2$ respectively.
To understand this from the brane web description, we take the 5d factorization limit
$q\sim y_3^2\rightarrow 0$, and also consider the limit
$v,\tilde{v}\rightarrow\infty$
to realize the sector with $k_1=k_3=0$. One finds that Fig.~\ref{232-5d}(b)
decomposes into two $SU(1)$ theories in this limit since $(11)\rightarrow 0$,
thus void. Each factor of Fig.~\ref{232-5d}(a) becomes the left side of 
Fig.~\ref{tilde-E1}, since $(2),(4),(7),(9)\rightarrow 0$.
After an $SL(2,\mathbb{Z})$ transformation, it becomes
the right side of Fig.~\ref{tilde-E1}. This is the standard brane configuration
for the 5d $\tilde{E}_1$ theory \cite{Morrison:1996xf}.  It is
the 5d $U(2)$ theory at Chern-Simons level $1$.
From these studies, one can identify the Kahler parameters (3), (8)
of Fig.~\ref{232-5d}. Note that in (\ref{SO7-test-tilde}), the leading
term at $U^1$ order is $\frac{U}{W_i\tilde{W}_i}$ (with $i=1,2$) for
the two 5d $\tilde{E}_1$ factors. This is the Kahler parameter for the bottom
horizontal line on the right side of Fig.~\ref{tilde-E1}, since the leading
BPS states come from the strings stretched along this line. So one finds
$(3)=\frac{U}{W_1\tilde{W}_1}=e^{-u}\frac{y_1y_2y_3^2}{\sqrt{q}}$, $(8)=\frac{U}{W_2\tilde{W}_2}=e^{-u}\frac{y_1y_2^2y_3^3}{w\tilde{w}\sqrt{q}}$,
which were already shown in Fig.~\ref{232-5d}. Once we know (3) and (8),
one can determine $(11)$ from the gluing condition
$(11)=(2)(3)^2(4)=(7)(8)^2(9)$, again already
shown in Fig.~\ref{232-5d}. Thus we fixed all Kahler parameters
of Fig.~\ref{232-5d} in terms of our 6d fugacities.

We have in fact made a nontrivial test of our elliptic genera of section 2,
for the $SO(7)$ instanton strings at $n_{\bf 8}=2$, using the 5-brane web
description, from (\ref{SO7-test-tilde}).
Although apparently we tested the elliptic genera in a 5d
factorizing limit, this is different from the tests made in section 3.
This is because the `5d limit' here scales other massive parameters and keeps
a different slice of BPS states in its zero momentum sector. Indeed, using the
original 6d variables, (\ref{SO7-test-tilde}) is a nontrivial series
in $Q_1=\frac{q}{y_1y_2y_3^2}\sim q$, acquiring contributions from the 6d KK tower.
So this provides an independent nontrivial test of our results in section 2.

\hspace*{-0.65cm}{\bf \underline{More general sectors}:}
We shall continue to study the scaling limit of the elliptic genera for more general winding sectors, at $(k_1, k_2, k_3) = (1,1,0),  (1,2,0), (1,1,1),(1,2,1)$.

In the first three sectors, Fig.~11(b) factorizes to two `5d $SU(1)$' factors
which are void, as these sectors are realized by $(4), (9), (11) \rightarrow 0$ for $(k_1, k_2, k_3) = (1,1,0),(1,2,0)$ and $(11)  \rightarrow 0$ for $(k_1, k_2, k_3) = (1,1,1)$. So we expect the factorization of the single particle index into two identical pieces, each representing a non-Lagrangian 5d SCFT engineered
by Fig.~11(a) in a particular limit. In all cases, we find
exact factorizations of $f_{\rm rel}$ into two functions of identical form, as follows:
\begin{align}
	\label{eq:index-110}
	(1,1,0):&&f_{\rm rel} &= e^{-v-u}\, \frac{y_1y_2y_3^2}{\sqrt{q}}\cdot \frac{(1+q/y_1y_2y_3^2)^2}{(1-q/y_1y_2y_3^2)^2} + e^{-v-u}\,\frac{w y_1y_2y_3^2}{\tilde{w}\sqrt{q}}\cdot\frac{(1+{y_2}/{y_1})^2}{(1-{y_2}/{y_1})^2}\nonumber\\
	(1,2,0):&&f_{\rm rel} &= e^{-v-2u}\,  \frac{y_1^2y_2^2y_3^4}{q}\cdot\frac{-10\,(q/y_1y_2y_3^2)^2 \, (1+q/y_1y_2y_3^2)}{(1-q/y_1y_2y_3^2)^6} \nonumber\\
&&&\hspace{0.5cm}+ e^{-v-2u}\, \frac{y_1^2 y_2^3 y_3^5}{\tilde{w}^2 q}\cdot\frac{-10\, (y_2/y_1)^2\, (1+{y_2}/{y_1})}{(1-{y_2}/{y_1})^6} \nonumber\\
	(1,1,1):&&f_{\rm rel} &= e^{-v-u-\tilde{v}}  \, \frac{y_1y_2y_3^2}{\sqrt{q}}\frac{(1+q/y_1y_2y_3^2)^3}{(1-q/y_1y_2y_3^2)^2} + e^{-v-u-\tilde{v}}  \,\frac{w\tilde{w}y_1y_3}{\sqrt{q}}\frac{(1+{y_2}/{y_1})^3}{(1-{y_2}/{y_1})^2}.
\end{align}
Each term is a product of the prefactor $(2)^{k_1} (3)^{k_2} (4)^{k_3}$ or $(7)^{k_1} (8)^{k_2} (9	)^{k_3}$ and a function of K\"ahler parameter $(1)=(5)$ or $(6)=(10)$, respectively. To test these results, extracted from the elliptic genera in
Section~4.1, we shall independently do the topological vertex calculus for the 5d SCFT of Fig.~11(a).

The topological vertex \cite{Aganagic:2003db} computes all genus topological
string amplitudes, which is equivalent to the logarithm of the 5d Nekrasov partition function on Omega-deformed $\mathbf{R}^4\times S^1$ \cite{Gopakumar:1998jq}. Here we refer to \cite{Bao:2013pwa,Hayashi:2013qwa} for its detailed description. We select an orientation of every edge in the 5-brane web. Each internal edge is associated with a Young diagram. We also assign an empty Young diagram to every external edge. The 5d partition function is given by a sum over all combinations of Young diagrams. The summand is a product of factors coming from every edge and vertex.	
We turn off $\epsilon_+= 0$ to simplify the formulae. When all three edges are outgoing from a given vertex, the vertex factor is given by
(where $u=e^{-\epsilon_-}$, $||\mu||^2=\sum_i\mu_i^2$)
\begin{align}
C_{\lambda\mu\nu}(u)&=u^{\frac{||\mu||^2+||\nu||^2-||\mu^t||^2}{2}}\, \prod_{s\in\nu}(1-u^{l_{\nu}(s)+a_{\nu}(s)+1})^{-1} \cdot \sum_{\eta} s_{\lambda^t/\eta}(u^{-\rho}u^{-\nu})s_{\mu/\eta}(u^{-\rho}u^{-\nu^t})\ .
\end{align}
$\lambda, \mu, \nu$ are Young diagrams associated to the edges. For an incoming edge, the assigned Young diagram should be transposed. The skew-Schur function $s_{\lambda/\eta}(\mathbf{x})$ depends on a possibly infinite vector $\mathbf{x}$, which in above is $u^{-\rho}u^{-\nu}\equiv (u^{\frac{1}{2}-\nu_1}, u^{\frac{3}{2}-\nu_2}, u^{\frac{5}{2}-\nu_3}, \cdots)$.
The functions $l_{\nu}(s)$ and $a_{\nu}(s)$ are defined by  $l_{\nu}(s)=\nu_i-j$ and $a_{\nu}(s)=\nu_j^t-i$, where $i, j$ represent the horizontal and vertical positions of the box $s$ from the upper-left corner of $\nu$. It is known that  $C_{\lambda\mu\nu}(u)$ is invariant under the cyclic permutation of $\lambda, \mu, \nu$ using Schur function identities \cite{Aganagic:2003db}. An internal edge glues a pair of vertices by multiplying the edge factor and summing over the assigned Young diagram. Denoting its K\"ahler parameter by $Q$, the edge factor
is given by
\begin{align}
	\includegraphics{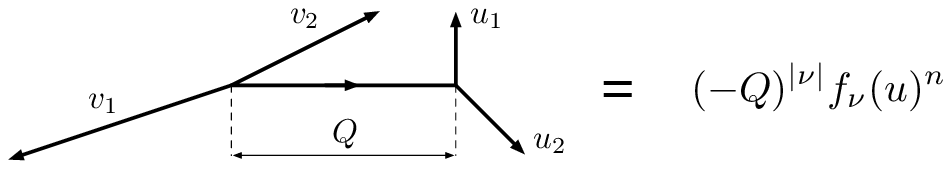}
\end{align}
where ${f}_{\nu}(u)=(-1)^{|\nu|}\,u^{\frac{||\nu^t||^2-||\nu||^2}{2}}$ and $n = \det(u_1,v_1)$. Applying these rules, one obtains the following partition function
of 5d SCFT engineered from the brane web of Fig.~11(a),
\begin{align}
\label{eq:tv-fig11a}
\includegraphics[scale=0.7]{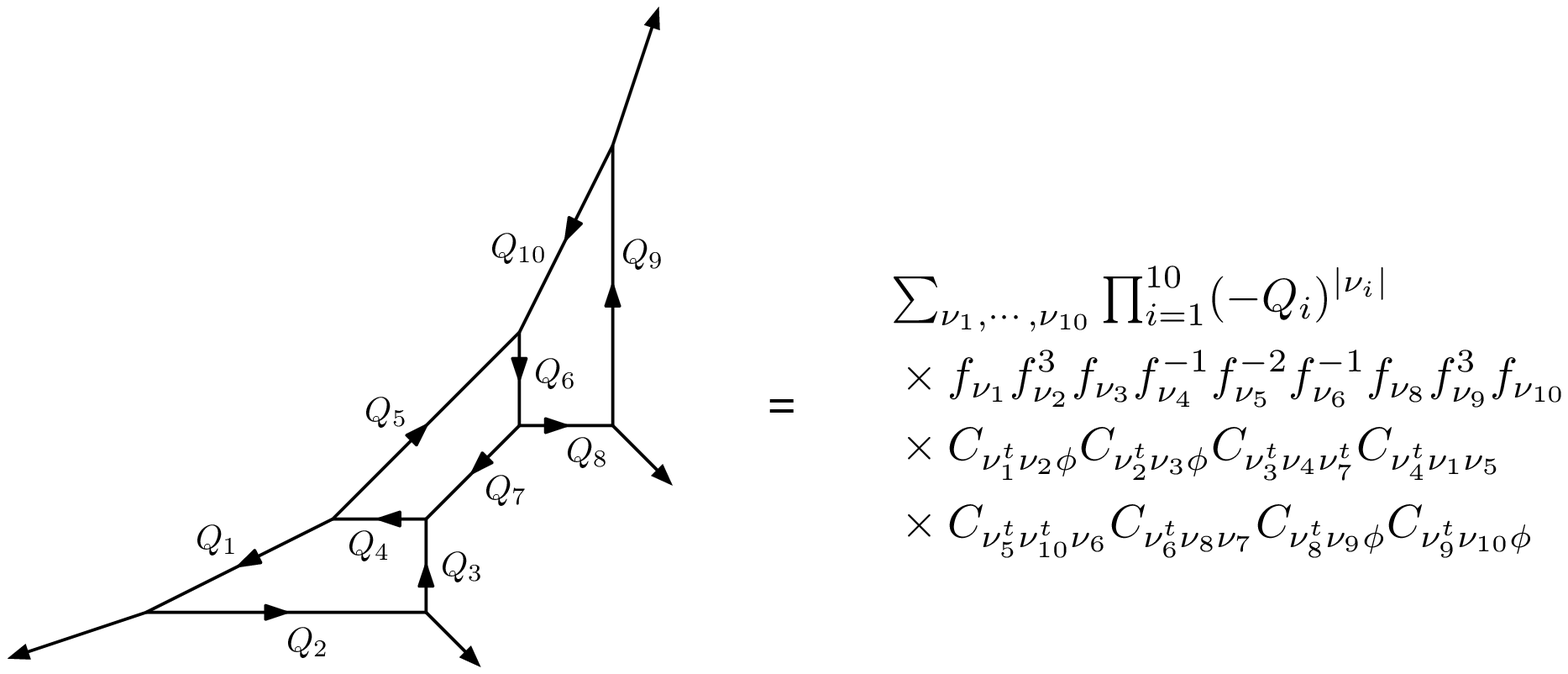}\ .
\end{align}
$Q_1, \cdots, Q_{10}$ are identified with the K\"ahler parameters in Fig.~11(a)
as
\begin{align}
Q_1= Q_3 = \alpha,\ Q_4=Q_6 = \beta,\ Q_7=\gamma,\ Q_8=Q_{10}=\delta,
\ Q_2=\alpha^2 \beta,\ Q_5=\beta \gamma,\ Q_9=\beta \delta^2
\end{align}
where $(\alpha, \beta, \gamma, \delta) = (e^{-v}, \frac{q}{y_1y_2y_3^2}, e^{-u}\frac{y_1y_2y_3^2}{\sqrt{q}},e^{-\tilde{v}})$ or $(\frac{w^2e^{-v}}{y_2y_3}, \frac{y_2}{y_1}, e^{-u}\frac{y_1y_2^2y_3^3}{w\tilde{w}\sqrt{q}}, \frac{\tilde{w}^2e^{-\tilde{v}}}{y_2y_3})$ respectively.
To derive the single particle spectrum of $(k_1, k_2, k_3)$  sector, we perform the sum \eqref{eq:tv-fig11a} over Young diagrams until $|\nu_1| + 2|\nu_2| + |\nu_3| \leq k_1$, $|\nu_5| + |\nu_7| \leq k_2$, $|\nu_8| + 2|\nu_9| + |\nu_{10}| \leq k_3$ and take the Plethystic logarithm.
To compare with \eqref{eq:index-110}, we further multiply $-(2\sinh\frac{\epsilon_-}{2})^2$ on it and take the limit
$\epsilon_- \rightarrow 0$. After these manipulations, one obtains
\begin{align}
	\label{eq:tp-110}
	(1,1,0):&&f_{\rm top} &= \alpha \gamma \cdot\left(1+4\beta+8\beta^2+12\beta^3+16\beta^4+20\beta^5 + \mathcal{O}(\beta^6)\right) \nonumber\\
	(1,2,0):&&f_{\rm top} &= \alpha \gamma^2 \cdot\left(-10\beta^2-70\beta^3-270\beta^4-770\beta^5 +  \mathcal{O}(\beta^6)\right) \nonumber\\
	(1,1,1):&&f_{\rm top} &= \alpha \gamma \delta \cdot\left(1+5\beta+12\beta^2+20\beta^3+28\beta^4+36\beta^5 +  \mathcal{O}(\beta^6)\right).
\end{align}
These agree with $f_{\rm rel}$ in \eqref{eq:index-110}, testing
our elliptic genera in Section~4.1.

We finally consider the $(1,2,1)$ sector. $f_{\rm rel}$
in the factorization limit is given by
\begin{align}
	\label{eq:index-121}
	f_{\rm rel} = e^{-v-2u-\tilde{v}} \frac{y_1^2y_2^2y_3^4}{q} \cdot \Bigg[& \frac{-2y_1y_2}{(1-y_1y_2)^2} + \Big(\frac{-2 k \left(k^4+4 k^3+30 k^2+4 k+1\right)}{(k-1)^6}\Big)\bigg|_{k=\frac{y_2}{y_1}}\nonumber\\
	&+ \Big(\frac{-2 k \left(k^4+4 k^3+30 k^2+4 k+1\right)}{(k-1)^6}\Big)\bigg|_{k=\frac{q}{y_1y_2y_3^2}}  + (-2)\Bigg]\ .
\end{align}
The common prefactor $e^{-v-2u-\tilde{v}} {y_1^2y_2^2y_3^4}/{q}$ is
$(2)(3)^2(4) = (7)(8)^2(9) = (11)$. The first term agrees with the 1 instanton partition function of 5d pure $SU(2)$ gauge theory, if we identify $\sqrt{y_1y_2}$ as the fugacity of the $SU(2)$ electric charge. It belongs to the 5d $E_1$ SCFT of Fig.~11(b). The next two terms take the same functional form,
respecting the $\mathbb{Z}_2$ symmetry of the two factors. To test this function,
we performed the topological vertex calculus for (\ref{eq:tv-fig11a}). We first sum
over all Young diagrams with $|\nu_1| + 2|\nu_2| + |\nu_3| \leq 1$, $|\nu_5| + |\nu_7| \leq 2$, $|\nu_8| + 2|\nu_9| + |\nu_{10}| \leq 1$ and take the Plethystic logarithm. We then subtract the extra factor $\alpha \gamma^2 \delta (2\sinh\frac{\epsilon_-}{2})^{-2}$ that arises
because the strings can propagate along the parallel 5-branes \cite{Bao:2013pwa,Hayashi:2013qwa}. Dividing out the center-of-mass factor $-(2\sinh\frac{\epsilon_-}{2})^{-2}$ and turning off $\epsilon_- \rightarrow 0$, the topological string partition function becomes
\begin{align}
	\label{eq:tp-121}
	(1,2,1):&&f_{\rm top} &= \alpha \gamma^2 \delta \cdot\left( - 2\beta - 20\beta^2 - 150\beta^3 - 648\beta^4 - 2010\beta^5 +  \mathcal{O}(\beta^6)\right).
\end{align}
It agrees with the second and third terms of $f_{\rm rel}$ in \eqref{eq:index-121}.
The final $(-2)$ comes from the perturbative $SU(2)_g$ vector multiplet. 
Again, this result gives a non-trivial independent  test of our elliptic 
genera in Section~4.1.

\subsection{$3,2$ and $3,2,2$: $G_2\times SU(2)$ gauge group}

We construct 2d quivers for the strings of other 6d SCFTs in Table \ref{other}.
The tests we can provide about them are weak (e.g. anomalies).
We keep the presentations rather brief.

\hspace*{-0.65cm}{\bf \underline{$3,2,2$ SCFT strings}:} The strategy is
similar to that of section 4.1. We first consider the limits in which all except
one gauge symmetry are ungauged in 6d, and take three factors of
ADHM(-like) quivers. We then combine these quivers by locking certain symmetries,
and introducing bi-fundamental matters of the form of (\ref{bif-ADHM-1}).
To be more precise, we have no 6d gauge
group associated with the `$2$' node on the right.
Although the notion of ungauging is absent for this node, we can still
take the tensor VEV associated with
this node to infinity. Whenever a node has a 6d gauge group, its inverse coupling
is proportional to the tensor VEV $\langle\Phi\rangle$, so taking
$\langle\Phi\rangle\rightarrow\infty$ ungauges the symmetry.

If one takes all tensor VEVs to infinity except the `$3$' node, one obtains
the 6d $G_2$ theory at $n_{\bf 7}=1$.
This is because the 6d matter in $\frac{1}{2}({\bf 7},{\bf 2})$
behaves like one full hypermultiplet in ${\bf 7}$, while
$\frac{1}{2}({\bf 1},{\bf 2})$ is neutral in $G_2$ and invisible in the
gauge dynamics. So with a $G_2$ theory at $n_{\bf 7}=1$,
its $k_1$ $G_2$ instanton strings are described by the
2d $U(k_1)$ gauge theory explained in section 2.2, with fields given by
(\ref{G2-matter-1}), (\ref{G2-matter-2}), (\ref{G2-matter-3}) at
$n_{\bf 7}=1$. The ungauged $SU(2)\sim Sp(1)$ acts as the flavor symmetry
of the 6d hypermultiplet. In the ADHM-like quiver at general $n_{\bf 7}$,
one may have as big as $U(2n_{\bf 7})$ flavor symmetry which rotates Fermi multiplets.
But the coupling to bulk fields only allowed $U(n_{\bf 7})$ part,
which we further expected to enhance to $Sp(n_{\bf 7})$. This is similar to the flavor
symmetries of $SO(7)$ ADHM-like theory at $n_{\bf 8}\neq 0$. In the current
context, again like the $2,3,2$ quiver, we should couple the system to different
bulk fields. At $n_{\bf 7}=1$, one can classically have as big as
$U(2n_{\bf 7})\rightarrow U(2)$
flavor symmetry. We restrict it to $SU(2)$ which rotates
$\Psi,\tilde\Psi^\dag$ of (\ref{G2-matter-3}) as a doublet.
Also, as explained in section 2.2, only $SU(3)\subset G_2$ is visible in this
quiver. More formally, it will be  convenient to regard the fields
$q_i,\tilde{q}^i,\phi_i,\phi_4$ as
transforming in $SU(3)\times SU(1)\subset SU(4)$.

When $G_2$ is ungauged and the tensor VEV for the right `$2$' node is sent to
infinity, we have 6d $SU(2)$ theory at $n_{\bf 2}=4$. Its ADHM quiver is explained
around (\ref{SU2-Nf=4-ADHM}). In this limit, $G_2$ is enhanced to
$SO(7)$ flavor symmetry rotating the four hypermultiplets in ${\bf 8}$ of
$SO(7)$, but only $SU(4)\subset SO(7)$ is visible in the UV ADHM, as explained
in section 4.1. $SO(7)$ will later be broken to $G_2$ by gauging. In our
ADHM-like quiver, which only sees $SU(3)\subset G_2$, $SU(4)$ will be
broken to $SU(3)\times SU(1)$, locked with the $G_2$ ADHM of the previous
paragraph.

\begin{figure}[t!]
    \centering
    \subcaptionbox{}{
	\includegraphics[width=9.5cm]{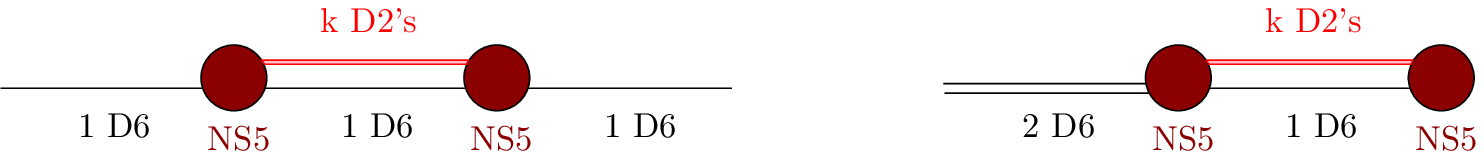}}\hspace{1cm}
    \quad
    \subcaptionbox{}{
    \includegraphics[width=5cm]{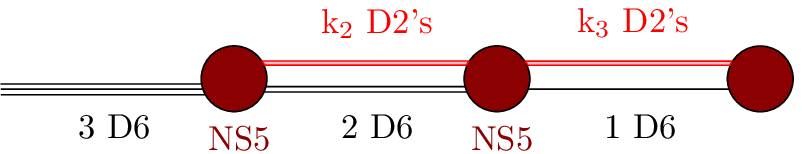}}
	\caption{Brane configurations for: (a) 6d $(2,0)$ SCFT of type $A_1$;
    (b) 6d $3,2,2$ SCFT in the limit with ungauged $G_2$}
	\label{2-22-brane}
\end{figure}
We finally ungauge $G_2\times SU(2)$, leaving one tensor VEV for the right `$2$' node
finite. One then obtains the 6d $\mathcal{N}=(2,0)$ SCFT of $A_1$ type,
geometrically engineered on the $O(-2)\rightarrow\mathbb{P}^1$ base with no
associated gauge group. Although the strings of this SCFT in the tensor branch
lacks the instanton string interpretation, one still knows the UV 2d
gauge theory description \cite{Haghighat:2013gba}. For $k$ strings, this is
a $U(k)$ gauge theory. The 2d fields are given by
\begin{eqnarray}\label{SU1-Nf=2-ADHM}
  (A_\mu,\lambda_0,\lambda)&:&\textrm{vector mutiplet in }({\bf adj},0)\\
  q_{\dot\alpha}=(q,\tilde{q}^\dag)&:&\textrm{hypermultiplet in }
  ({\bf k},-1)\nonumber\\
  a_{\alpha\dot\beta}\sim(a,\tilde{a}^\dag)&:&
  \textrm{hypermultiplet in }({\bf adj},0)\nonumber\\
  \Psi_a&:&\textrm{Fermi multiplet in }({\bf k},0)\ ,\nonumber
\end{eqnarray}
where $a=1,2$. We showed the representation and charge of
the classical symmetry $U(k)\times U(1)$, where one should further restrict
$U(1)\rightarrow SU(1)$ due to mixed anomaly. This formally
takes the form of the ADHM instanton strings of `6d $SU(1)$ theory' with two
charged quarks. The $SU(2)_F$ flavor symmetry which rotates $\Psi_a$ is identified
with the enlarged R-symmetry group of the 6d $(2,0)$ theory. Namely, we expect
that $SU(2)_R$ of 6d $(1,0)$ SCFT enhances to $SO(5)_R$. In the tensor branch,
this is broken to $SO(4)\sim SU(2)_R\times SU(2)_L$, where the latter $SU(2)_L$
is realized as $SU(2)_F$ in the 2d quiver.
The 6d $A_1$ $(2,0)$ theory and the above 2d gauge theory
admit D-brane engineerings. Using D2-D6-NS5, one can use either of
Fig. \ref{2-22-brane}(a), in IIA or massive IIA string theory
\cite{Hanany:1997gh,Brunner:1997gf}.

\begin{figure}[t!]
    \centering
	\includegraphics[width=9.5cm]{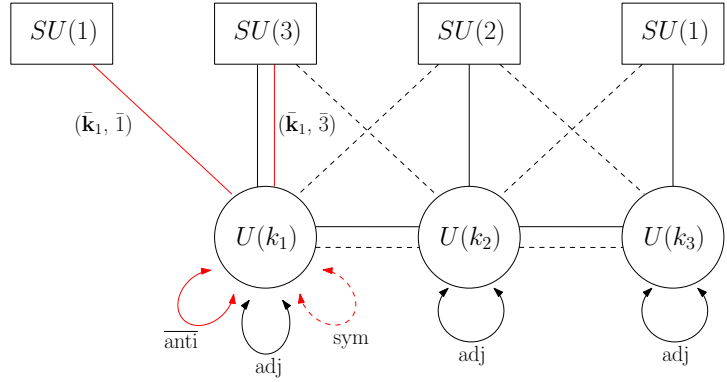}
	\caption{2d quiver for the strings of 6d $3,2,2$ SCFT}
	\label{322-quiver}
\end{figure}

Before fully combining the three ADHM(-like) quivers, we note
that the combination of two `$2$' nodes (with $G_2$ ungauged) is dictated by
a D-brane setting. This is given by the brane configuration of
Fig.~\ref{2-22-brane}(b) in the massive IIA theory. The 2d quiver is given by
Fig.~\ref{322-quiver} at $k_1=0$. The quiver and the brane system only
has manifest $SU(3)\times SU(2)\times U(1)$ symmetry,
where the last $U(1)$ is a combination of three overall $U(1)$'s in
$U(3)\times U(2)\times U(1)$ which survive the mixed anomaly cancelation with
$U(k_2)\times U(k_3)$. More precisely, taking the overall $U(1)$ generators
$Q_i$ for $SU(i)$, $i=1,2,3$, only $Q_1+Q_2+Q_3$ is free of the mixed anomaly.
(This $U(1)$ is not shown in Fig.~\ref{322-quiver}, as it will be irrelevant
generally at $k_1,k_2,k_3\neq 0$.)
One can see that the 2d quiver exhibits
$SU(3)\times U(1)\rightarrow SO(7)$ symmetry enhancement, say by studying the
elliptic genera. This should be the case since one has 6d $SU(2)$ theory at
$n_{\bf 2}=4$. Just to be sure, we tested the $SO(7)$ enhancement
of the elliptic genus at $k_2=k_3=1$.

Now we keep $k_1\neq 0$, with $G_2$ gauged. In our UV GLSM, we can
only see $SU(3)\subset G_2$, which we lock with the $SU(3)$ symmetry of the
quiver in the previous paragraph. The resulting
$U(k_1)\times U(k_2)\times U(k_3)$ quiver is given by
Fig.~\ref{322-quiver}. The potentials can be written down in a similar manner as
the $2,3,2$ quiver of section 4.1. We skip the details here.

As a small test of our quiver, we compute the 2d anomalies. We first
compute it from inflow. The Green-Schwarz
part of the 6d anomaly 8-form is given by $I_{GS}=\frac{1}{2}\Omega^{ij}I_iI_j$
with
\begin{equation}\label{322-GS-anomaly}
  I_i=\left(\begin{array}{c}
  \frac{1}{4}{\rm Tr}(F_{G_2}^2)+\alpha_1c_2(R)+\alpha_2p_1(T)\\
  \frac{1}{4}{\rm Tr}(F_{SU(2)}^2)+\beta_1c_2(R)+\beta_2p_1(T)\\
  \gamma_1c_2(R)+\gamma_2p_1(T)
  \end{array}\right)\ \ ,\ \ \
  \Omega^{ij}=\left(\begin{array}{ccc}
    3&-1&0\\-1&2&-1\\0&-1&2
  \end{array}\right)\ ,
\end{equation}
where $(\alpha_1,\beta_1,\gamma_1)=(\frac{17}{7},\frac{23}{7},\frac{15}{7})$,
$(\alpha_2,\beta_2,\gamma_2)=(\frac{3}{28},\frac{1}{14},\frac{1}{28})$.
We explain how to get this result. \cite{Ohmori:2014kda}
uses two methods to compute $I_{GS}$. One is applicable when
all nodes have gauge symmetries. In this case, one demands that $I_{GS}$ cancels
all terms in $I_{\textrm{1-loop}}$ containing dynamical fields.
This is the method we used so far in this paper.
When some nodes do not have gauge symmetries, this method alone cannot
completely determine $I_{GS}$. We use the following strategy to compute
(\ref{322-GS-anomaly}). Firstly, we compute the 1-loop anomaly containing
the dynamical $G_2\times SU(2)$ gauge fields, and demand that this part is
completely canceled by $I_1,I_2$ part of $I_{GS}$ \cite{Intriligator:2014eaa}.
Then one obtains $I_i$ of the form (\ref{322-GS-anomaly}), where the six
coefficients $\alpha_{1,2},\beta_{1,2},\gamma_{1,2}$ are constrained only
by the following four equations,
\begin{equation}\label{coefficient-eqn-1}
  3\alpha_1-\beta_1=4\ ,\ \ 3\alpha_2-\beta_2=\frac{1}{4}\ ,\ \
  2\beta_1-\alpha_1-\gamma_1=2\ ,\ \ 2\beta_2-\alpha_2-\gamma_2=0\ .
\end{equation}
To further constrain them, we consider the limit in which two tensor VEVs
are sent to infinity so that $G_2\times SU(2)$ are ungauged.
In this limit, we can use the known expression for $I_{GS}$ for the
$A_1$ $(2,0)$ theory in the tensor branch \cite{Intriligator:2014eaa,Ohmori:2014kda},
\begin{equation}\label{(2,0)-GS}
  I_{GS}=\frac{1}{2}\Omega\left(\frac{1}{2}(c_2(R)-c_2(L))\right)^2\ ,
  \ \ \Omega=2\ ,
\end{equation}
with enhanced $SU(2)_R\times SU(2)_L=SO(4)\subset SO(5)$ R-symmetry.
After taking this limit, we can set
$\frac{1}{4}{\rm Tr}(F_{SU(2)}^2)=c_2(L)$ by identifying the ungauged $SU(2)$
with $SU(2)_L$. To take this limit, consider the vector kinetic terms
proportional to
$\mathcal{L}_{\rm v}\sim\Omega^{ij}\Phi_i{\rm Tr}(F_j^2)\equiv \Phi^i{\rm Tr}(F_i^2)$.
We keep $\Phi^3=2\Phi_3-\Phi_2$ finite, while taking
$\Phi^1=3\Phi_1-\Phi_2$ and $\Phi^2=2\Phi_2-\Phi_1-\Phi_3$ to $+\infty$,
to ungauge $G_2\times SU(2)$. To properly do so,
note that the kinetic terms for $\Phi_i$ are proportional to
$\mathcal{L}_{\rm t}\sim\Omega^{ij}\partial^\mu \Phi_i\partial_\mu\Phi_j$.
This is diagonalized by taking, say,
$\Phi_3=a+\chi$, $\Phi_2=2a$, $\Phi_1=b+\frac{2a}{3}$, since
$\mathcal{L}_t\sim\frac{14}{3}(\partial a)^2+3(\partial b)^2+2(\partial \chi)^2$.
So one holds the scalars $a,b$ very large and fixed,
unaffected by the dynamical $\chi$ and its superpartner.
More precisely, $a,b$ can be hold fixed, given by infinite constant plus
a finite background function given by the background gauge fields.
$\chi$ is a dynamical scalar associated with
the right `$2$' node with normalization $\Omega=2$.
In this parameterization $\chi,a,b$ of tensor multiplet scalars, one can
similarly show that the superpartners $H_a,H_b$ of $a,b$ can be consistently
taken to be fixed background functions, unaffected by dynamical $\chi$ and
its superpartner $H_\chi$. Now consider the equation of motion for $H_\chi$.
The coupling between $B_\chi$ and the dynamical/background vector fields
is given by
\begin{equation}
  \Omega^{ij}B_i\wedge I_j\rightarrow B_\chi \Omega^{3i}I_i=B_\chi\wedge (2I_3-I_2)\ .
\end{equation}
We used $\Omega^{ij}B_j=
(\cdots,-B_\chi+\cdots,2B_\chi)$, where $\cdots$ depend on $B_a,B_b$,
so that it depends on $B_\chi$ as $\Omega^{i3}B_\chi$.
From the equation of motion for $B_\chi$, one obtains
\begin{equation}
 d\star H_\chi =(\Omega^{33})^{-1}\Omega^{3i}I_i=I_3-\frac{1}{2}I_2\ .
\end{equation}
By comparing this with (\ref{(2,0)-GS}), one obtains
$I_3-\frac{1}{2}I_2=\frac{c_2(R)-c_2(L)}{2}$ with
$c_2(L)=\frac{1}{4}{\rm Tr}(F_{SU(2)}^2)$.
This leads to two more equations for $\alpha_{1,2},\beta_{1,2},\gamma_{1,2}$,
\begin{equation}\label{coefficient-eqn-2}
  2\gamma_1-\beta_1=1\ ,\ \ 2\gamma_2-\beta_2=0\ .
\end{equation}
The unique solution of (\ref{coefficient-eqn-1}), (\ref{coefficient-eqn-2}) is
the one stated right below (\ref{322-GS-anomaly}).\footnote{In fact,
expanding the arguments of this paragraph, one can compute $I_{GS}$ if
one knows the Green-Schwarz anomalies of all individual rank $1$ nodes
before combining them. The general rule is as
follows. Suppose that $I^{(i)}_{\rm GS}=\frac{1}{2}\Omega^{ii}(I_i)_{\rm single}^2$
(no sum of $i$) when only $i$'th node is kept. Then defining
$I^i\equiv (\Omega^{ii})^{-1}(I_i)_{\rm single}$ (no sum of $i$), one finds
$I_{\rm GS}=\frac{1}{2}(\Omega^{-1})_{ij}I^iI^j$. $I_i$ that we computed in
(\ref{232-GS-anomaly}) and (\ref{322-GS-anomaly}) are given by
$I_i=(\Omega^{-1})_{ij}I^j$.}

This leads to the anomaly 4-form on the strings from inflow,
(\ref{inflow-anomaly}), given by
\begin{equation}\label{anomaly-322-string}
  I_4=\left(k_1k_2+k_2k_3-\frac{3}{2}k_1^2-k_2^2-k_3^2\right)\chi(T_4)
  +k_1(I_2-3I_1)+k_2(I_1+I_3-2I_2)+k_3(I_2-2I_3)\ .
\end{equation}
This is computed from our 2d gauge theory as follows.
We again decompose the anomaly into contributions
$I^{(1)}_4$ from the $G_2$ ADHM-like quiver, $I^{(2)}_{4}$ from the
middle `$2$' node (6d $SU(2)$ theory at $n_{\bf 2}=4$),
$I^{(3)}_4$ from the right `$2$' node, and
$I^{\rm bif}_4=(k_1k_2+k_2k_3)\chi(T_4)$. $I^{(1)}_4$ and $I^{(2)}_4$ are
given by (\ref{G2-2d-anomaly}), and (\ref{SU2-Nf=4-anomaly})
replacing $F_{SO(7)}\rightarrow F_{G_2}$.
$I^{(3)}_4$ is given by eqn.(5.21) of  \cite{Kim:2016foj} at $N=1$,
\begin{equation}
  I^{(3)}_4=-\frac{k_3}{2}{\rm Tr}(F_{SU(1)}^2)
  +\frac{k_3}{4}{\rm Tr}(F_{SU(2)}^2)-k_3c_2(R)-k^2\chi(T_4)\ ,
\end{equation}
where one should set $F_{SU(1)}=0$.
Adding $\sum_{i=1}^3I^{(i)}_4+I^{\rm bif}_4$, one precisely reproduces
(\ref{anomaly-322-string}).

\hspace*{-0.65cm}{\bf \underline{$3,2$ SCFT strings}:}
This SCFT can be obtained from
the previous $3,2,2$ SCFT by taking the tensor VEV of the right `$2$' node
to infinity. The corresponding 2d quiver for its strings can be obtained from
our previous quiver for the $3,2,2$ model, by taking $k_3=0$. All the
discussions made for the $3,2,2$ string quivers apply here as well.

\section{Conclusion and remarks}

In this paper, we first proposed 1d ADHM-like gauge theories for Yang-Mills
instantons for 5d $SO(7)$ theories with $n_{\bf 8}\leq 4$ matters in spinor
representation, and for $G_2$ theories with $n_{\bf 7}\leq 3$ matters in ${\bf 7}$.
At $n_{\bf 8}=2$ for $SO(7)$ and at $n_{\bf 7}=1$ for $G_2$, where anomaly-free
6d gauge theories exist, our gauge theories uplift to 2d for
instanton strings. These ADHM strings can be used to construct
the 2d quivers for the `atomic' non-Higgsable 6d SCFTs of Table \ref{other}.
These gauge theories do not describe the symmetric phase physics of instantons,
but we propose them to compute the Coulomb phase partition functions correctly.
Although the worldvolumes of instantons host
$\mathcal{N}=(0,4)$ SUSY (or its 1d reduction), our gauge theories are made of
$\mathcal{N}=(0,2)$ supermultiplets, and some of their interactions only exhibit
$\mathcal{N}=(0,1)$ SUSY. We expect various symmetry enhancements in 1d/2d.

We tested our 1d/2d gauge theories by computing their Witten indices or
elliptic genera using various other methods. Firstly, for 5d $G_2$ theory
without matters,
$n_{\bf 7}=0$, we used the results of \cite{Cremonesi:2014xha}, which
uses 3d Coulomb branch techniques. We tested our results for instanton
numbers $k\leq 3$, but the comparisons can in principle be made
for arbitrary high $k$'s.
In section 3, we developed another D-brane-based
method to study the instantons of 5d $SO(7)$ theories at $n_{\bf 8}=1,2$, and
related $G_2$ theories at $n_{\bf 7}=0,1$. This method provides
a much more elaborate computational procedure, which however does not require
guesswork. We used this method to successfully test our results for
$n_{\bf 8}=2$ at $k=1$, and for $n_{\bf 8}=1$ at $k=1,2$. Finally, we used
the 5-brane web description of \cite{Hayashi:2017jze} to test the $SO(7)$ instanton
strings
at $n_{\bf 8}=2$, and also the strings of the 6d $2,3,2$ SCFT of Table \ref{other}.
All the methods that we used to test our results exhibit manifest $(0,4)$ SUSY.
So the agreements of our indices with these alternative calculus are indirect
signals that our systems exhibit $(0,4)$ SUSY enhancement.

As we alluded to at the beginning of section 2, we have made similar trials
to construct ADHM-like gauge theories with other gauge groups,
e.g. some of them in Table \ref{subgroup}. For technical reasons,
we focused on the $E_7$ case, using a formalism which only sees
manifest $SU(8)\subset E_7$. We managed to build a model which exhibits
the correct anomaly polynomial 4-form for $E_7$ instanton strings, which
also `closely' (but not precisely) reproduces the 1-instanton Hilbert series of
the $E_7$ instanton particle. For instance, keeping $t=e^{-\epsilon_+}$ fugacity
only and multiplying the center-of-mass factor $2\sinh\frac{\epsilon_{1,2}}{2}$,
the correct Hilbert series \cite{Hanany:2012dm} and the index of
our trial gauge theory are given as follows
(up to $0$-point energy factor):
\begin{eqnarray}\label{E7-hilbert}
  \hat{Z}_{k=1}^{E_7}&=&1 + 133 t^2 + 7371 t^4 + 238602 t^6 + 5248750 t^8 +
  85709988 t^{10} + 1101296924 t^{12} \\
  &&+ 11604306012 t^{14} +103402141164 t^{16} + 797856027500 t^{18} 
  + 5431803835220 t^{20}+\cdots\ ,\nonumber\\
  \hat{Z}_{k=1}^{\rm trial}&=&1+133 t^2 + 7300 t^4 + 234689 t^6
  + 5143821 t^8 + 83863116 t^{10} + 1077066537 t^{12}\nonumber\\
  &&+ 11349844981 t^{14}+101164274246 t^{16} + 780860775912 t^{18}
  + 5317874678676 t^{20}+\cdots\ .\nonumber
\end{eqnarray}
The coefficient of $t^{2n}$ for the correct result $\hat{Z}^{E_7}_{k=1}$ is
the $n$'th symmetric product of the $E_7$ adjoint representation ${\bf 133}$.
The index $\hat{Z}^{\rm trial}_{k=1}$ is
close to $\hat{Z}^{E_7}_{k=1}$ at low orders in $t$ (e.g. exact at $t^2$),
but slightly deviates and converges to $\approx -1.928\%$ error asymptotically
at large orders. We think this failure provides helpful lessons on
our ADHM-like trials, and possible subtle points (some
stated at the beginning of section 2).

First of all, the gauge theory we constructed which yields
$\hat{Z}^{\rm trial}_{k=1}$ has two branches of moduli spaces. The first
branch has the
$SU(8)$ instanton moduli space as a subspace, and has the right complex
dimension $2kc_2(E_7)=36k$ for $E_7$ instantons.
The second branch meets the first one at a point,
the small instanton singularity, and arises from the extra matters we
added to the $SU(8)$ ADHM. We find that the second branch cannot be
eliminated (with given extra matters) by turning on $\mathcal{N}=(0,1)$ potentials.
We suspect that there may be a contribution to $\hat{Z}^{\rm trial}_{k=1}$
from the second branch, which spoils the results. Since we are doing a UV
computation in which two branches are not separated, we do not know how/whether
one can separate the contributions from two branches. Certainly this may be
one reason for the deviation. In case this is the dominant reason for the deviation,
there should be large enough contributions from massless fermions on
the second branch, as $\hat{Z}^{\rm trial}_{k=1}$ is always no greater
than $\hat{Z}^{E_7}_{k=1}$.

Secondly, there is another reason why we suspect the deviation happens.
To explain this, note that ${\bf 133}$ at $t^2$ order is correctly reproduced
in our trial gauge theory. From the branching rule
${\bf 133}\rightarrow{\bf 63}\oplus{\bf 70}$, two different contributions
make this to happen. ${\bf 63}$ is nothing but the $t^2$ order contribution
from the $SU(8)$ ADHM fields. ${\bf 70}$ comes from gauge invariant operators
of the $SU(8)$ ADHM fields and extra fields we added.
The next order $t^4$ contains the irrep. ${\bf 7371}$ of $E_7$, which is
rank $2$ symmetric product of ${\bf 133}$. We are missing some states
in $\hat{Z}^{\rm trial}_{k=1}$, which is ${\bf 70}\oplus{\bf 1}$,
the rank $4$ antisymmetric representation and a singlet of $SU(8)$.
If we blindly take the operators in our trial gauge theory which successfully
reproduced ${\bf 133}$, and take rank $2$ symmetric product of them, we find
that the representation ${\bf 70}\oplus{\bf 1}$ is missing due to the
compositeness of ${\bf 70}$ that appeared at $t^2$ order. This is like 
the quarks of QCD accounting for the plethora of gauge-invariant mesons at low
energy, with less microscopic degrees of freedom, while quarks manifest
themselves at high energy. So it appears that
we should add more extra fields to make up for these missing states.
We have found several possible combinations of extra supermultiplets
which one may add to the $SU(8)$ ADHM, satisfying very strong constraints
of the correct 2d anomaly. But we have not managed yet to construct the model which
exhibits the right Hilbert series of (\ref{E7-hilbert}).

Finally, while restricting to $SU(8)\subset E_7$ subset of moduli space,
we deleted one simple root so that we lost extra possibility of
embedding $SU(2)$ single instanton. Note that the deleted root has same
length as the roots kept in $SU(8)$, so that we might have lost extra
small instanton saddle points residing in the deleted $SU(2)$. This may
be related to the observations in the previous paragraph.
In any case, if the basic idea of this paper is applicable to other
exceptional instantons, we have many strong constraints which may eventually
guide us to the correct ADHM-like models. We hope to come back to this problem
in the future.

One important spirit of our construction is to take advantage of
symmetry enhancements of gauge theories after RG flow. One is the
enhancement of global symmetries from a classical group to an exceptional one.
Another is the SUSY enhancement to $\mathcal{N}=(0,4)$. Allowing
less number of SUSY in UV provides more room to
engineer the desired system. Being in 1d or 2d, we can have as little as
$1$ Hermitian SUSY in UV, the minimal number which admits any computation
relying on SUSY. Still the requirement of using gauge theories put some constraints.
However, we are not fully aware of whether we have overlooked the possibilities
of subtler supermultiplets or more general interactions, even within gauge theory.

It may also be interesting to further study the physics of exceptional instanton
strings from recent 4d $\mathcal{N}=1$ gauge theory descriptions for
$\mathcal{N}=2$ SCFTs, by suitably reducing them to 2d \cite{Gadde:2015xta,Maruyoshi:2016tqk}.
Many aspects of these constructions are different from ours.
For instance, different symmetries are manifest in UV,
so it may be helpful to compare the two approaches.

Some of our 1d/2d gauge theories are not tested with their
Witten indices and elliptic genera, simply because we have not thoroughly
thought about alternative approaches. The recent
engineering of 5d SCFTs (e.g. see \cite{Jefferson:2017ahm,DelZotto:2017pti})
will allow more geometric/brane realizations. We may be able to test our models
relying on these developments, perhaps using topological vertices.
Also, for studying 6d strings, one can use the topological string approach
(e.g. see \cite{Haghighat:2014vxa} and references therein) or the modular
bootstrap like approach \cite{DelZotto:2016pvm,DelZotto:2017mee,klp}.

The strategy of this paper was just to write down UV models and test them
empirically when data from alternative descriptions are available.
It will be nice to have a more conceptual understandings
of the ADHM-like models, either from string theory or by other means.
Here we feel that, compared to the simple Young diagram sums for the indices and
elliptic genera, the microscopic explanations in terms of $\mathcal{N}=(0,1)$ UV
gauge theories look less elegant, although practically useful
and flexible. In particular, at least at the moment, it is hard for
us to imagine a viable string theory engineering of our
gauge theories. It is not clear to us whether we are making UV uplifts
intrinsically beyond the territory of string theory, or whether there are
nicer reformulations which may allow string theory embeddings.

Finally, it will be interesting to see if our ADHM-like gauge theories can
be used to study other observables in the Coulomb branch. For instance,
study of the Wilson loops or other defect operators will be interesting.
See, e.g. \cite{Kim:2016qqs,Gaiotto:2014ina,Bullimore:2014awa,Gaiotto:2015una}
and references therein. As many of these constructions rely
on D-brane settings, one should see
if employing similar prescriptions without D-brane engineering will work
(as we did for the partition function in this paper).

\vskip 0.5cm

\hspace*{-0.8cm} {\bf\large Acknowledgements}
\vskip 0.2cm

\hspace*{-0.75cm} We thank Amihay Hanany, Kimyeong Lee, Gabi Zafrir, 
and especially Sung-Soo Kim and Kantaro Ohmori for helpful discussions.
HK is supported in part by NSF grant PHY-1067976. SK and KL are supported
in part by the National Research Foundation of Korea (NRF) Grant 2015R1A2A2A01003124.
JP is supported in part by the NRF Grant 2015R1A2A2A01007058.

\end{document}